\documentclass[conference]{IEEEtran}
\IEEEoverridecommandlockouts
\usepackage{cite}
\usepackage{amsmath,amssymb,amsfonts}
\usepackage{algorithmic}
\usepackage{graphicx}
\usepackage{textcomp}
\usepackage{xcolor}
\usepackage{hyperref}

\def\BibTeX{{\rm B\kern-.05em{\sc i\kern-.025em b}\kern-.08em
    T\kern-.1667em\lower.7ex\hbox{E}\kern-.125emX}}
\begin{document}

\title{Designing Artificial Intelligence Equipped Social Decentralized Autonomous Organizations for Tackling Sextortion Cases\\ Version 0.71
}

\author{\IEEEauthorblockN{Alex Norta}
\IEEEauthorblockA{\textit{Tallinn University, Tallinn, Estonia}\\
\textit{Dymaxion O{\"U}, Tallinn, Estonia} \\
alex.norta.phd@ieee.org}
\and
\IEEEauthorblockN{Sotiris Makrygiannis}
\IEEEauthorblockA{\textit{aialab.com} \\ 
\textit{Paphos, Cyprus}\\
sotiris@philotimo.eu}
}

\maketitle


\begin{abstract}
With the rapid diffusion of social networks in combination with mobile phones, a new social threat of sextortion has emerged, in which vulnerable young women are essentially blackmailed with their explicit shared multimedia content. The phenomenon of sextortion is now widely studied by psychologists, sociologists, criminologists, etc. The findings have been translated into scattered help from NGOs, specialized law enforcement units, and therapists, who usually do not coordinate their efforts among each other. This paper addresses the gap of lacking coordination systems to effectively and efficiently use modern information technologies that align the efforts of scattered and non-aligned sextortion help organizations. Consequently, this paper not only investigates the goals, incentives, and disincentives for a system design and development that not only governs effectively and efficiently diverse cases of sextortion victims, but also leverages artificial intelligence in a targeted manner. It explores how AI and, in particular, autonomous cognitive entities can improve victim profiles analysis, streamline support mechanisms, and provide intelligent insight into sextortion cases. Furthermore, the paper conceptually studies the extent to which such efforts can be monetized in a sustainable way. Following a novel design methodology for the design of trusted blockchain decentralized applications, the paper presents a set of conceptual requirements and system models based on which it is possible to deduce a best-practice technology stack for rapid implementation deployment.
\end{abstract}


\section{Introduction}
\label{sec:introduction}


Addressing the growing issue of online sexual exploitation and its consequences is becoming an ever more prominent issue. Briefly, sextortion~\cite{hendry2021sextortion} describes the abuse of power by those in authority who demand sexual favors in exchange for personal benefits. It is a global phenomenon that affects vulnerable individuals and has far-reaching implications for gender equity, democratic governance, economic development, and peace and stability. Very typically, sextortion incidents involve young women~\cite{amundsen2023turn} who share explicit pictures with a person of trust who then starts to threaten to openly publish the files unless the young woman engages in various acts of favor. As a recent case shows, also young men are vulnerable victims of sextortion to the point of tragically committing suicide\footnote{https://abcnews.go.com/US/2-suspects-accused-running-illegal-sextortion-ring-extradited/story?id=102236971}. 
Often, there is no suitable, quick and well-coordinated counseling, therapy, legal support, law enforcement available~\cite{alsoubai2022friends} to rescue sextortion victims before tragic events occur, such as suicide.


Given the sharp increase in sextortion cases worldwide~\cite{o2023short}, many studies~\cite{paradiso2023image} are conducted on this tragic issue to improve the understanding of this phenomenon. The current state of the art in related research for the domain of sextortion involves studies exploring the characteristics and dynamics of sextortion~\cite{finkelhor2023dynamics}, the psychological and emotional impacts on victims~\cite{champion2022examining}, the role of technology~\cite{gamez2022technology} in facilitating sextortion, and efforts toward prevention and intervention strategies~\cite{o2023minor}, including educational resources~\cite{doi:10.1177/10790632221145925}, reporting mechanisms~\cite{cross2023pay}, and support services~\cite{rajanikanth2023cyber}. 


The state of the art in sextortion research has focused on understanding the motivations and dynamics of sextortion, as well as exploring potential interventions to address the issue. However, there is a gap in the state of the art in investigating the degree to which modern technologies are utilized most effectively and efficiently to organize and coordinate psychological support, legal aid, and policing activities to quickly address and resolve sextortion incidents in the best interest of victims. Consequently, this paper addresses the gap by asking the main research question of how to employ modern advanced governance technologies for quickly investigating and responding in a multifaceted way to sextortion incidents in the best interest of the victim. To establish a separation of concerns and achieve a manageable complexity in answering the main question, we deduce three subquestions. What are the main goals and stakeholders affected by a sextortion emergency governance system? What are the static architecture components with blockchain use affiliated to the system goals? What is the dynamical system behavior involving a legally relevant set of transactions for swiftly coordinating victim releave? 
The arrangement of the subquestions form a logical line of reasoning with the aim of developing a diagnostic conceptual understanding for a sextortion governance system that is initially independent of detailed technology choices. 


The remaining paper structure\footnote{We emphasize that this technical report draft version is generated with the support of generative AI tools.} first gives a hypothetical running case in Section~\ref{sec:presuppositions} together with background literature. In Section~\ref{sec:goals}, we model the goals of the sextortion system and their associations with the affected stakeholders. Section~\ref{sec:architecture} gives the sextortion governance architecture model deduced from the goal model. Section~\ref{sec:dynamic} gives a set of important transactions that occur in the sextortion system, and subsequently these transactions are positioned into the dynamic information exchange protocols that occur in the architecture. 
Next, in Section~\ref{sec:evaluationdiscussion}, we present and discuss a proof-of-concept prototype for a feasibility evaluation of the results. Finally, Section~\ref{sec:conclusion} concludes this paper with conclusions and future work.

\section{Presuppositions}
\label{sec:presuppositions}

Presuppositions are assumptions that are taken for granted in a given context. They are often implicit and can be difficult to identify, yet they are essential for understanding the meaning of a text. Thus, we first present the running case to evaluate the cross-blockchain connection in Section~\ref{sec:runningcase}. Next, Section~\ref{sec:background} provides more literature with concepts that are relevant to follow the subsequent sections.
 
\subsection{Running case}
\label{sec:runningcase} 
We depict conceptually below in Figure~\ref{fig:runningcase} a running case for the sextortion process that consists of three phases. The sequence of events outlined in Phase 1 of Figure~\ref{sec:runningcase} within a sextortion case can exhibit variability based on specific circumstances, yet several common elements are recurrent in many instances. Initially, there is the phase of initial contact, where the victim and perpetrator might have connected online or held some preexisting relationship. The perpetrator often gains the victim's trust by assuming a false identity, often posing as a romantic interest or friend. Subsequently, a request for sexually explicit content may emerge, ranging from direct appeals to manipulation tactics, potentially involving threats against the victim or their loved ones. Once explicit material is shared, the perpetrator might use it to force more content or favors. This could escalate into increasingly demanding or menacing requests, causing the victim to feel trapped. In some cases, blackmail enters the equation, as the perpetrator threatens to disseminate explicit content to the victim's social circle unless their demands are met. The ramifications can be profound, inducing severe emotional turmoil such as anxiety, depression, and even suicidality, and in extreme scenarios, physical harm might occur, such as self-harm or injury. Reporting such cases is hindered by fear or shame, but when victims report, legal action can be taken against the perpetrator. Importantly, the delineated sequence of events is not universal, and some cases involve multiple culprits or alternative forms of exploitation. Furthermore, victims are not exclusively minors or individuals with vulnerabilities; adults can also be subjected to sextortion.

\begin{figure*}[htpb]
    \vspace{0.2cm}
    \begin{center}
        \includegraphics[scale=0.69]{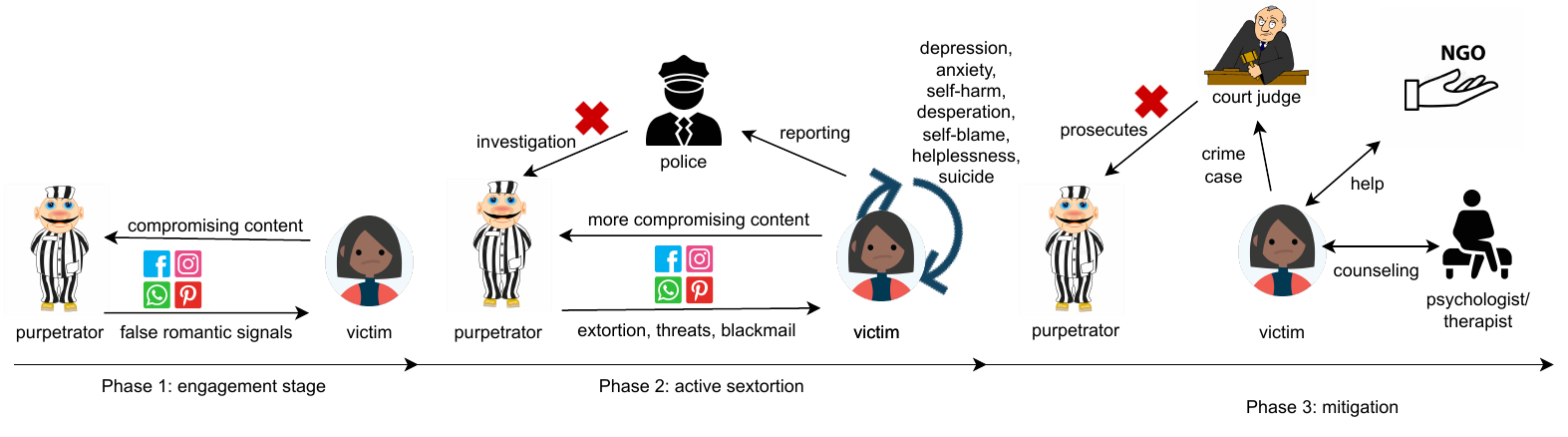}
        \caption{A hypothetical scenario of sextortion divided into three phases.}
        \label{fig:runningcase}
    \end{center}
    \vspace{-0.5cm}
\end{figure*}

Despite the significant harm caused by sextortion, victims often encounter obstacles in receiving adequate assistance, a scenario depicted in Phase 2 of Figure~\ref{sec:runningcase}. Several reasons contribute to this as follows. Stigma and shame cast a shadow, and victims grapple with feelings of embarrassment and humiliation about their ordeal. This emotional burden may deter them from seeking help, driven by fears of being judged or ostracized by others. The intricate nature of sextortion, characterized by a gradual escalation of demands and manipulative tactics, complicates victim recognition of exploitation and impedes identification and appropriate response from others. A lack of awareness compounds the issue, affecting not only victims but also law enforcement personnel and service providers who may be unfamiliar with the dynamics of sextortion and thus lack the tools to offer effective assistance and support. Establishing the crime's occurrence can be intricate; even when victims report sextortion, evidential requirements can be extensive and challenging to meet. This complexity in proving the crime can impede the prosecution of perpetrators and discourage victims from reporting. The fear of retaliatory actions is high. Victims may worry that reporting the crime could lead the perpetrator to retaliate by disseminating explicit material to others, magnifying their distress. Additionally, victims often struggle with limited access to support services, compounded by a lack of awareness of where to seek help in such situations.

Within the running case of Figure~\ref{fig:runningcase}, Phase 3 delineates the common avenues of assistance available to victims of sextortion, encompassing the following: First, reaching out to local law enforcement is a crucial step, which involves reporting the crime to the police. Second, victims may find comfort in seeking support from a counselor or therapist, which fosters emotional resilience in the aftermath of the incident. Third, specialized organizations designed to help victims of sextortion, such as the National Center for Missing and Exploited Children (NCMEC), are potential allies. If the victim is under 18 years old, the participation in the CyberTipline of NCMEC can be especially relevant. Fourth, legal consultation with an attorney could be warranted, guiding the victim through the complexities of the legal system while safeguarding his rights. Importantly, the individuals and entities offering help must approach the situation with compassion and without judgment, collaborating with the victim to construct strategies for regaining control. Therefore, sextortion is a complex and often unreported crime that inflicts severe emotional turmoil and physical harm. To better champion sextortion victims, it is imperative to address underlying obstacles such as stigma, ignorance, and fear of reprisal. Equally important is ensuring that victims have access to appropriate support services. Additionally, increasing social awareness and fostering proactive actions to curb such crimes are vital elements of a comprehensive approach to addressing sextortion.

\subsection{Background Literature}
\label{sec:background} 

We begin by discussing the background literature related to sextortion in Section~\ref{sec:sexbackground}, then move on to the technology-oriented background in Section~\ref{sec:techbackground}.

\subsubsection{Sextortion Background Literature}
\label{sec:sexbackground} 

Sextortion is a form of online sexual exploitation in which individuals, usually minors, are blackmailed or bullied into providing sexually explicit material or performing sexual acts. This material is then used as leverage to extract more material or favors from the victim. Sextortion is often carried out by individuals with whom the victim has had prior contact, such as an online acquaintance or a romantic partner, and can lead to significant emotional distress and physical harm.
A notable case of sextortion is that of Amanda Todd\footnote{https://www.bbc.com/news/world-us-canada-62326780}, a Canadian teenager who committed suicide in 2012 after being bullied and blackmailed about explicit material she had shared online. Her story brought attention to the problem of sextortion and the devastating consequences it can have on its victims.
Research on sextortion~\cite{hong2020digital} shows that it often involves a power imbalance between the perpetrator and the victim and that victims can be particularly vulnerable due to their age or a history of exploitation. Studies~\cite{nilsson2019understanding,walsh2022if} have also found that sextortion can lead to a variety of negative outcomes, including anxiety, depression, and suicidality.
A study~\cite{wolak2018sextortion} found that victims of sextortion tend to be young, with the majority between the ages of 12 and 17. They also tend to have a prior relationship with the perpetrator and may have initially consented to sharing the explicit material.
Another study~\cite{ojeda2022lines} finds that sextortion prevention and intervention efforts should focus on education and awareness raising, as well as on empowering victims to report abuse and seek support.

\subsubsection{Technology Background Literature}
\label{sec:techbackground} 



For the dApp design, security is an important aspect given the sensitive nature of the sextortion-case related information. Thus, the aim is to integrate a multifactor challenge set self-sovereign identity authentication (MFSSIA)~\cite{nortablockchain22} system.

\begin{figure}[htpb]
    \vspace{0.2cm}
    \begin{center}
        \includegraphics[scale=0.35]{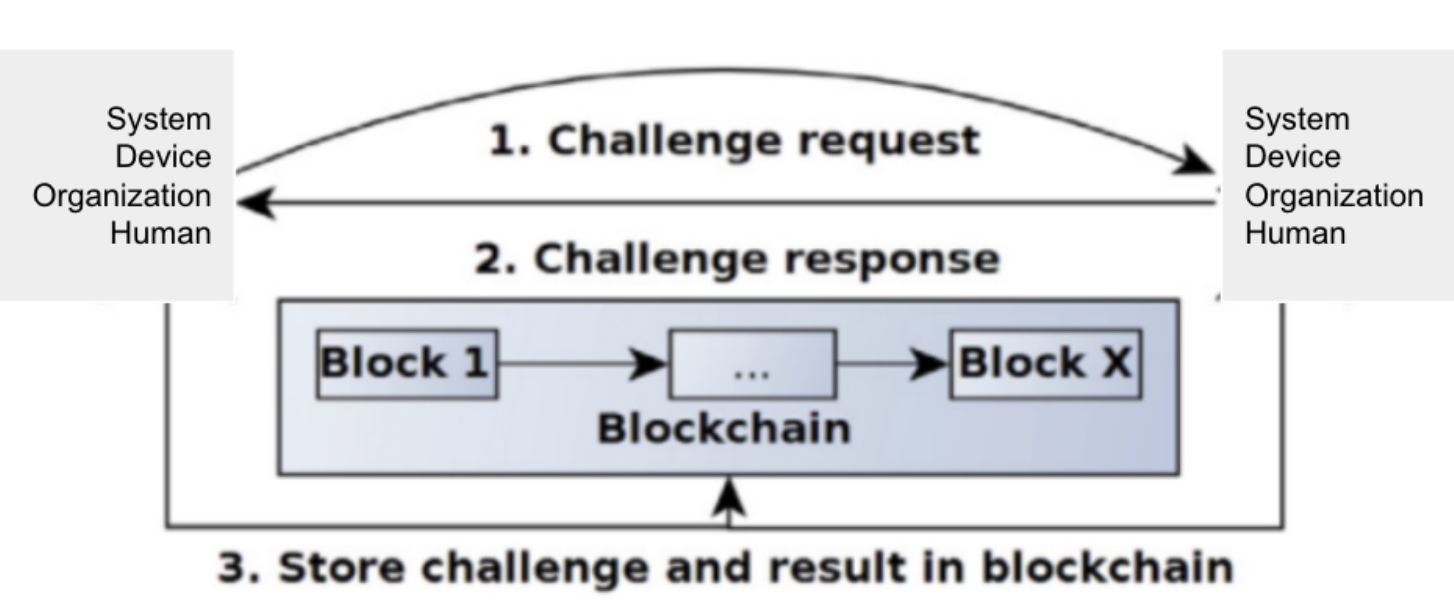}
        \caption{Conceptual depiction of the MFSSIA lifecycle for challenge-resposne management (see~\cite{nortablockchain22}).}
        \label{fig:mfssialifecycle}
    \end{center}
    \vspace{-0.5cm}
\end{figure}



\noindent The representation of the MFSSIA life cycle, as illustrated in Figure~\ref{fig:mfssialifecycle}, underscores the dynamic nature of identity authentication. It is important to note that challenges and their respective responses can have diverse origins, spanning systems, devices (e.g., mobile devices), organizations, or individuals. To maintain an unchangeable record of the entire lifecycle, both the challenge set and the corresponding responses are stored on a blockchain. The evaluation of responses subsequently decides the success or failure of identity authentication within a particular context. If responses are found to be unsatisfactory, it leads to the conclusion of a communication to prevent insecure information transfer.



It should be noted that, in addition to challenge-response life-cycle management, MFSSIA incorporates two other essential blockchain technologies. According to Figure~\ref{fig:mfssialifecycle}, MFSSIA utilizes blockchain decentralized knowledge graphs (DKG)~\cite{vide2021designing,wang2019decentralized} to create a unified view in business collaborations where data sets are distributed and diverse. The integrated DKG in MFSSIA is an open and collaborative network that combines semantic data networks with blockchain technology to form an immutable trust network. The DKG employed\footnote{https://docs.origintrail.io/} facilitates the integration of knowledge graphs and utilizes blockchain technology for improved trust and traceability.


In MFSSIA, blockchain technology is further integrated through the use of oracles. These oracles bridge the gap between smart contract blockchain systems and external data sources. Still, this integration yields the "oracle problem" that is the risk of exposing trusted blockchains to the untrustworthy off-chain data sources. To address this issue, MFSSIA implements decentralized oracles that provide users with increased trust assurance. These oracles require human or nonhuman agents, who are computing entities providing computing resources, to adhere to a Proof-of-Contribution (PoCo) consensus protocol~\cite{ding2020blockchain} to validate their contributions.

\begin{figure}[htpb]
    \vspace{0.2cm}
    \begin{center}
        \includegraphics[scale=0.2]{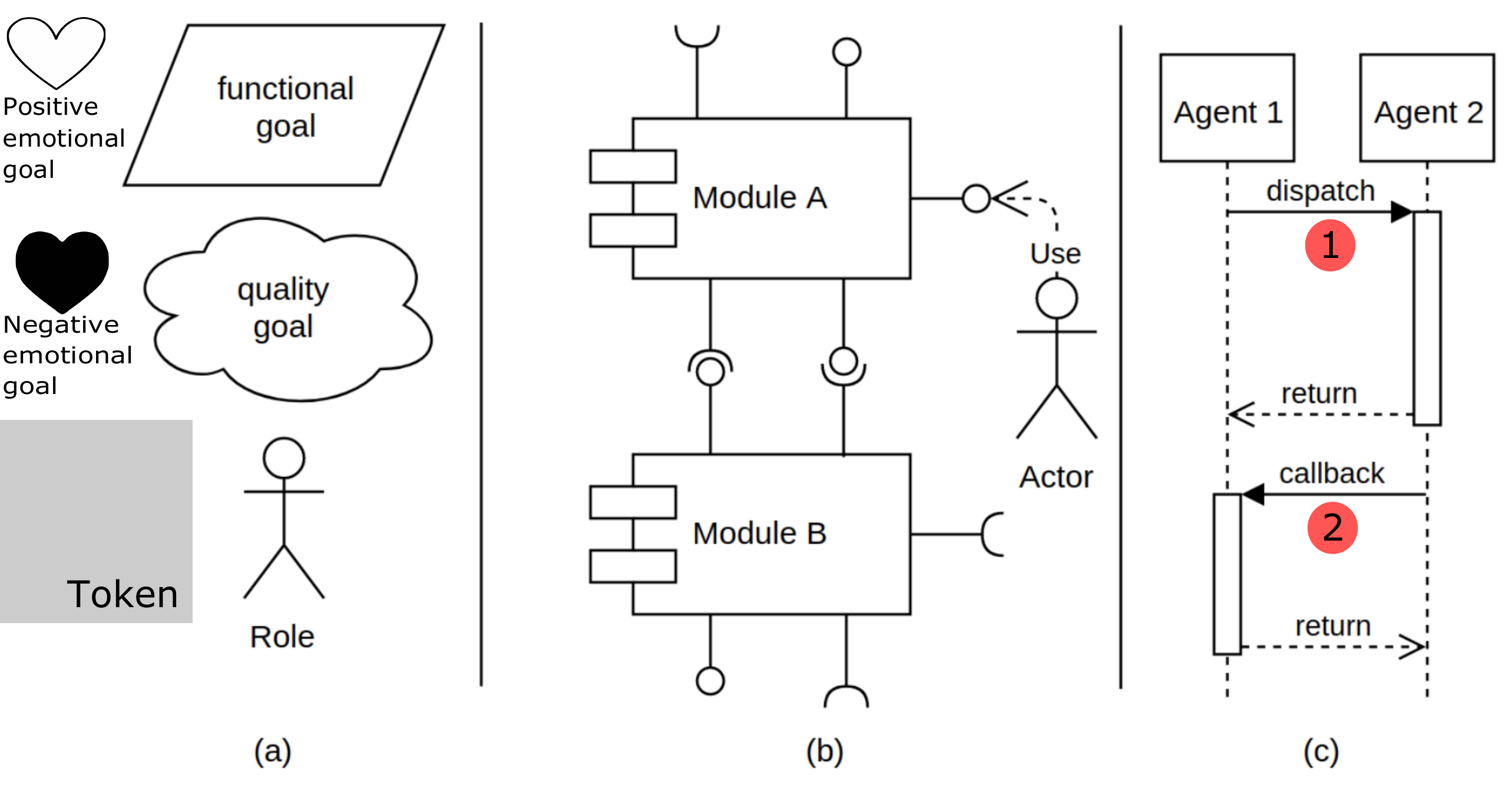}
        \caption{Goal-model notation in (a), followed by component-diagram notation in (c) and finally, BPMN notation in (c).}
        \label{fig:daomnotation}
    \end{center}
    \vspace{-0.5cm}
\end{figure}

This paper seeks to introduce a dApp system that follows the principles of distributed blockchain applications~\cite{udokwu2018exploration,udokwu2018state,udokwu2020evaluation,udokwu2021deriving,udokwu2021designing}. The approach used is a multi-modeling notation, as illustrated in Figure~\ref{fig:daomnotation}, which displays three different notations in a row.

\noindent In this study, goal models are used, which are part of the agent-oriented modeling (AOM) approach~\cite{sterling2009art}, to specify the requirements of the dApp. Figure~\ref{fig:daomnotation}(a) of our notation displays the goal models consisting of functional goals represented as parallelograms, "quality goals" or synonymously non-functional requirements depicted as clouds, and agents with designated roles, whether human or artificial, represented as stick figures. Additionally, gray boxes with labels are used to enclose certain functional goals, indicating the parts of the system that involve the implementation of blockchain technology. The label within each gray box denotes the token name that serves as the basis for the subsequent development of a token economy. Briefly, a token economy~\cite{davis2023token} is a system of behavior modification and reinforcement that uses crypto tokens or symbolic rewards as a form of positive reinforcement for desired behaviors. It is often used in various settings, including education, therapy, and institutional settings, to encourage people to exhibit specific behaviors or to discourage undesirable ones. 

For agents in Figure~\ref{fig:daomnotation}, the label assigns a role to show if it is filled by a human, while the label $<$ AI agent$>$ with a role label underneath indicates an artificial agent. Furthermore, the specific label $<$Oracle$>$ is used to indicate a connection between a specific oracle type and a functional goal. The root of the hierarchically decomposed goal model is a 'value proposition', represented as a functional goal that signifies the overall goal of the system. Quality goals and roles are attached to functional goals at various levels within the hierarchy. Finally, the hearts at the left top of Figure~\ref{fig:daomnotation}(a) denote positive and negative emotional goals that are assigned to roles in their correlation with respective functional goals. Therefore, the functional goal must ensure that positive emotional goals are reinforced and negative emotional goals are mitigated.  


We use a simple mapping calculation to derive a UML component diagram~\cite{uml01} from the goal model to describe the static architectural structure of the social DAO system employing the trustable dApp modeling (T-DM) method~\cite{udokwu2022modellingphd}. Figure~\ref{fig:daomnotation}(b) of the T-DM notation shows the UML elements, with the components represented as labeled rectangles and equipped with the provided and required interfaces. The required interfaces are represented as lines with a circle, whereas the provided interfaces are depicted as lines with a cup at the end. Component diagrams are created from functional goals, and components related to blockchain technology and specific token types are shaded gray to indicate their implementation. In addition, the goal model agents are also included in the component diagram. Component diagrams also specify the data-exchange channels between components and between components and agents. Further details on the T-DM method and the mapping heuristics from goal models to component diagrams and sequence diagrams can be found in~\cite{udokwu2022modellingphd}.


To illustrate the dynamic behavior of the social DAO system, we employ UML sequence diagrams~\cite{alhazmi2021learning}. Figure~\ref{fig:daomnotation}(c) of our T-DM notation shows two entities, Agent 1 and Agent 2, with their timelines represented by dashed lines that decrease. The communication between them is represented by straight arrows, and the activation bars indicate processing threads. Additionally, red circles with numerical values are displayed along the arrows, which signify on-chain transactions that are included in the sequence diagram.

Finally, we aim to employ AI agents for the multi-faceted interaction with the teenagers, victims, and so on. Conceptually, these AI agents are a sophisticated digital entity designed to operate with high levels of autonomy. The goal of these AI agents is to outsource human labor and perform tasks that require complex cognition, strategic decision-making, and ethical alignment. 

\begin{figure}[htpb]
    \vspace{0.2cm}
    \begin{center}
        \includegraphics[scale=0.18]{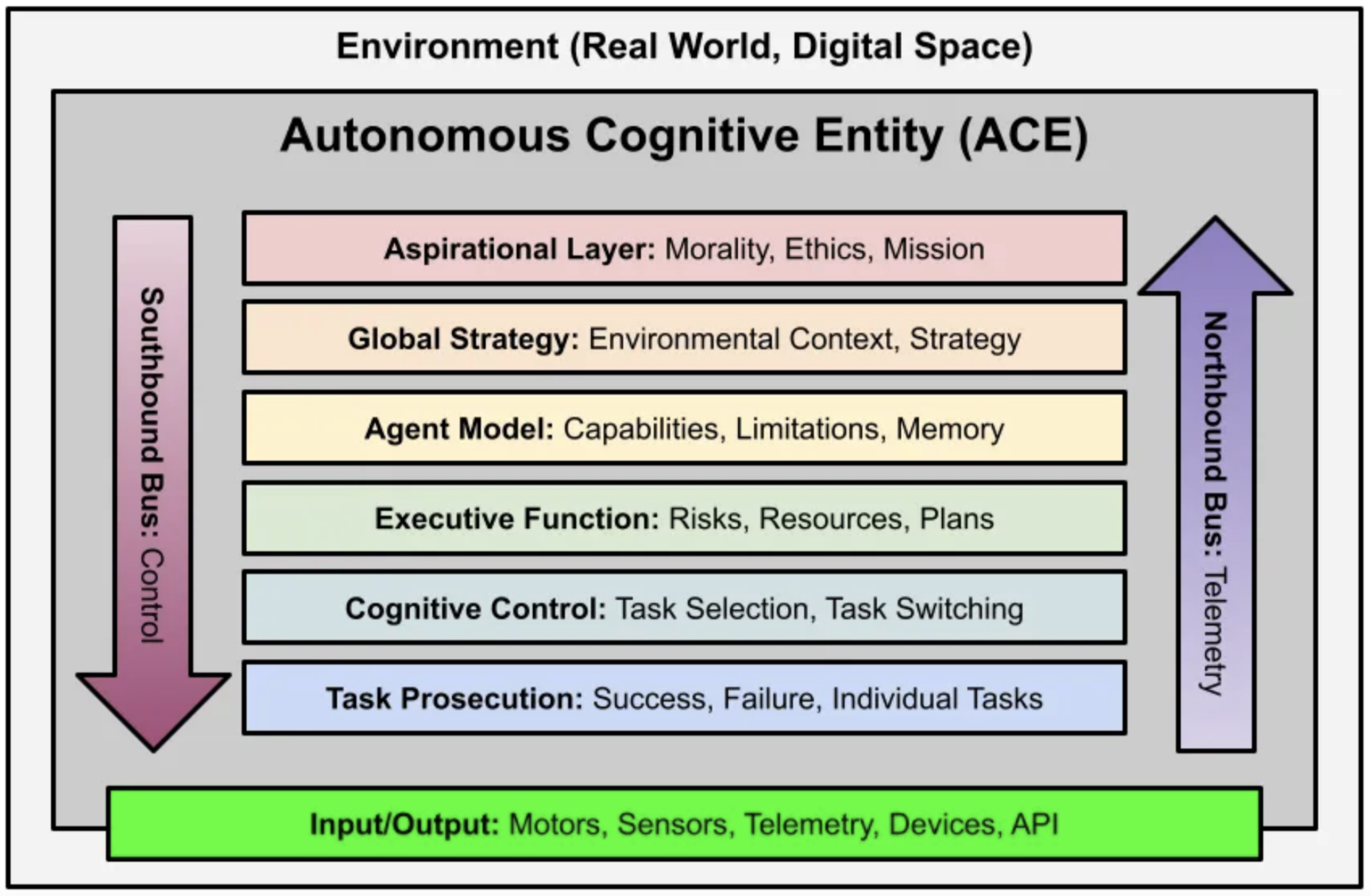}
        \caption{The layered ACE framework into which AI agents are embedded (Source: \url{https://github.com/daveshap/ACE_Framework}).}
        \label{fig:aceframework}
    \end{center}
    \vspace{-0.5cm}
\end{figure}

The AI agents are part of the framework of autonomous cognitive entities (ACE)\footnote{https://medium.com/@dave-shap/autonomous-agents-are-here-introducing-the-ace-framework-a180af15d57c} that provides a structured layered architecture to achieve this, consisting of six hierarchical layers, as Figure~\ref{fig:aceframework} shows. Briefly, the Aspirational Layer acts as an ethical compass, aligning the agent's actions with a constitution based on ethical principles and human values. The Global Strategy Layer uses real-world data to contextualize high-level goals and strategic plans. The Agent Model Layer maintains a self-model that outlines the agent's capabilities and limitations, enabling it to reason effectively. The Executive Function Layer translates these strategic directions into detailed project plans and allocates resources accordingly. The Cognitive Control Layer dynamically selects and switches tasks based on environmental and internal state in real time. Finally, the Task Prosecution Layer executes these tasks, either digitally or physically, while monitoring the results to determine success or failure.

\section{The Sextortion Governance Goals and Stakeholders}
\label{sec:goals}

The goal of sextortion governance is to ensure that all stakeholders are aware of the risks associated with sextortion and are able to take appropriate measures to protect themselves and their data. This includes understanding the different types of sextortion, the potential consequences of sextortion, and the various stakeholders involved in the governance of sextortio cases. By understanding the goals and stakeholders of sextortion governance, organizations can better protect themselves and their data from malicious actors. Therefore, in Section~\ref{sec:valuepropqual}, we give the value proposition and define the associated
quality goals in the context of this paper. Next, Section~\ref{sec:goalrefine} presents
the further details of the goal model to refine the value
proposition. Note that the explanations for the value proposition and the first of the four contained refining goal models are more expanded and detailed to convey the intricacies of this SocialDAO dApp conceptually. The remaining three subsequent refinement goal models are described in a more compact way, as the intricacies of the value proposition and the first goal model repeat themselves. Finally, Section~\ref{sec:tokeneconfound} explains the foundational aspects of the token economy specified in the goal model. 

\subsection{Root Value Proposition and Associated Quality Goals}
\label{sec:valuepropqual}

\begin{figure}[htpb]
    \vspace{0.2cm}
    \begin{center}
        \includegraphics[scale=0.69]{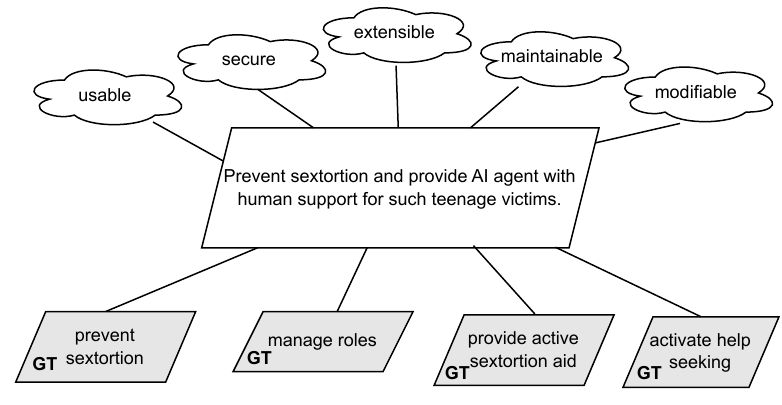}
        \caption{The root value proposition model and associated quality goals, including the first-level functional-goal refinements.}
        \label{fig:rootvalue}
    \end{center}
    \vspace{-0.5cm}
\end{figure}

The purpose of the goal model in Figure~\ref{fig:rootvalue} is to prevent sextortion and support teenage victims. This value proposition is represented by a root parallelogram in the center of the figure, and the assigned quality goals are symbolized by clouds. This root association implies that the quality goals are applicable to the entire refinement of the goal model discussed in this section that comprises properties that partially intersect.

\noindent We briefly explain the respective meanings of the value proposition-associated quality goals in the context of SocialDAO tackling sextortion. 

\textit{Usable} refers to the system's ability to provide an intuitive, user-friendly, and efficient user experience. Usability is a crucial aspect of software design and development, as it directly impacts the ease with which users can interact and navigate the application. Several key aspects of usability are explained next. The dApp should have an intuitive user interface (UI) and user experience (UX), meaning that users can easily understand how to use the application without the need for extensive training or guidance. Elements such as clear navigation, straightforward workflows, and familiar design patterns contribute to intuitiveness. Usability also involves ensuring that users can perform their tasks efficiently. This means minimizing unnecessary steps, reducing the time required to complete actions, and optimizing the overall workflow. The application should not require users to perform complex or time-consuming processes to achieve their goals. Usability extends to ensuring that the dApp is accessible to users with diverse needs, including those with disabilities. This includes providing features for screen readers, keyboard navigation, and other accessibility enhancements to make the application inclusive of all users. The dApp's design and behavior should be consistent throughout the application. Consistency in the placement of buttons, the use of terminology, and the overall look and feel of the UI helps users feel more comfortable and confident when using the system. The application should provide clear feedback to users about the results of their actions. Additionally, it should handle errors gracefully by providing informative error messages and guidance on how to resolve issues. Usability is achieved through a user-centered design approach, where the preferences, needs, and feedback of end-users are taken into account during the development process. User testing and feedback collection are essential for refining the dApp's usability. As the dApp is designed for mobile devices, usability also involves ensuring that the application is responsive and functions well on various screen sizes and orientations. While usability primarily focuses on functionality and ease of use, the visual design and aesthetics of the dApp should also be appealing to users, making them more likely to engage with the application.

\textit{Secure} implies that the application must prioritize robust measures to protect user data, privacy, and overall integrity of the system. Security in this context encompasses several key aspects as follows. The dApp must employ strong encryption and access controls to protect user data from unauthorized access, breaches, or data leaks. Personal and sensitive information related to sextortion cases, user profiles, and communications must be stored and transmitted securely.  Secure user authentication methods, such as MFSSIA, should be implemented to ensure that only authorized users have access to the application. Proper authorization mechanisms must be in place to restrict users' actions based on their roles and permissions. The dApp must adhere to data protection regulations, especially in the European Union (e.g., GDPR\footnote{https://gdpr.eu/}). This includes obtaining explicit consent for data processing, allowing users to exercise their data rights, and providing transparent data handling practices. All communication between users and the dApp, as well as any external services or APIs, should be encrypted and protected against eavesdropping or intercept. Secure communication protocols such as HTTPS should be enforced. Continuous security assessments, code reviews, and vulnerability scans should be conducted to identify and address potential security weaknesses. Regular software updates and patch management are essential to mitigate known vulnerabilities. Measures should be in place to protect the dApp from Distributed Denial of Service (DDoS) attacks, ensuring its availability even during high-traffic or attack scenarios. If the dApp involves smart contracts on a blockchain, these contracts should undergo thorough security audits to prevent vulnerabilities or exploits that could compromise user funds or the integrity of the system. A well-defined incident response plan must be in place to address security breaches or incidents promptly, minimize impact, and notify affected parties as required by law. The dApp should have policies in place for the retention and deletion of user data to ensure that the data are not stored longer than necessary. Staff and stakeholders involved in the development and operation of the dApp should receive regular security training and awareness programs to mitigate human-related security risks.

\textit{Extensible} implies that the application must be designed and built in a way that allows for easy and seamless expansion of its functionality, features, and capabilities in the future. This extensibility is essential to accommodate evolving user needs, emerging technologies, and potential changes in the scope of the application. Key aspects of extensibility in this context include the following. The dApp's architecture should be modular, allowing for the addition or removal of modules, components, or plugins without disrupting the core functionality. New features can be developed as separate modules and integrated when needed. The dApp should provide well-documented and flexible APIs that enable external services, applications, or third-party developers to integrate with the system easily. This promotes interoperability and future integrations. The dApp's infrastructure should be designed to scale horizontally or vertically to handle increased user load or additional features. This ensures that the system can grow as user demand or requirements expand. Users or administrators should have the ability to customize certain aspects of the dApp to tailor it to their specific needs or preferences. This may include configuring user interface elements or adjusting settings.  The dApp should support smooth and secure software updates and version upgrades. This ensures that users can benefit from new features, bug fixes, and security enhancements without major disruptions. The dApp should anticipate potential technological advancements or changes in user requirements. Design choices should avoid unnecessary constraints that could hinder future development. Comprehensive and up-to-date documentation should be available to guide developers and administrators in extending or customizing the dApp's functionality. New extensions or features should be designed with compatibility in mind, ensuring that they do not negatively impact existing functionality or disrupt user experiences. 

\textit{Maintainable} refers to the ability of the application to be easily and cost-effectively maintained and updated over its lifecycle. It encompasses several key aspects that contribute to the application's maintainability. The dApp's source code should be well-structured, clean, and adhere to coding standards and best practices. High-quality code is easier to understand, modify, and maintain. Comprehensive and up-to-date documentation should be provided for both the codebase and the system architecture. This documentation assists developers, administrators, and maintainers in understanding and troubleshooting the application. The dApp should be designed with a modular architecture that allows for the isolation of different components. This modularity simplifies maintenance by allowing developers to update or replace specific modules without affecting the entire system. Source code should be managed using version control systems such as Git, which facilitates tracking changes, collaboration among developers, and the ability to revert to previous versions if necessary. Robust testing practices should be in place, including unit testing, integration testing, and automated testing. Automated testing helps identify and address issues quickly and efficiently during maintenance. A system should be established to track and manage bugs, issues, and feature requests. This allows for organized prioritization and resolution of problems. The dApp's architecture should allow for easy scalability to accommodate increased user load or growth in data volume without requiring significant code changes. Dependencies on external libraries and components should be well documented and kept up to date. This ensures that security vulnerabilities are addressed promptly. Regularly apply security updates and patches to the dApp's dependencies and underlying infrastructure to mitigate vulnerabilities and protect against security threats. Code should include meaningful comments and annotations to explain complex or critical sections, making it easier for developers to understand and modify the code during maintenance.  Implement formal change management processes to assess and approve proposed changes, updates, or enhancements to the dApp to minimize risks and disruptions. Ensure knowledge transfer within the development team and to new team members to maintain continuity in maintaining and improving the dApp. 

\textit{Modifiable} refers to the application's ability to undergo changes, updates, and enhancements with relative ease and minimal disruption. Modifiability is essential to ensure that the dApp can adapt to evolving user requirements, technological advancements, regulatory changes, and emerging threats. Key aspects of modifiability in this context include the following. The dApp's architecture and design should be flexible and open to changes. It should allow developers to introduce new features, components, or functionalities without requiring a complete system overhaul. A modular structure must be in place, allowing developers to add, remove, or modify modules or components independently, reducing the risk of unintended consequences. The dApp should offer well-defined and documented APIs that enable seamless integration with external systems, services, or plugins. This promotes extensibility and the ability to easily add new integrations. The dApp should separate concerns, such as presentation, business logic, and data storage, to facilitate changes in one area without affecting others. Changes and updates should be designed with backward compatibility in mind to ensure that existing users and data are not negatively impacted. Well-maintained, high-quality code with appropriate comments and documentation contributes to modifiability, making it easier for developers to understand and modify the codebase. Using version control systems allows for tracking changes, managing code branches, and collaborating on modifications while maintaining a stable version of the application. Implementing continuous integration and continuous deployment (CI/CD) pipelines streamlines the process of testing, validating, and deploying changes quickly and efficiently. Comprehensive testing practices, including automated testing, ensure that modifications do not introduce defects or regressions. Automated testing allows for rapid feedback on changes. It is prudent to establish formal change management processes to evaluate, prioritize, and approve proposed modifications or enhancements based on their impact and importance. The development team should also maintain comprehensive documentation that reflects the current state of the dApp, including design decisions, APIs, and best practices for making modifications. These measures facilitate knowledge transfer within the development team to ensure that team members are well equipped to make and understand modifications.

In Figure~\ref{fig:rootvalue}, the value proposition is further refined at the first level with functional goals to establish a separation of concerns. These functional goals are \textit{prevent sextortion}, \textit{manage roles}, \textit{provide active sextortion aid}, and \textit{activate help seeking}. These functional goal in their refinements are explained below.

\subsection{Further Goal-Model Refinements}
\label{sec:goalrefine}

For the four functional goals of the first-level value proposition that refine functional goals, we first explain the refinement hierarchy, followed by expanding on the assigned quality goals and roles. Next, the emotional goals are explored, and finally, the token economy is presented.

\subsubsection{Functional Goal Prevent Sextortion}
\label{sec:preventsext}

\begin{figure*}[htpb]
    \vspace{0.2cm}
    \begin{center}
        \includegraphics[scale=0.55]{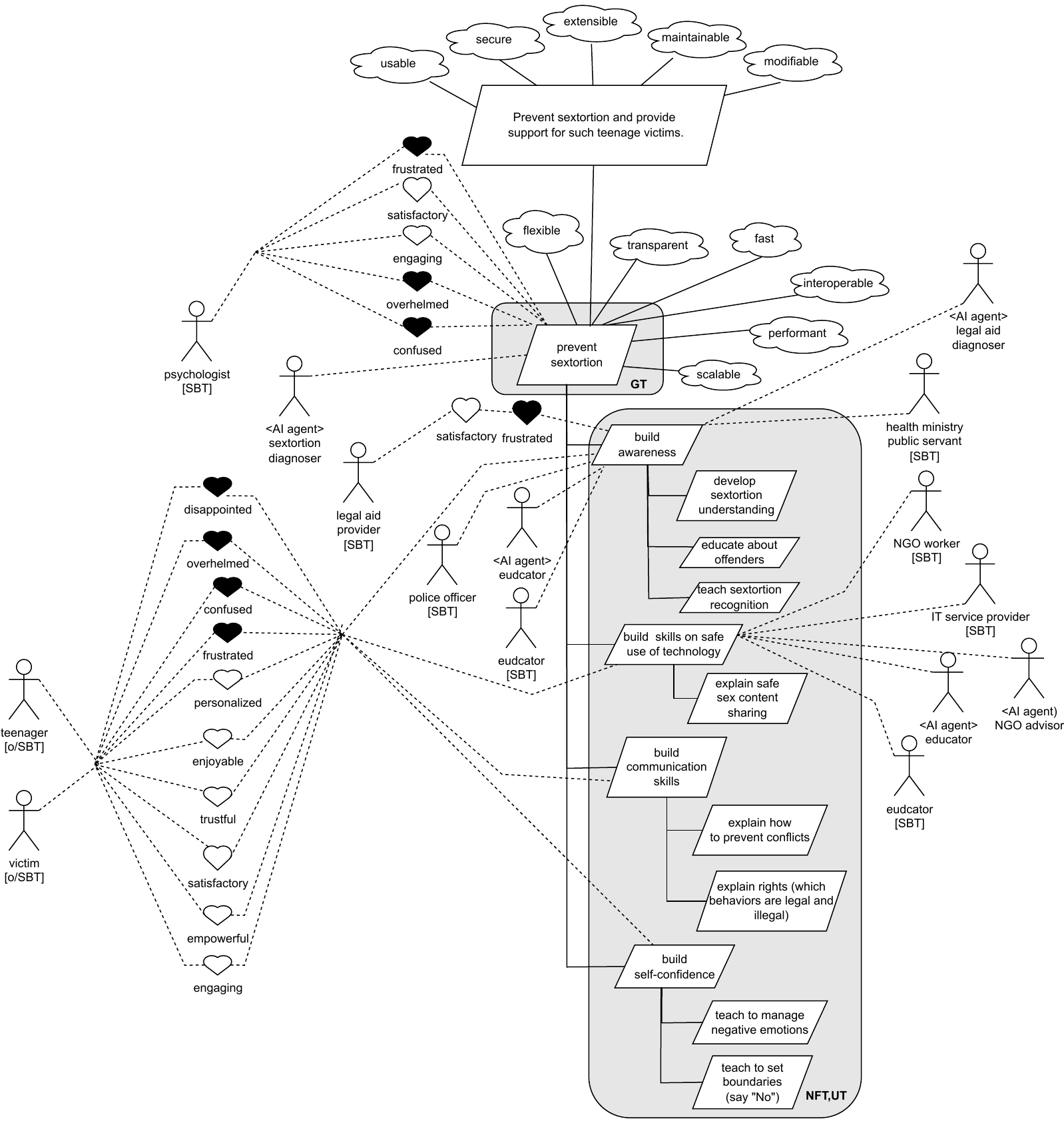}
        \caption{The goal-model refinement for preventing sextortion.}
        \label{fig:goalpreventsextortion}
    \end{center}
    \vspace{-0.5cm}
\end{figure*}

The goal model in Figure~\ref{fig:goalpreventsextortion} shows the refinement of the functional goal \textit{prevent sextortion} to which several quality goals are assigned, as we explain below. 

\textit{Flexible} means that the system's prevention mechanisms and strategies are adaptable and versatile to address different sextortion scenarios and evolving threats. A more detailed explanation means that the system can adapt its preventive measures based on the specific characteristics of each sextortion case. It can recognize and respond to various tactics used by perpetrators, taking into account the unique circumstances of each victim. The system is designed to handle different scales of sextortion incidents, from individual cases to larger-scale threats. It can scale its preventive efforts up or down as needed to effectively address the scope of the problem.  The roles involved can customize and configure preventive measures to suit their specific needs and preferences. This customization allows for tailoring the prevention approach to the unique requirements of different situations. The system can receive and integrate real-time updates and threat intelligence to stay up to date with the latest trends and sextortion tactics. It can adjust its prevention strategies based on this up-to-date information.  Instead of relying on a single fixed method, the system employs a multifaceted approach to prevention. Combining various techniques, such as user education, content analysis, threat monitoring, and reporting mechanisms, sextortion is effectively addressed from different angles. The system can seamlessly integrate with other relevant technologies and organizations involved in the prevention of sextortion. It can share data and coordinate efforts with external entities to improve its flexibility in sextortion control. Finally, the system is designed for ongoing improvement and refinement. It can incorporate feedback and lessons learned from previous sextortion cases to improve its preventive measures over time.

\textit{Transparent} means that the prevention efforts and systems processes are open, clear, and easily understandable to both users and stakeholders. The prevention mechanisms and strategies of the system are presented in a clear and straightforward manner. Users can easily understand how the system works to prevent sextortion, the steps involved, and the rationale behind its actions. The system is transparent about its data collection, analysis, and usage practices. Users are informed about what data is collected, why it is collected, and how it is used for prevention purposes. This includes explicit consent and disclosure of data handling practices, especially if personal information is involved. The system provides access to its policies, guidelines, and rules related to sextortion prevention. These policies are readily available for users to review, ensuring that they understand the system's approach to addressing sextortion. The system has mechanisms in place to track and report on its prevention activities. Users can see the results of prevention efforts, such as the number of incidents prevented or actions taken to address threats. Users have the means to provide feedback and report concerns about the system's prevention measures. The system communicates its response to reported incidents, ensuring transparency in addressing user concerns. The system is clear about how it handles user data, especially in cases where reporting or evidence collection is involved. Users understand how their data is processed, stored, and secured in the context of sextortion prevention. The system maintains comprehensive audit trails that record its prevention activities. These logs are accessible to authorized parties and provide transparency into the system's actions and decision-making processes. If the system integrates with external services or organizations for prevention purposes, these integrations are transparently disclosed. Users are informed about the roles and responsibilities of these third parties in preventing sextortion. The system offers educational resources to help users understand sextortion prevention, recognize potential threats, and take proactive measures. These resources contribute to user awareness and empowerment.

\textit{Fast} means that the system's prevention measures and response times are rapid and efficient in addressing threats to sextortion and protecting potential victims. The system can quickly identify and detect threats of extortion, including suspicious activities or content that may indicate a potential case. This detection should occur in near real time or without significant delay. Once a threat of sextortion is detected, the system responds promptly to mitigate the risk. This may include blocking or flagging harmful content, initiating protective measures, or notifying relevant authorities. The system continuously monitors sextortion-related activities and updates its threat assessment in real time. It does not rely on periodic or manual checks but actively watches for signs of sextortion. Users should be able to report sextortion incidents quickly and easily through the system. The reporting process is streamlined to ensure that potential victims can seek help without delay. When possible, the system automates actions to prevent sextortion or provide immediate support. For example, it can automatically block or filter explicit content or provide resources and guidance to potential victims. The system facilitates rapid communication between users, support organizations, and authorities involved in sextortion prevention. Communication channels are designed to minimize response times. The system is optimized for performance, ensuring that it can handle large volumes of data and user interactions without becoming sluggish or unresponsive. In cases where real-time communication is critical, such as crisis intervention, the system minimizes latency to ensure that responses are almost instantaneous. Resources, such as computing power and network bandwidth, are allocated efficiently to prioritize and accelerate prevention efforts. Users receive timely notifications and updates regarding the status of their reported sextortion incidents or the progress of prevention actions.

\textit{Interoperable} denotes that the system is designed to work seamlessly and exchange information with other relevant systems, organizations, or stakeholders involved in the prevention of sextortion. Interoperability ensures that the system can collaborate effectively with external entities, share data, and coordinate efforts to improve sextortion prevention. Thus, the system can exchange data, information, and reports with other organizations, such as law enforcement agencies, victim support groups, and relevant government entities, to facilitate coordinated efforts to prevent sextortion. It uses standardized data formats and protocols that are widely accepted and recognized within the sextortion prevention community, making it easier to share and interpret information. The system provides well-documented and accessible application programming interfaces (API) that allow external systems and services to integrate with it. This enables seamless data flow and communication between the system and external stakeholders. It is compatible with different platforms and technologies used by various organizations involved in sextortion prevention, ensuring that data can be exchanged without compatibility issues.  Interoperability is achieved while maintaining data security and privacy. Data transferred between the system and external entities is encrypted and protected to prevent unauthorized access. The system consolidates information from various sources and organizations involved in sextortion prevention, providing a comprehensive view of the sextortion landscape and enabling more effective decision-making. It supports collaborative workflows and information sharing among different stakeholders, facilitating efforts to prevent sextortion cases and providing support to victims. The system can coordinate actions and share information between multiple agencies or organizations simultaneously, enhancing the overall effectiveness of sextortion prevention efforts. The SocialDAO facilitates collaboration and data sharing across international borders when sextortion cases involve multiple jurisdictions or countries. The system adheres to the relevant data protection and security standards, ensuring that data exchange and interoperability are carried out in a compliant and secure manner.  Integration with external systems is designed to be user-friendly, allowing organizations to easily connect their tools and resources with the system.

\textit{Performant} as a quality goal for sextortion prevention is optimized for high performance and efficiency in its sextortion prevention efforts. It is able to perform its functions quickly and effectively, while utilizing system resources efficiently. The system responds quickly to detected threats or incidents of extortion, minimizing the time it takes to take preventive actions or provide support to potential victims. It can scale its operations to handle a large volume of sextortion-related activities, cases, and user interactions without significant performance degradation. The system utilizes hardware and network resources efficiently, ensuring that it operates smoothly without causing resource bottlenecks or excessive resource consumption. For real-time interactions, such as crisis intervention or reporting, the system minimizes latency to ensure that users receive immediate responses and support. It can handle a substantial number of simultaneous tasks, such as processing reports, analyzing content, and coordinating with external entities, without slowdowns or delays. The system is designed to minimize downtime, ensuring that it remains available and responsive to users and stakeholders at all times, even during maintenance or updates. It employs efficient algorithms and data structures for tasks such as content analysis, threat detection, and data processing, reducing computational overhead. Load balancing mechanisms are in place to evenly distribute incoming traffic and tasks across multiple servers or resources, optimizing performance and reliability. Frequently accessed data or content are cached to reduce the need for repeated processing, resulting in faster response times. The system continually monitors its performance metrics and can adjust its resource allocation or configuration to maintain optimal performance levels. While ensuring high performance, the system does not compromise security. It effectively protects user data and maintains robust security measures. Provides real-time analytics and insights into sextortion prevention activities, allowing stakeholders to assess the effectiveness of the system and make data-driven decisions.

\textit{Scalable} requires that sextortion prevention be designed to handle an increasing volume of sextortion prevention activities, cases, and user interactions without a significant decrease in performance or functionality. Scalability ensures that the system can grow and adapt to meet the growing demands and challenges of sextortion prevention effectively. The system can accommodate a larger number of incidents and activities related to sextortion as necessary. It can expand its resources and capabilities to handle the increased workload.  It exhibits elasticity, allowing for dynamic resource allocation and de-allocation based on demand. Resources can be scaled up or down in response to changing needs, ensuring efficient resource utilization. The system efficiently distributes incoming workloads, such as processing reports, analyzing content or managing user interactions, to prevent bottlenecks and ensure smooth operation. Scalability can be achieved through horizontal scaling, where additional servers, nodes, or infrastructure components can be added to the system's architecture to distribute the load effectively. As the system scales, it maintains consistent performance levels, ensuring that users and stakeholders continue to experience reliable and responsive services. Scalability extends to geographic areas, allowing the system to serve users and stakeholders in different regions or jurisdictions effectively.  The system can handle an increasing volume of data, including user profiles, reports, and threat intelligence, while maintaining efficient data storage and retrieval processes. It can manage a higher level of concurrent user interactions, such as real-time reporting, support chats, or content analysis, without performance degradation. Scalability includes redundancy and failover mechanisms to ensure system availability and data integrity, even in the event of hardware failures or disruptions. Scalability is achieved in a cost-effective manner, ensuring that the system can grow while effectively managing infrastructure and operational costs. Scalable systems include monitoring and alerting capabilities to detect performance bottlenecks or resource limitations and take proactive measures to address them. Scalability extends to the ability to integrate with external systems, organizations, or resources to improve the overall capacity and effectiveness of sextortion prevention efforts.

Quality goals are inherited through the hierarchy of functional goals. Therefore, at the second refinement level, functional goals denote \textit{building awareness}, \textit{building skills on safe use of technology} with the third-level refinement \textit{explain safe sex content sharing}, \textit{build communication skills}, \textit{build self-confidence}, and finally, a functional goal to \textit{promote the application and services} which we will detail in further detail below. The third refinement level for \textit{build awareness} comprises the functional goal \textit{develop sextortion understanding}, \textit{educate about offenders}, and \textit{teach sextortion recognition}.  The goal to \textit{build communication skills} is further refined by the functional goals \textit{explain how to prevent conflicts} and \textit{explain rights} to raise awareness of legal issues.  Finally, for the goal \textit{build self-confidence}, the functional goal refinements are \textit{teach to manage negative emotions} and \textit{teach to set boundaries}. Note that the technical implementation of the conceptual functional goals and also the quality goals are studied in Section~\ref{sec:evaluationdiscussion} where a rapid deployment technology stack is presented. 

Associated to the functional goal \textit{prevent sextortion} is the \textit{psychologist} role, which means this role is also involved in the lower-level functional-goal refinements of Figure~\ref{fig:goalpreventsextortion}. Psychologists are trained professionals with expertise in understanding human behavior, emotions, and mental health. Their involvement in the goal of preventing sextortion signifies that they can provide psychological support to individuals who may be at risk of becoming victims or those who have already experienced sextortion. Psychologists can identify individuals, particularly teenagers or vulnerable individuals, who may be susceptible to sextortion due to emotional or psychological vulnerabilities. They can intervene early to provide counseling, guidance, and strategies to build resilience against potential sextortion threats. Furthermore, psychologists can assess the psychological well-being of individuals involved in sextortion cases. By understanding their emotional state and vulnerabilities, psychologists can help create personalized prevention plans to reduce the risk of victimization. Psychologists can also design and implement educational programs or workshops aimed at increasing awareness about sextortion risks, safe online behavior, and healthy relationships. These programs are crucial for adolescents and young adults who are often targets of sextortion.  In cases where sextortion has already occurred, psychologists can provide immediate crisis intervention and psychological support to victims. They help victims cope with the emotional trauma and provide strategies for recovery. Psychologists can provide insights into the behavioral patterns of potential sextortion offenders. By understanding the psychological motives behind such behavior, preventive measures can be tailored more effectively. Psychologists can offer guidance and support to the families of sextortion victims. They help family members understand the emotional impact of sextortion and provide strategies to support their loved ones. Psychologists often collaborate with other professionals, such as law enforcement, legal aid providers, and educators, to create a multidisciplinary approach to sextortion prevention. Their expertise complements the efforts of various stakeholders. Psychologists can work with individuals to enhance their emotional resilience, self-esteem, and decision-making skills. Empowered individuals are less likely to engage in risky online behavior or fall victim to sextortion. Finally, psychologists can conduct research and analysis to better understand the psychological factors contributing to sextortion and to identify effective prevention strategies.

The association between the psychologist and the functional goal for preventing sextortion in Figure~\ref{fig:goalpreventsextortion} is described by positive and negative emotional goals. Focusing first on positive emotional goals in Figure~\ref{fig:goalpreventsextortion}, \textit{satisfactory}  refers to the psychologist's experience of contentment, fulfillment and a sense of accomplishment when engaging with the dApp function to prevent sextortion. It signifies that the psychologist should find the use of the dApp in their role as a rewarding and fulfilling experience. The software feature that can be implemented is a comprehensive and user-friendly dashboard designed specifically for psychologists. This dashboard provides several functionalities and features that contribute to the psychologist's sense of satisfaction. The dApp offers a robust case management system where psychologists can easily access and review information related to sextortion cases they are handling. This includes details about the victim, the nature of the incident, progress in providing support, and any ongoing communication.  The dashboard provides real-time analytics and insights into the impact of the psychologist's interventions. This includes statistics on the number of cases they have worked on, successful prevention outcomes, and positive feedback from victims. The dApp offers collaboration tools that facilitate communication and coordination between psychologists and other stakeholders, such as legal aid providers or law enforcement. This ensures a seamless exchange of information and a collective effort in preventing sextortion. The dApp provides a repository of resources, best practices, and guidelines for psychologists involved in sextortion prevention. This empowers them with the necessary knowledge and tools to provide effective support to victims. Psychologists can provide feedback on the dApp's usability, effectiveness, and any suggestions for improvement. This feedback loop allows them to have a direct impact on the continuous enhancement of the system. The dashboard is designed with a user-friendly and intuitive interface, minimizing the learning curve and ensuring that psychologists can navigate the system effortlessly. The dApp acknowledges the contributions of psychologists in preventing sextortion by providing recognition, badges, or achievement milestones for their successful interventions. This recognition reinforces their sense of accomplishment. Psychologists can feel satisfied knowing that the dApp prioritizes the ethical and secure handling of sensitive information, ensuring the privacy and well-being of victims. The dApp offers opportunities for psychologists to enhance their skills and knowledge in dealing with sextortion cases through training modules and access to expert insights.

The positive emotional goal \textit{engaging} refers to the psychologist's experience of being captivated, immersed, and deeply involved when participating in the dApp function to prevent sextortion. It signifies that the psychologist should find the use of the dApp to be highly engaging, stimulating their interest and active participation. To reinforce this positive emotional goal of \textit{engaging}, the software feature that can be implemented is a dynamic and interactive virtual environment within the dApp. This virtual environment is designed to provide an engaging and immersive experience for psychologists while they work to prevent sextortion cases. The dApp offers a virtual collaboration space where psychologists can interact with other stakeholders involved in sextortion prevention, such as lawyers, white-hat hackers, and victims. This virtual space is designed to mimic a real-world meeting environment, complete with avatars and interactive elements.  Psychologists can engage in real-time text, voice, or video communication with other users in the virtual environment. This fosters collaboration and discussion, making prevention efforts more engaging and interactive. The dApp provides interactive case simulations in which psychologists can work through realistic sextortion scenarios. These simulations challenge their problem-solving skills and decision-making abilities, keeping them engaged in active learning. The virtual environment incorporates gamified elements, such as challenges, quests, and rewards. Psychologists can earn points, badges, or achievements as they successfully prevent sextortion cases or provide support to victims. These elements of the game make the experience more engaging. Psychologists can collaborate with other users to solve complex sextortion cases. They can brainstorm solutions, share insights, and participate in role-playing exercises to develop effective prevention strategies. The dApp encourages psychologists to share their expertise and insights through knowledge-sharing sessions or webinars. These sessions are interactive and allow for engaging discussions on best practices and emerging trends in sextortion prevention. The dApp offers personalized learning paths and recommendations based on the psychologist's interests and areas for skill development. This keeps them engaged in continuous learning and improvement. Engaging multimedia content, such as videos, infographics, and interactive presentations, is available within the virtual environment. Psychologists can access and use these resources to improve their understanding of prevention of sextortion. Psychologists can contribute their own content, case studies, or success stories to the virtual environment, fostering a sense of ownership and involvement within the community.

The negative emotional goal \textit{frustrated} refers to the psychologist's experience of feeling annoyed, discouraged, or hindered when engaging the dApp function to prevent sextortion. It signifies that the psychologist should not encounter unnecessary obstacles, difficulties, or sources of irritation that hinder their effectiveness in preventing sextortion. As a mitigation , several software features can be implemented within the dApp. The dApp features an intuitive and user-friendly interface that is easy for psychologists to navigate. Clear menus, well-organized content, and straightforward workflows minimize confusion and frustration. The dApp provides comprehensive training materials and resources to help psychologists become familiar with its functions and features. Accessible guides, tutorials, and onboarding assistance reduce frustration related to the learning curve. The dApp offers a streamlined case management system that allows psychologists to efficiently access and update information related to sextortion cases they are handling. Quick search and filtering options make it easy to find and work on specific cases. Psychologists receive real-time notifications for critical updates, such as new sextortion reports or urgent support requests. This ensures that they stay informed without having to constantly check the system. The dApp includes collaboration tools that facilitate communication and coordination with other stakeholders involved in sextortion prevention. Psychologists can easily reach out to lawyers, white-hat hackers, or victims, reducing frustration related to communication barriers. The dApp is optimized for performance to ensure that it operates smoothly without lags or delays. Psychologists can work efficiently without being hindered by slow response times. Psychologists have the option to provide feedback on their experiences with the dApp. This feedback loop allows them to express concerns or suggest improvements, addressing the sources of frustration over time. The dApp is transparent about its data handling policies and security measures. Psychologists have clear insights into how data is handled, reducing concerns and frustrations related to data privacy. The dApp allows psychologists to easily import and export relevant data, documents, or reports. This feature streamlines data management and reduces frustration when transferring information. Psychologists can customize certain aspects of their dashboard or user preferences, allowing them to customize the dApp to their specific needs and preferences. The dApp provides informative error messages and access to support resources in case psychologists encounter problems or require assistance. This minimizes frustration by offering clear paths to solve the problem. 

The next negative emotional goal \textit{overwhelmed} refers to the psychologist's experience of feeling extremely stressed, burdened, or inundated when engaging with the dApp function to prevent sextortion. It signifies that the psychologist should not be faced with an excessive workload or complexity that hinders their effectiveness and well-being. Several software features can be implemented within the dApp as mitigation measures. The dApp includes intelligent workload management algorithms that distribute cases and tasks evenly among psychologists. This prevents any single psychologist from being overwhelmed with too many cases at once. Psychologists have access to tools to prioritize sextortion cases based on severity, urgency, or specific criteria. This helps them focus on the most critical cases first and manage their workload effectively. Routine and repetitive tasks are automated within the dApp, reducing the administrative burden on psychologists. This includes features such as automated report generation or data entry. The dApp facilitates real-time collaboration with other stakeholders, such as lawyers or white-hat hackers. Psychologists can seek assistance or share responsibilities, preventing them from becoming overwhelmed with complex cases. The dApp provides intelligent recommendations for external resources or support services that psychologists can refer victims to. This reduces the need for psychologists to handle every aspect of support themselves. Comprehensive and easily accessible guidelines are available within the dApp, offering step-by-step instructions and best practices for handling sextortion cases. This reduces confusion and uncertainty. Psychologists can use scheduling tools to manage their time effectively, ensuring that they have dedicated time for case management, communication, and self-care. Psychologists can provide feedback on the performance of the dApp and their workload. This feedback helps optimize the system to prevent overwhelming situations. The dApp includes emergency protocols and procedures for high-risk cases, ensuring that psychologists know how to respond to critical situations without feeling overwhelmed. Resources for self-care and stress management are available within the dApp. The dApp offers ongoing training and skill development opportunities to help psychologists adapt to the evolving challenges in the prevention of sextortion. Psychologists can easily track their progress in resolving cases and providing support. This helps them stay organized and avoid feeling overwhelmed by the number of open cases.

The final negative emotional goal \textit{confused} refers to the psychologist's experience of feeling disoriented, uncertain, or lacking clarity when engaging the dApp function to prevent sextortion. It signifies that the psychologist should not encounter complexities or ambiguities that hinder their understanding and effectiveness. To mitigate this negative emotional goal, the dApp features a clear and intuitive user interface with well-defined menus, icons, and navigation paths. This minimizes confusion when psychologists interact with the system. The dApp offers guided workflows with step-by-step instructions for handling sextortion cases. Psychologists can follow these structured processes, reducing the risk of confusion. Contextual help resources are available within the dApp, providing on-demand assistance and explanations when psychologists encounter unfamiliar terms or procedures. A comprehensive knowledge base is accessible, offering detailed information on sextortion prevention strategies, legal aspects, and psychological support techniques. Psychologists can refer to this resource for clarity. Psychologists can maintain detailed case notes and documentation within the dApp. This ensures that information is organized and accessible, reducing confusion when revisiting cases. The dApp facilitates real-time collaboration with other stakeholders, such as legal experts or white hat hackers. Psychologists can seek clarification and guidance, eliminating uncertainties. The dApp incorporates data validation checks to prevent errors and inaccuracies in case information. Psychologists receive prompts and corrections to maintain data accuracy. Psychologists can provide feedback on the dApp's usability and clarity. This feedback contributes to continuous improvement and the elimination of confusing elements. Clear and transparent data handling policies and security measures are communicated to psychologists, ensuring they understand how data is managed within the system. Psychologists have access to role-specific dashboards that present information relevant to their responsibilities. This focused view reduces unnecessary complexity.  The dApp offers training modules and resources to help psychologists become proficient in using the system effectively. Ongoing training reduces confusion over time. Psychologists can generate standardized reports with ease, streamlining their reporting process and reducing the likelihood of errors.

An association of another role with the functional goal \textit{ prevents sextortion} is the AI agent \textit{sextortion diagnoser}. This association signifies the deployment of a state-of-the-art, autonomous cognitive entity designed to play a pivotal role in preventing sextortion through its advanced cognitive capabilities. Briefly, for the aspirational layer at the highest level, the AI agent \textit{sextortion diagnoser} is aligned with ethical principles and human values related to preventing harm and ensuring online safety. Its overarching goal is to contribute to the well-being of individuals, particularly teenagers, by actively combating sextortion. The global strategy layer provides the AI agent with a contextual understanding of the current state of sextortion threats and trends across digital platforms. It gathers real-time data and intelligence, allowing the agent to form strategic plans for effective prevention. On the agent model layer, the agent maintains a comprehensive self-model that includes its cognitive capabilities, domain expertise in sextortion detection, and awareness of its limitations. This self-awareness ensures that the agent operates within its actual abilities and can accurately assess its capacity to prevent sextortion. On the executive function layer, the AI agent translates its high-level strategic goals into detailed plans and resource allocation. It decides how to allocate computational resources and adapt its algorithms to optimize sextortion prevention efforts. The cognitive control layer enables the agent to dynamically select tasks and make real-time decisions based on the evolving environment and internal assessments. For example, it may decide to intensify monitoring on specific platforms or trigger alerts when certain risk factors are detected. On the task prosecution layer, the AI agent executes tasks related to sextortion prevention, such as monitoring online communications, analyzing content for explicit material, and assessing the level of coercion or threats involved. It also initiates interventions and support measures when necessary. The ACE framework ensures that the AI agent operates at the highest levels of autonomy, making complex cognitive decisions and strategic interventions to prevent sextortion. It continually learns and adapts to new sextortion tactics and online platforms, staying ahead of emerging threats. Additionally, the ACE framework's emphasis on ethical alignment and corrigibility ensures that the AI agent operates ethically, respects privacy, and is responsive to feedback and oversight.

The association of the role \textit{legal aid provider} with the functional goal \textit{to raise awareness} indicates that legal aid providers play a crucial role in raising awareness regarding certain legal aspects or issues related to the prevention of sextortion or related matters. Legal aid providers typically have expertise in various aspects of the law, including cyber laws, privacy regulations, and criminal justice systems. Their role in "building awareness" involves using their legal knowledge to educate individuals, organizations, or the public about legal rights, responsibilities, and implications related to sextortion. Legal aid providers may engage in awareness campaigns or initiatives that aim to inform the public about the legal consequences of sextortion. This could include organizing online workshops, seminars, or online webinars to educate individuals on how to recognize sextortion, report it, and seek legal recourse when necessary. Sextortion cases often involve complex legal considerations, such as evidence collection, filing legal complaints, and understanding the legal avenues available to victims. Legal aid providers can offer guidance and information on these aspects, helping individuals understand their legal options. Legal aid providers can raise awareness about the rights and remedies available to sextortion victims. They may explain legal protections in place, such as restraining orders or legal actions against perpetrators, empowering victims to seek justice. Legal aid providers may engage in advocacy efforts to influence policies related to sextortion prevention and legal responses. By participating in awareness-building activities, they can draw attention to gaps in existing laws or advocate for legal reforms to better address sextortion cases. Legal aid providers often collaborate with other stakeholders, such as psychologists, law enforcement, or victim support organizations, to build awareness collectively. This multidisciplinary approach ensures a comprehensive understanding of sextortion issues. In addition to awareness campaigns, legal aid providers may offer their services to victims directly. They can help victims navigate the legal process, provide legal representation when needed, and ensure that their rights are protected throughout legal proceedings. Legal aid providers may disseminate informative materials, brochures, or online resources that clarify legal aspects related to sextortion. These materials serve as educational tools to enhance awareness.

The association qualification between the role \textit{legal aid provider} and the functional goal \textit{build awareness} with the positive emotional goal \textit{satisfactory} is attainable with specific software measures. First, ensure that the dApp has a user-friendly and intuitive interface that makes it easy for legal aid providers to access and navigate awareness-building tools and resources. A well-designed interface contributes to user satisfaction. Create a library of comprehensive resources, including legal documents, templates, guides, and educational materials. Legal aid providers should find it satisfying to access a rich repository of materials they can use in their awareness-building efforts. Allow legal aid providers to customize awareness-building content to align with their target audience and specific legal topics. Customization options give them a sense of control and satisfaction in tailoring materials to their needs. Provide performance analytics and reports that show the impact of their awareness-building efforts. Metrics such as reach, engagement, and feedback can reinforce satisfaction by demonstrating the effectiveness of their work. Implement collaboration tools that enable legal aid providers to work together on awareness campaigns and initiatives. Collaborative features foster a sense of teamwork and accomplishment, contributing to satisfaction. Include a feedback mechanism that allows legal aid providers to gather input from their audience or stakeholders. Positive feedback and recognition can enhance their satisfaction with the dApp's effectiveness. Offer task management and scheduling tools that help legal aid providers plan, organize, and execute awareness-building activities efficiently. Completing tasks on time can lead to a sense of satisfaction. Keep legal aid providers informed with automated updates about relevant legal changes, policy updates, or news related to the areas they focus on. Staying well-informed contributes to their professional satisfaction. Consider implementing a recognition and rewards system within the dApp to acknowledge the contributions and achievements of legal aid providers in building awareness. Recognizing their efforts enhances satisfaction. Provide responsive user support to address any issues or questions legal aid providers may have. Effective support services contribute to their overall satisfaction with the dApp. Offer user training and onboarding resources to help legal aid providers become proficient with the dApp's features. Competence and confidence enhance satisfaction. Demonstrate a commitment to continuous improvement by actively seeking user feedback and implementing enhancements based on their input. Legal aid providers will find it satisfying to see their suggestions implemented.

The negative emotional goal \textit{frustrated} between the role \textit{legal aid provider} and the functional goal \textit{build awareness} can be mitigated with the following software features. Provide legal aid providers with tools for easy content creation, editing, and organization. A user-friendly content management system reduces frustration by simplifying the process of updating and maintaining awareness materials. Offer pre-designed content templates that legal aid providers can customize for their awareness campaigns. Templates ensure consistency and save time, reducing the frustration of starting from scratch. Implement collaboration features that allow legal aid providers to work on awareness materials collaboratively. Version control ensures that changes are tracked, preventing version conflicts and associated frustration. Enable legal aid providers to schedule the automatic publication of awareness content. This feature reduces the stress of manual timing and coordination. Provide options for automated distribution of awareness materials through various channels, such as social media or email marketing. Automation reduces the frustration of manual distribution efforts. Offer user-friendly analytics tools that provide insights into the reach and impact of awareness campaigns. Clear and accessible data helps legal aid providers assess their effectiveness and make data-driven decisions without frustration. Include a feedback mechanism for legal aid providers to receive input from their target audience. This helps them refine their awareness materials and campaigns, reducing potential frustration due to unclear messaging. Offer training resources and tutorials within the dApp to enhance legal aid providers' digital literacy and marketing skills. Improved competence reduces frustration when dealing with technology. Provide responsive user support channels that legal aid providers can turn to when encountering issues or challenges. Prompt assistance minimizes frustration and keeps them on track. Allow legal aid providers to personalize awareness content based on the needs and preferences of their target audience. Personalization options reduce frustration by ensuring that messages resonate effectively. Use AI-driven content recommendation engines to suggest relevant topics, articles, or updates that can be incorporated into awareness campaigns. This feature assists legal aid providers in content curation, reducing frustration related to content discovery. Automate repetitive tasks, such as content distribution or reporting, to save time and reduce the frustration of manual, time-consuming processes. Finally, maintain clear and easily accessible documentation that legal aid providers can refer to when using the dApp. Comprehensive documentation minimizes frustration related to navigating the platform.    

The association between the role of the AI agent \textit{legal aid diagnoser} and the functional goal \textit{build awareness} signifies a specialized and strategic partnership aimed at enhancing awareness related to legal aspects and resources available to prevent and address sextortion incidents. This association involves several key components. The AI Agent \textit{legal aid diagnoser} is designed to provide intelligent and context-aware assistance regarding legal matters related to sextortion. It possesses a deep understanding of relevant legal frameworks, procedures, and resources.The AI agent \textit{legal aid diagnoser} actively disseminates accurate and up-to-date legal information related to sextortion. This includes explaining relevant laws, procedures for reporting incidents, and the rights of victims. It ensures that this information is accessible and understandable to users. The AI agent provides personalized legal guidance to individuals based on their specific situations. For example, it may ask questions to assess the circumstances of a victim and then offer tailored advice on the appropriate legal actions to take. In addition to legal advice, the AI agent recommends relevant legal aid providers, lawyers, or organizations that can assist victims. It ensures that victims have access to the necessary support for legal action. Legal processes can be complex and filled with jargon. The AI agent simplifies legal language and explanations, making it easier for users, particularly victims, to understand their legal options. The AI agent helps individuals determine if they qualify for legal aid services and guides them through the application process if applicable. The AI agent may also contribute to awareness campaigns by creating and disseminating legal awareness materials. These materials can be distributed through various channels, including social media, to reach a broader audience. Recognizing the sensitivity of legal matters, the AI agent ensures the privacy and confidentiality of users' interactions. It follows strict data protection measures and ethical guidelines.

The association of the role \textit{health ministry public servant} with the functional goal \textit{build awareness} denotes a collaborative effort between government health ministries or public health agencies and the broader community to increase awareness regarding health-related aspects, including those related to sextortion. Public servants working within health ministries typically have access to expertise and resources related to public health, mental health, and well-being. Their involvement in the functional goal "build awareness" signifies leveraging this expertise to address health aspects associated with sextortion. Sextortion can have significant mental and emotional health implications for victims. The health ministry public servants may focus on raising awareness about the psychological and emotional toll of sextortion, including issues such as anxiety, depression, and trauma. They can provide information on available mental health resources and support. Public servants in health ministries may contribute to awareness campaigns that emphasize preventative health measures. This could include educating individuals, especially teenagers, on maintaining good mental and emotional health, recognizing signs of distress, and seeking help when needed. The association suggests collaboration between health ministries and other stakeholders, such as mental health professionals, educators, and support organizations. This collaboration can lead to the development of comprehensive awareness initiatives that address both the psychological and physical well-being of sextortion victims. Health ministry public servants may play a role in crafting and disseminating public health messaging related to sextortion. These messages can be designed to reach a wide audience and encourage individuals to seek help for mental health issues arising from sextortion experiences. They may work on making mental health and counseling resources more accessible to victims. This could involve initiatives to reduce stigma around seeking mental health support and increasing the availability of counselors and therapists with expertise in trauma and victim support. Health ministries can contribute to research efforts related to the health impact of sextortion. They may collect and analyze data to better understand the health consequences and inform evidence-based awareness programs.

The association of the role \textit{police officer} as an organization with the functional goal \textit{build awareness} signifies that law enforcement agencies play a crucial role in raising awareness about sextortion-related issues and promoting safety within the community. Police officers have authority and credibility in the eyes of the public. When they engage in awareness-building activities related to sextortion, their messages carry weight and are more likely to be taken seriously by the community. Police offciers can actively engage in prevention and education efforts aimed at informing individuals and communities about the risks associated with sextortion. They can conduct workshops, seminars, and awareness campaigns to educate the public on recognizing and avoiding sextortion threats. Police officers can provide insights into the legal aspects of sextortion, including relevant laws, reporting procedures, and the consequences for offenders. This knowledge is essential for individuals to understand their rights and the legal recourse available. Police officers can promote and facilitate reporting mechanisms for sextortion cases. They can explain how victims or witnesses can report incidents, ensuring that law enforcement is informed and can take appropriate action. Police officers can collaborate with other stakeholders, such as educators, NGOs, and legal aid providers, to create a comprehensive and coordinated approach to building awareness. Collaborative efforts enhance the effectiveness of awareness campaigns. Promoting awareness of sextortion contributes to overall community safety. Police officers can emphasize the importance of safe online behavior, recognizing threats, and reporting suspicious activities to protect individuals from victimization.  Police officers can provide information about their role in responding to sextortion cases. Victims and the community need to know how law enforcement can assist and investigate these crimes. Police officers can create and distribute informative online materials, brochures, or online resources that educate individuals on sextortion risks and safety measures. These materials serve as educational tools for the public. Police officers can raise awareness about cybersecurity best practices, highlighting the importance of strong passwords, secure online communication, and protecting personal information. In cases where legal action is required, police officers can refer victims to legal aid providers or other relevant support services. They play a role in connecting victims with the necessary assistance. Police officers can engage with the community through various channels, including social media, community meetings, and school presentations. Engaging directly with the community builds trust and encourages open communication.

The association of the role \textit{educator} as part of an organization with the functional goal \textit{build awareness} indicates that educators play a vital role in raising awareness about sextortion-related issues and promoting a safe and informed online environment. Educators, particularly in schools and colleges, have direct access to students and young individuals who are often the target demographic for sextortion. They can engage with students to educate them about the risks and consequences of sextortion. Educators can integrate awareness about sextortion into their curricula, ensuring that students receive formal education on recognizing and addressing online threats. This integration helps create a culture of awareness and safety. Educators can organize online workshops, seminars, and awareness sessions led by experts or law enforcement agencies. These events provide students with firsthand knowledge and guidance on staying safe online. Educators can implement digital literacy programs that teach students about responsible online behavior, privacy protection, and the importance of not sharing explicit content. Such programs empower students to make informed decisions. Educators can initiate awareness campaigns within their campuses, encouraging students to be vigilant and report any sextortion-related incidents. These campaigns create a sense of responsibility among students.  Educators can involve parents by organizing awareness sessions or distributing informational materials that guide parents on safeguarding their children online. Parental support is crucial in preventing sextortion among young individuals. Educators can establish and enforce safe internet usage policies that promote responsible online behavior among students. Clear policies set expectations and boundaries. Educators can collaborate with law enforcement agencies, NGOs, and other organizations to create a comprehensive approach to building awareness. Collaborative efforts amplify the impact of awareness initiatives.  Institutions can offer student support services, including counseling or guidance, for those who may have been affected by sextortion or need assistance in coping with online threats. Educators can create and distribute awareness materials, such as posters, brochures, or online resources, to students, parents, and staff. These materials reinforce key messages about sextortion prevention. Institutions can establish clear reporting mechanisms for sextortion incidents within their campus community. This ensures that cases are promptly addressed and appropriate actions are taken. Finally, educators can provide training sessions or workshops for teachers and staff to ensure they are equipped to address and prevent sextortion incidents among students.

The roles of \textit{teenager} and \textit{victim} are distinct in the context of the functional goal \textit{build awareness} with respect to sextortion prevention. We next discuss the differences between these roles and their respective associations with the goal of building awareness. The \textit{teenager} role represents young individuals, often adolescents or young adults, who are at risk of encountering sextortion threats due to their age, online presence, and vulnerability to manipulation. On the other hand, the \textit{victim} role represents individuals who have already fallen victim to sextortion. These individuals have experienced the negative consequences of sextortion and may require support and assistance.

The association of the \textit{teenager} role with the functional goal \textit{build awareness} emphasizes the importance of educating teenagers about the risks of sextortion and empowering them with knowledge to prevent falling victim to such threats. Teenagers can play a pivotal role in raising awareness among their peers. They are more likely to trust and engage with messages delivered by fellow teenagers. Their association with the goal signifies the need to create awareness programs tailored to this age group. Awareness initiatives targeting teenagers aim to provide education on responsible online behavior, recognizing potential sextortion threats, and understanding the consequences of sharing explicit content. These programs equip teenagers with the skills to protect themselves. Teenagers can also act as sources of support for their peers who may have encountered sextortion. They can share information about reporting mechanisms and encourage victims to seek help.

The association of the \textit{victim} role with the functional goal \textit{build awareness} underscores the importance of providing support, guidance, and resources to victims of sextortion. The focus rests on helping victims recover and navigate the aftermath of such incidents. Victims may choose to share their experiences as part of awareness campaigns. Their stories can serve as powerful tools to educate others about the real dangers of sextortion and the importance of prevention. Awareness initiatives related to victims emphasize educating individuals about their legal rights, reporting mechanisms, and the support available to them. This knowledge helps victims take appropriate actions and seek help.  Victims may require psychological support to cope with the emotional trauma of sextortion. Awareness programs for victims may include information on accessing counseling or therapy services.

In the context of the functional goal \textit{build awareness} and its association with the positive emotional goal \textit{personalized}, the roles of \textit{teenager} and \textit{victim} each experience the concept of \textit{personalized} awareness differently. We next explain  the differences between these roles concerning the positive emotional goal \textit{personalized}. For teenagers, the positive emotional goal of \textit{personalized} awareness means receiving tailored and age-appropriate educational content. Awareness initiatives should be designed to resonate with the unique needs, interests, and online behaviors of teenagers. They should find the awareness materials and messages relevant to their experiences and potential risks. Personalization in this context means delivering content that speaks directly to their concerns and challenges. Teenagers are more likely to engage with awareness efforts when the content feels personalized to their demographic. Interactive and relatable materials can capture their attention and encourage active participation. Personalized awareness efforts may include stories or testimonials from peers who have faced or prevented sextortion risks. Hearing from individuals of a similar age group can make the awareness message more relatable and impactful. Emphasize the importance of privacy and consent in online interactions, ensuring that teenagers understand how to protect their personal information and navigate online spaces safely. Awareness materials for teenagers should address digital literacy skills, helping them make informed decisions about sharing content and recognizing potential threats. On the other hand, victims of sextortion who experience the positive emotional goal of \textit{personalized} awareness require tailored support and resources. The awareness efforts should acknowledge their unique experiences and challenges. For victims, personalization means awareness initiatives that are centered on their recovery and well-being. This may include information on seeking professional help, legal guidance, and emotional support. Victims need to feel understood and supported. Awareness campaigns should convey empathy and compassion, letting them know that they are not alone and that help is available. Victims should be made aware of their legal rights, such as reporting options, pursuing legal action against perpetrators, and protecting themselves from further harm. Personalized awareness for victims may include guidance on accessing psychological support or counseling services to address the emotional trauma caused by sextortion. Victims should receive personalized advice on enhancing their online safety and security to prevent future incidents. Awareness efforts can connect victims with local or online community resources, support groups, or organizations specializing in assisting victims of sextortion. 

For teenagers, the positive emotional goal of \textit{enjoyable} awareness implies that the awareness initiatives should deliver content and messages in an engaging and captivating way. Awareness materials should be designed to hold their interest and attention. Awareness efforts targeting teenagers may incorporate interactive elements, such as quizzes, gamified scenarios, or online challenges. These elements make the learning experience fun and enjoyable. Encourage teenagers to participate in awareness activities with their peers, fostering a sense of community and shared learning. Group discussions, workshops, or awareness events can be enjoyable social experiences. Use creative and innovative approaches to convey awareness messages. This might include multimedia content, storytelling, animations, or user-generated content, which teenagers are more likely to find enjoyable. Awareness materials can present real-world scenarios and case studies that resonate with teenagers' lives, making the content relatable and enjoyable to learn from. Implement feedback mechanisms and provide rewards or incentives for active participation. Positive reinforcement can make the awareness process more enjoyable. Ensure that awareness materials are easily accessible on digital platforms and mobile devices, aligning with teenagers' preferences for online content consumption. On the other hand, victims of sextortion who experience the positive emotional goal of \textit{enjoyable} awareness require a supportive and empathetic environment. Awareness efforts for victims can include interactive resources that guide them through the steps of recovery or legal processes. Interactive checklists, self-assessment tools, or chatbots can make the journey more engaging. Offer victims personalized support, where they can access resources and information at their own pace and comfort level. This approach can reduce stress and make seeking help more enjoyable.  Victims may find it enjoyable to connect with support groups or communities of individuals who have faced similar experiences. Peer support can provide a sense of belonging and comfort. mphasize positive outcomes and success stories of individuals who have overcome sextortion. These stories can provide hope and motivation, making the awareness journey more enjoyable. Provide user-friendly tools and resources that victims can use independently or with the assistance of professionals. Easy-to-navigate interfaces and clear instructions contribute to a more enjoyable experience. Ensure that victims have access to emotional support services, such as counseling or helplines, in a way that respects their privacy and emotional well-being. While the primary focus for being \textit{enjoyable} awareness is on support and recovery, the process can be made more enjoyable by creating a so-called safe space~\cite{o2023minor,pevac2022tertiary}. This refers to creating an environment where the victim feels secure, protected, and free from judgment or further harm. This concept of a safe space is crucial for victims, as they have experienced a traumatic and violating incident, and their emotional well-being and recovery are of utmost importance. Victims should have confidence that their information and conversations are kept private and confidential. This includes secure communication channels and protocols to protect their identity and sensitive details. A safe space is built on trust and empathy. Support providers, whether they are counselors, therapists, or support groups, should convey understanding, compassion, and genuine concern for the victim's well-being. Victims should not feel re-victimized or subjected to further harm in any way. This means that the environment is free from threats, harassment, or any form of intimidation. Victims should have control over their participation and choices in the recovery process. They should not be pressured into actions or decisions they are not comfortable with. A safe space ensures that victims have easy access to the resources, information, and support services they need. This may include helplines, counseling, legal advice, and educational materials. Victims should have access to emotional support, whether through one-on-one counseling, group therapy, or peer support networks. Emotional support helps them cope with the trauma and regain a sense of control. The safe space is centered around the victim's recovery journey. It offers guidance, resources, and tools that aid in their healing process and helps them regain their confidence and well-being. Victims should be empowered with knowledge and skills to protect themselves from future sextortion threats. Education on online safety and consent is an essential part of this empowerment. A safe space is inclusive and does not discriminate based on gender, age, race, or any other factors. It is a place where all victims, regardless of their background, can find support and understanding. The concept of a safe space extends beyond immediate crisis intervention. It involves ongoing support to help victims rebuild their lives and address the long-term emotional and psychological effects of sextortion. 

\textit{Trustful} awareness for teenagers involves transparent communication. They should have access to clear and honest explanations about the risks of sextortion, the consequences, and the preventive measures they can take. Teenagers require awareness initiatives that provide accurate, reliable, and trustworthy information about sextortion. They should feel confident that the information they receive is credible and backed by experts in the field. Teenagers benefit from knowing that the awareness materials are developed with the guidance of professionals, such as psychologists, educators, and cybersecurity experts. This assurance contributes to trust in the content. Trustful awareness efforts respect teenagers' privacy and ensure that their personal information remains confidential. They should not feel that their participation in awareness campaigns jeopardizes their privacy. Awareness materials should be inclusive, respecting the diversity of teenagers' experiences and backgrounds. Trust is established when teenagers see themselves represented in the content. Teenagers should believe that the awareness initiatives have no hidden agendas or ulterior motives. They should not feel manipulated or coerced into taking specific actions. Trustful awareness materials are presented on user-friendly platforms and websites. Teenagers should find it easy to navigate and access information without encountering obstacles or intrusive elements. On the other hand, \textit{trustfu}l awareness for victims is built on sensitivity and empathy. They should feel that the materials and support services understand the emotional and psychological toll of sextortion. Victims must trust that their interactions with awareness and support providers will remain confidential. This assurance encourages victims to seek help without fear of their experiences being disclosed. Victims require access to awareness materials and resources that have been developed with the input of experts, including therapists, legal professionals, and victim support specialists. Knowing that professionals are involved instills trust. Trustful awareness efforts emphasize the victim's recovery journey and provide guidance on how to access the necessary resources for healing. This includes information on therapy, counseling, and legal support. Victims should be educated about their legal rights, and they should trust that the information provided is accurate and up-to-date. This ensures they can make informed decisions about pursuing legal action. Trustful awareness services offer non-judgmental support, ensuring that victims feel safe sharing their experiences and concerns without fear of blame or criticism. Victims trust awareness initiatives that demonstrate a long-term commitment to their well-being. Continuous support and resources contribute to building this trust. Trustful awareness materials empower victims by providing them with the knowledge and tools to protect themselves and prevent future incidents. This empowerment builds confidence and trust.

\textit{Satisfactory} awareness for teenagers means that educational materials should be engaging, interactive, and enjoyable. They should find the learning process interesting and satisfactory, promoting their understanding of sextortion risks and prevention. Awareness efforts targeting teenagers should make them feel that the content is relevant to their daily lives and experiences, ensuring that what they learn can be applied practically. A \textit{satisfactory} experience for teenagers involves feeling empowered with knowledge and skills to protect themselves from sextortion risks. They should be satisfied with the tools provided to safeguard their online presence. Satisfactory awareness initiatives for teenagers often encourage peer interaction and discussions, allowing them to share insights and learn from each other's experiences. Teenagers may find it satisfactory to track their progress in understanding and implementing safety measures, reinforcing a sense of achievement. Satisfactory awareness can incorporate positive reinforcement mechanisms, such as badges or rewards, to motivate teenagers to actively participate in the learning process. Teenagers should have access to user-friendly resources and platforms that make it easy for them to navigate and find information, contributing to their satisfaction. On the other hand, for victims, the \textit{satisfactory} awareness  involves receiving empathetic and supportive information that acknowledges their emotional trauma. They should feel satisfied with the emotional support provided. Victims should be satisfied with the awareness materials' focus on their recovery and well-being. The content should provide guidance on accessing necessary support and resources. A \textit{satisfactory} experience for victims means receiving clear and actionable guidance on how to cope with the aftermath of sextortion. They should feel satisfied with the practical advice offered. Victims should be satisfied with the assurance of privacy and confidentiality in their interactions with awareness and support services, allowing them to share their experiences without fear. Satisfactory awareness initiatives for victims include comprehensive information about their legal rights and options. They should feel satisfied with their understanding of the legal aspects. Victims may find satisfaction in connecting with supportive communities or support groups where they can share experiences, fostering a sense of belonging. Satisfactory awareness can involve measuring progress in terms of emotional healing and recovery, allowing victims to track their journey and find satisfaction in their achievements.

\textit{Empowerful} awareness for teenagers means that awareness initiatives should empower them through education. They should receive comprehensive information about sextortion risks, consequences, and prevention strategies to make informed decisions. Awareness efforts targeting teenagers should equip them with practical skills and knowledge to protect themselves online. This includes understanding the importance of consent, recognizing warning signs, and knowing how to respond to potential threats. Empowering teenagers involves building their confidence in navigating the digital world safely. They should feel capable of making responsible choices and knowing where to seek help or support when needed. Awareness materials should encourage critical thinking and digital literacy among teenagers. They should be empowered to assess online content critically, verify information, and be aware of potential risks. Teenagers should be empowered to establish and maintain respectful and healthy online relationships. This includes understanding the boundaries of consent and recognizing manipulative behaviors. Empowerment for teenagers can also involve educating them on how to be empathetic bystanders who can support their peers and report instances of sextortion or online harassment. On the other hand, for victims, an \textit{empowerful} awareness means that awareness materials should empower them through the recovery process. They should be empowered to take control of their healing and well-being.  Empowering victims includes providing them with knowledge about available resources and support services. Victims should feel empowered to seek help from professionals, support groups, or helplines. Victims should be informed about their legal rights and options for seeking justice or redress. Empowerment in this context involves understanding the legal aspects of their situation. Awareness efforts for victims should focus on building emotional resilience and coping skills. They should feel empowered to manage the emotional challenges associated with sextortion. Empowerment for victims also includes regaining a sense of privacy and control over their online presence. They should learn strategies to protect their personal information and digital identity. Victims should be empowered to connect with supportive networks, whether it is through support groups, therapy, or counseling. Building a support system is a crucial aspect of their empowerment. Victims should be empowered with knowledge about how to prevent future sextortion incidents and protect themselves from similar threats.

\textit{Engaging} awareness for teenagers means that educational materials should be interactive and enjoyable. They should have opportunities to actively participate in learning, such as quizzes, games, and interactive scenarios. Awareness efforts targeting teenagers should be visually appealing, with engaging graphics, videos, and multimedia content. Eye-catching visuals and modern design elements contribute to engagement. Content should be relatable to teenagers' experiences, interests, and concerns. It should use language and examples that resonate with their daily lives. Engagement for teenagers often involves opportunities for peer involvement. Awareness initiatives may encourage them to discuss topics with friends, share insights, or participate in group activities. Awareness materials should present real-world scenarios and case studies that teenagers can relate to. These scenarios help them understand the potential risks and consequences of sextortion. Engagement is enhanced when teenagers can personalize their learning experience. This may include choosing topics of interest, setting goals, and tracking progress. On the other hand, for victims, \textit{engaging} awareness means that materials should be sensitive to their emotional state. Awareness initiatives should convey empathy, understanding, and support for their unique experiences. Engagement for victims revolves around materials that are focused on their recovery journey. They should feel engaged with content that offers guidance on healing and regaining control. Awareness efforts should provide victims with accessible and easy-to-understand resources. Engaging materials may include step-by-step guides, FAQs, and clear explanations of available support services. Victims may find engagement through interactive support mechanisms, such as chatbots or helplines. These tools can provide immediate assistance and answers to their questions. Engagement for victims involves assuring them of privacy and confidentiality. They should feel safe when seeking information or support, knowing that their interactions are discreet. Awareness materials should engage victims in discussions about their emotional well-being and coping strategies. Encouraging them to express their feelings and concerns can foster engagement. Victims may find engagement in connecting with support communities or peer groups where they can share their experiences and learn from others who have faced similar challenges.

In the context of the negative emotional goal \textit{disappointed} associated with the functional goal \textit{build awareness}, both the roles \textit{teenager} and \textit{victim} may experience disappointment, but the reasons for their disappointment and the technological mitigation means to address it differ. Teenagers may feel disappointed if awareness content is not relevant to their age group or lacks real-world applicability. To mitigate this, technological solutions should include personalized content recommendations and age-appropriate examples. If awareness materials are not engaging or fail to hold their attention, teenagers may become disappointed. Technological mitigation includes the use of interactive multimedia, gamification elements, and user-friendly interfaces to keep them engaged. Outdated or generic information may lead to disappointment. Technology can help by ensuring that content is up-to-date, and regular updates are pushed to users. Teenagers often prefer learning alongside their peers. A lack of opportunities for peer interaction can be disappointing. Technological solutions can include discussion forums, social media integration, or virtual classrooms to facilitate peer engagement. If awareness materials are not easily accessible on the devices and platforms teenagers commonly use, they may feel disappointed. Technological mitigation involves optimizing content for various devices, including smartphones, and ensuring cross-platform compatibility. On the other hand, victims may experience disappointment if awareness materials lack emotional sensitivity and fail to address their unique needs. Technological mitigation includes incorporating empathetic language, providing emotional support chatbots, or connecting victims with online support communities. If awareness initiatives do not lead to tangible solutions or do not help victims cope with their situation, they may feel disappointed. Technological solutions should offer practical resources, crisis helplines, and access to professional support. Victims highly value their privacy and security. If awareness efforts compromise these aspects, disappointment can arise. Technological mitigation involves robust data protection measures, secure communication channels, and clear privacy policies. Victims may be disappointed if they perceive a lack of real-time support. Technology can address this by providing 24/7 support chatbots or connecting victims with trained counselors via secure video conferencing. Overly complex or technical information can lead to disappointment. Awareness materials should be presented in a straightforward and understandable manner, leveraging multimedia formats and plain language. 

Teenagers may feel \textit{overwhelmed} in their awareness building when presented with too much information at once. Technological mitigation includes providing bite-sized, digestible content and allowing users to set their own learning pace. Awareness materials that are overly complex or technical can lead to overwhelming. Technology can simplify content through clear visuals, infographics, and multimedia presentations. Peer pressure and social expectations can make teenagers feel overwhelmed by the need to conform. Technology can offer peer support networks, anonymous forums for discussion, and resources on handling peer pressure. Teenagers have diverse learning styles. Technological solutions should cater to visual, auditory, and kinesthetic learners by offering a variety of content formats, such as videos, podcasts, and interactive quizzes. Excessive screen time and digital fatigue can contribute to overwhelm. Technology can implement features like screen-time reminders, offline modes for learning, or integrations with physical activities. Victims of sextortion, on the other hand, may experience emotional trauma, making them vulnerable to overwhelm. Technological mitigation includes providing access to mental health resources, crisis helplines, and online support groups. Victims may be overwhelmed by privacy concerns, fearing further exposure. Technology should prioritize robust data protection measures, secure communication channels, and clear privacy policies. Dealing with legal aspects of sextortion cases can be overwhelming. Technological solutions can offer access to legal aid providers, simplified legal guidance, and resources on reporting incidents. Victims may feel overwhelmed when trying to navigate available resources. Technology can provide user-friendly interfaces, search functionalities, and chatbots to guide them to the right resources. Victims may feel isolated in their experiences. Technology can connect them with online communities or forums where they can share their stories and receive support from others who have faced similar situations. Concerns about personal safety can contribute to overwhelm. Technology can offer safety planning tools and emergency contact features.

In the context of the negative emotional goal \textit{confused} associated with the functional goal \textit{build awareness}, both the roles \textit{teenager} and \textit{victim} may experience confusion, but the reasons for their confusion and the technological mitigation means to address it differ. Teenagers may become confused when presented with complex or technical information. Technological mitigation means include simplifying content through clear visuals, plain language, and interactive elements. Without proper guidance, teenagers may not know where to start or how to navigate awareness materials effectively. Technology can provide user-friendly interfaces, step-by-step guides, and navigation cues. The abundance of resources on the internet can overwhelm teenagers and lead to confusion about which sources are reliable. Technological solutions should curate trusted resources and provide source credibility indicators. Teenagers may become confused if the objectives of awareness efforts are unclear. Technology can use clear and concise messaging to outline the goals and expected outcomes of awareness campaigns. Conflicting information and differing perspectives on certain topics can create confusion. Technology can offer multiple viewpoints while emphasizing evidence-based information. Victims of sextortion, on the other hand, may find legal processes and requirements confusing. Technological mitigation means include providing access to legal aid providers, simplified legal guidance, and resources on reporting incidents. Concerns about privacy and data protection laws can lead to confusion. Technology should ensure transparent privacy policies and options for users to control their personal data. Victims in distress may have difficulty processing information. Technological solutions should incorporate emotional support resources, such as chatbots or helplines, to assist during moments of confusion. Victims may become confused when trying to find specific resources or assistance. Technology can offer intuitive search functions, resource categorization, and chatbots for resource recommendations. Victims may be confused about creating safety plans or understanding safety measures. Technology can provide interactive safety planning tools and access to professionals who can offer guidance.
    
In the context of the negative emotional goal \textit{frustrated} associated with the functional goal \textit{build awareness}, both the roles \textit{teenager} and \textit{victim} may experience frustration. Teenagers may become frustrated if awareness materials are not engaging or fail to capture their interest. Technological mitigation means include interactive content, gamification elements, and multimedia formats to enhance engagement. Frustration can arise if awareness materials are not relevant to teenagers' lives or do not address their specific concerns. Technology can provide personalized content recommendations based on user interests and needs. Static or unresponsive content can lead to frustration. Technology should enable interactive features such as quizzes, discussions, and simulations to encourage active participation. Teenagers may become frustrated if the content is overly complex or difficult to understand. Technological solutions involve presenting information in a clear, simple, and age-appropriate manner.  The lack of opportunities for peer interaction can be frustrating for teenagers who prefer learning from their peers. Technology can incorporate social features like discussion forums or group activities. Victims, on the other hand, may feel frustrated if awareness efforts do not provide the immediate support or solutions they need. Technological mitigation means include access to crisis helplines, chat support, and resources for immediate assistance. Legal complexities can cause frustration for victims trying to navigate legal avenues. Technology can offer simplified legal guidance, access to legal aid providers, and assistance with reporting incidents. Frustration can arise if victims feel their privacy is not adequately protected during awareness interactions. Technology should prioritize strong data protection measures, secure communication channels, and privacy controls. Victims may become frustrated when trying to locate specific resources or assistance. Technology can provide clear navigation, search functionality, and recommendations based on user needs. Victims experiencing emotional distress may feel frustrated if awareness materials do not address their emotional needs. Technology can incorporate empathetic chatbots and resources for emotional support.

For the functional sub-goal \textit{build skills on safe use of technology}, three organisations are associated being NGO workers, IT service providers and educators. Thus, the NGO worker is responsible for implementing educational initiatives aimed at building skills related to the safe use of technology. This includes creating and delivering training programs, workshops, or awareness campaigns focused on online safety and digital literacy. The NGO identifies and reaches out to specific target audiences or communities that may benefit from building these skills. This could include teenagers, vulnerable groups, or individuals at risk of encountering online threats like sextortion. The NGO may develop a structured curriculum or educational materials tailored to the needs of the target audience. These materials should cover topics such as recognizing online risks, protecting personal information, understanding consent, and responding to threats. Ensuring access to necessary resources is crucial. The NGO may provide educational materials, guides, online resources, and tools that individuals can use to enhance their technology-related skills and knowledge. Conducting workshops, training sessions, or webinars is part of the NGO's role. These events offer hands-on learning experiences and opportunities for participants to practice safe technology use. The NGO may launch awareness campaigns to promote the importance of safe technology use within the community or target audience. These campaigns aim to raise awareness, change attitudes, and encourage individuals to take proactive steps to protect themselves online. Collaboration with other stakeholders, such as schools, local authorities, law enforcement, and tech companies, may be essential. Working together can amplify the impact of the NGO's efforts and reach a broader audience. To ensure the effectiveness of their programs, the NGO should implement monitoring and evaluation mechanisms. This involves tracking the progress of participants, collecting feedback, and making necessary adjustments to improve the programs. Depending on the nature of the NGO's work, they may establish support systems or helplines for individuals who encounter issues related to unsafe technology use. These support channels can provide guidance and assistance to those in need.  In some cases, NGOs may engage in advocacy efforts to influence policies and regulations related to online safety. They can work to ensure that the legal framework supports the protection of individuals from online threats like sextortion. NGO workers may engage in research and data collection to better understand the evolving landscape of online risks and vulnerabilities. This information can inform the development of effective educational materials and strategies.  

For the \textit{IT service provider}, building skills on the safe use of technology entails bringing technical expertise to the initiative. They can offer in-depth knowledge about various aspects of technology, cybersecurity, and digital tools, which is essential for building skills in safe technology use. The IT service provider can conduct training sessions, workshops, or seminars focused on cybersecurity, safe online practices, and digital literacy. These sessions aim to educate individuals on how to protect themselves while using technology. IT service providers can tailor training programs to meet the specific needs of the target audience. This may involve creating customized content and learning materials that address the unique challenges faced by different groups. They can provide access to security tools, software, and resources that individuals can use to enhance their online safety. This may include antivirus software, firewalls, password managers, and encryption tools. IT service providers can offer practical demonstrations of security measures and best practices. These demonstrations help participants understand how to apply safe technology use in real-world scenarios. They may provide technical support and guidance to individuals who have questions or encounter issues related to technology and online safety. This support can include troubleshooting assistance and advice on secure configurations. The IT service provider can raise awareness about cybersecurity threats and vulnerabilities. This includes educating individuals about common cyber threats like phishing, malware, and sextortion. Ensuring data protection and privacy is a key responsibility. The IT service provider can teach individuals how to safeguard their personal and sensitive information while using digital platforms. They can promote a culture of continuous learning by offering ongoing training opportunities and resources. Technology evolves rapidly, so staying informed about new threats and protective measures is crucial. Some IT service providers offer monitoring services and alert systems to detect and respond to potential security breaches or suspicious activities. This proactive approach helps individuals stay safe online. Collaboration with other stakeholders, such as educators, NGOs, or government agencies, may be necessary to reach a broader audience. The IT service provider can work together with these partners to maximize the impact of their initiatives. They should ensure that their training and guidance align with relevant cybersecurity regulations and compliance standards to provide accurate and up-to-date information.

The association of the functional sub-goal \textit{build skills on safe use of technology} with an \textit{educator} signifies the role and responsibilities of the educator in achieving this sub-goal. Educators can integrate cybersecurity and safe technology use topics into their curriculum. This means developing courses or modules that teach students essential skills for navigating the digital world securely. They can offer digital literacy programs that cover topics such as recognizing online risks, protecting personal information, understanding consent, and responding to threats like sextortion. Educators are adept at tailoring training to different age groups, ensuring that the content is age-appropriate and engaging for students at various levels, from primary to higher education. The focus is on skill development. Educators can provide hands-on experiences and practical exercises to teach students how to apply safe technology practices effectively. They can organize awareness campaigns and events within the educational community to promote safe technology use. These campaigns can include workshops, seminars, and discussions on online safety. Educators can emphasize the concept of digital citizenship, teaching students about their rights and responsibilities in the digital world. This includes ethical online behavior and respecting others' digital privacy. Educators within the institution can receive training in cybersecurity and safe technology use to effectively impart knowledge and skills to students. Educators may provide students with access to resources such as educational materials, online tools, and software that can aid in safe technology use. Collaboration with cybersecurity experts, NGOs, or IT service providers can enhance the effectiveness of educational initiatives. Partnerships can bring in specialized knowledge and resources. Educators can monitor students' progress in acquiring safe technology use skills and assess the effectiveness of their programs through assessments and evaluations. In some cases, educators involve parents in educational efforts. They provide guidance to parents on how to support their children to use technology safely at home. Promoting lifelong learning is a core aspect. Educators can encourage students to stay up to date on cybersecurity and online safety even after graduation. Ensuring that educational programs align with relevant cybersecurity regulations and guidelines is essential to provide accurate and legally compliant information.

The association between the AI agent role \textit{educator} and the functional goal \textit{build skills on the safe use} signifies the pivotal role of AI-powered educational tools and resources in empowering individuals, particularly teenagers, with the knowledge and skills needed for safe and responsible online behavior. Educators, whether human or AI agents, possess expertise in pedagogy and instructional design. Their involvement in the goal "build skills on safe use" ensures that educational content is well-structured, engaging, and effective in conveying critical information about online safety. AI-powered educators can personalize learning experiences to meet the unique needs and learning styles of individual users. They adapt content delivery, pace, and complexity, ensuring that users can acquire skills at their own pace. AI-driven educational tools can offer interactive and engaging learning experiences. This can include simulations, quizzes, and real-life scenarios that help users practice safe online behaviors in a risk-free environment. AI agents can provide ongoing, up-to-date information about evolving online threats and safety measures. They can adapt content to address emerging risks, keeping users informed and prepared. AI educators can make educational resources accessible to a broad audience, including those with diverse learning needs. This inclusivity ensures that everyone, regardless of their background or abilities, can acquire essential online safety skills. AI-driven educational tools can reach a large number of users simultaneously, making them scalable for widespread awareness and skills-building campaigns. AI agents ensure that the educational content is delivered consistently, eliminating variability in teaching quality that can occur with human educators. AI agents can monitor users' progress and assess their skill development. This data can be used to provide feedback and recommendations for improvement. 

The association between the AI agent role \textit{NGO advisor} and the functional goal \textit{build skills on the safe use} is a strategic collaboration aimed to enhance individuals' skills and knowledge in safe and responsible online behavior. NGOs specializing in online safety and digital literacy often have a deep understanding of the challenges individuals, especially teenagers, face in the digital world. They have experience in developing educational resources and programs tailored to specific online safety issues. The AI agent \textit{NGO advisor} can work closely with these NGOs to access their educational materials, curricula, and resources. This ensures that the educational content provided to users is well-researched, up-to-date, and aligned with best practices in online safety. NGOs typically have extensive networks and partnerships with other organizations, including schools, community centers, and youth groups. This association can facilitate the distribution and implementation of online safety education programs to reach a broader audience. The AI agent can collaborate with NGOs to customize educational content to address specific online safety concerns and cultural nuances. This customization ensures that the content is relevant and effective for the target audience. NGOs often run awareness campaigns related to online safety. The AI agent can integrate these campaigns into the educational process, reinforcing the importance of safe online behaviors and providing practical guidance. NGOs may offer support services for individuals who have experienced online harassment or exploitation. The AI agent can connect users with these services when needed, ensuring a holistic approach to online safety. The AI agent can monitor users' progress in online safety education and provide feedback. If users encounter difficulties or have questions, the agent can connect them with NGO resources for further assistance.  NGOs often track the impact of their programs. The AI agent can collect data on users' skill development and knowledge gain, contributing to the assessment of the effectiveness of online safety education initiatives. By collaborating with NGOs, the AI agent can enhance user engagement through interactive and community-based learning experiences. This can include forums, peer support, and discussions facilitated by the NGO's expertise. The AI agent can work with NGOs to continuously update and improve educational content based on emerging threats and user feedback, ensuring that the skills-building process remains relevant and effective.

The association of the roles \textit{teenager} and \textit{victim} with the functional goal \textit{build communication skills} signifies a shared effort to enhance the ability of both teenagers and victims of sextortion to effectively communicate, both in seeking help and supporting one another. For teenagers, particularly those who may be vulnerable to sextortion, building communication skills is essential. It means equipping them with the ability to express themselves clearly, report incidents of sextortion, and seek support or assistance when they encounter troubling situations online. Victims of sextortion often face emotional distress and fear. By associating them with the goal of building communication skills, it implies providing them with the tools to communicate their experiences, emotions, and concerns effectively. This can be crucial when seeking help from authorities, support organizations, or therapists. The association also suggests the formation of peer support networks among teenagers. Building communication skills can involve teaching teenagers how to listen empathetically, provide emotional support to their peers who may have experienced sextortion, and encourage open and non-judgmental communication within their communities. Effective communication skills enable teenagers to report sextortion incidents promptly and accurately to relevant authorities or support organizations. This includes knowing how to provide essential details and evidence while maintaining their privacy and safety. The goal may include educational programs that teach teenagers and victims about the importance of communication in addressing sextortion. This can encompass understanding the psychological and emotional impact of sextortion and how communication can be a path toward recovery. Communication skills involve conflict resolution techniques. Victims may need assistance in resolving issues related to sextortion, such as confronting perpetrators, dealing with blackmail, or managing conflicts within their social circles. Victims often require immediate crisis intervention and counseling. The association implies that communication skills training may help them express their emotions and experiences to therapists or counselors, leading to effective therapeutic interventions. Building communication skills can empower teenagers and victims to speak out against sextortion and advocate for their rights and safety. This includes participating in awareness campaigns, sharing their experiences (if they choose), and influencing policy changes.  In today's digital world, communication often occurs online. Building communication skills includes educating teenagers and victims on safe and responsible online communication to prevent further victimization. The association underscores the importance of creating supportive communities where communication skills are valued. Such communities can provide a safe space for teenagers and victims to share their experiences, seek advice, and find solidarity. Finally, we infer that the emotional goals of teenagers and victims in association with the functional goal of building communication skills are very similar to the discussion above. Thus, the reiteration of the emotional goal discussion is omitted for space limitations.

The final association of the roles \textit{teenager} and \textit{victim} with the functional goal \textit{build communication skills} in Figure~\ref{fig:goalpreventsextortion} signifies a concerted effort to empower both teenagers and victims of sextortion with the necessary skills and mindset to enhance their self-esteem and self-assurance. This association is further refined by the functional goals of teaching to manage negative emotions and teaching to set boundaries, including the ability to say no. Building self-confidence is essential for individuals who have experienced sextortion or are vulnerable to it. It involves helping them recognize their self-worth and believe in their abilities to cope with challenges, make informed decisions, and seek help when needed. Teaching individuals, particularly victims, to manage negative emotions is an integral part of building self-confidence. This includes strategies for coping with the emotional impact of sextortion, such as fear, shame, guilt, and anxiety. Learning to identify, express, and regulate these emotions contributes to greater emotional resilience and self-confidence. Building self-confidence involves teaching individuals how to set and maintain personal boundaries. This includes the ability to say no assertively when faced with unwanted requests or pressure, a skill that can be especially valuable in sextortion prevention. Victims can benefit from learning how to assert their boundaries and protect their privacy and dignity. Self-confidence is closely tied to a positive self-image. By helping teenagers and victims develop a healthy and positive perception of themselves, they can feel more self-assured in their interactions and decisions. Building self-confidence also means fostering resilience, which is the ability to bounce back from adversity. Victims of sextortion may face various challenges during their recovery process, and self-confidence plays a crucial role in their ability to overcome these challenges and move forward. Support systems, whether from peers, professionals, or AI agents, can contribute to building self-confidence. Knowing that there is a network of understanding and empathetic individuals and resources available can boost individuals' confidence in seeking help and addressing the emotional toll of sextortion. Building self-confidence is an ongoing process of personal development and growth. It involves acquiring skills and knowledge, such as emotional intelligence, communication skills, and resilience-building techniques, that contribute to an individual's sense of self-assurance. Ultimately, the association aims to empower teenagers and victims to regain control over their lives and make decisions that prioritize their well-being. This empowerment is rooted in a strong sense of self-confidence and the ability to navigate challenges with resilience.

\subsubsection{Functional Goal Managing Roles}
\label{sec:managingroles}

\begin{figure*}[htpb]
    \vspace{0.2cm}
    \begin{center}
        \includegraphics[scale=0.55]{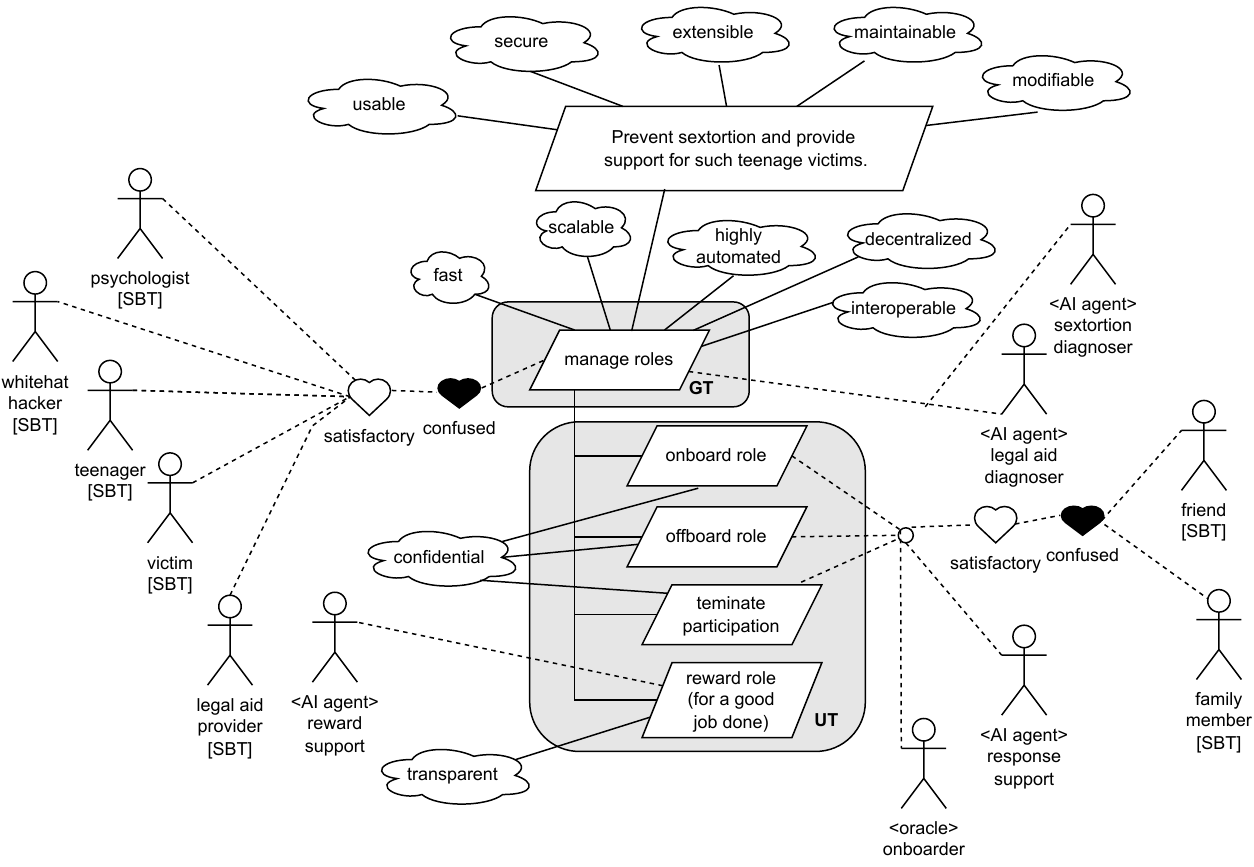}
        \caption{The goal-model refinement for managing roles.}
        \label{fig:goalmanageroles}
    \end{center}
    \vspace{-0.5cm}
\end{figure*}

The goal model in Figure~\ref{fig:goalmanageroles} shows the \textit{manage roles} functional goal with quality goals assigned in addition to those inherited from the level of the value proposition. Next, we briefly explain each associated quality goal. First, thew quality goal \textit{fast} refers to roles management swiftly and without unnecessary delays. It aims to minimize the time required to add, modify, or remove roles within the system. The system should promptly respond to requests related to role management. Users should experience minimal waiting times when assigning or updating roles. \textit{Fast} infers that role changes should take effect in real-time or near real-time. When a role is assigned or modified, the system should reflect these changes without significant delays. 

The quality goal \textit{scale} in Figure~\ref{fig:goalmanageroles} means that the role management system should be able to accommodate changes in the number of roles, users, or permissions without a significant decrease in performance or functionality. It adapts seamlessly to varying workloads and requirements. Scalability implies that the system can handle a growing number of roles and users efficiently. It should not experience bottlenecks or degradation in performance as the system expands.  Scalability also considers efficient resource allocation. The system should distribute resources appropriately to ensure that role management tasks can be performed without overloading specific components or servers. Scalable systems often incorporate load balancing mechanisms to evenly distribute workloads across available resources. This ensures that no single component becomes a performance bottleneck. Scalability may involve redundancy strategies to enhance reliability. Redundant servers or components can take over if one fails, minimizing downtime and disruptions in role management. The system should exhibit elasticity, meaning it can scale up or down based on demand. During periods of high demand, additional resources can be allocated to maintain performance.

Next, the quality goal \textit{high automation} signifies the emphasis on reducing manual intervention and increasing the level of automated processes in role management. High automation implies that role management tasks should require minimal manual effort. Many aspects of role assignment, modification, and removal should be automated to streamline processes. Automated role management is often essential for scalability. As the system grows and handles more roles and users, automation helps maintain performance and responsiveness. Automation can include features like scheduling role-related tasks or permissions changes, making it easier to implement changes at specific times or in response to certain events. Automated systems can include error handling and notifications to alert administrators or users when issues arise during role management. 

The quality goal \textit{decentralized} signifies that the role management processes within the system are distributed across multiple nodes or components rather than being centralized in a single location. A decentralized role management system distributes control and decision-making authority across various nodes or entities. This can include allowing individual users or departments to have some level of autonomy in managing their roles. Decentralization enhances system resilience. If one component or node fails, other decentralized nodes can continue to manage roles independently, reducing the risk of a single point of failure. In a decentralized setup, role management tasks can be performed locally, reducing latency associated with centralization. Users can interact with the system more responsively. Decentralization allows for easier adaptation to changes in organizational structure or role management requirements. New nodes can be added, and roles can be managed independently to accommodate evolving needs.

Finally, the quality goal \textit{interoperability} signifies the ability of the role management system to seamlessly exchange information and interact with other systems, software, or components. Interoperability ensures that the role management system can integrate with other systems and applications within an organization's technology ecosystem. This allows for the sharing of role-related data and functionalities.  Interoperability enables the exchange of role-related data with other systems, such as identity management, access control, or enterprise resource planning (ERP) systems. This exchange can include user attributes, permissions, and role assignments. Interoperability often involves adhering to common communication standards and protocols that facilitate data exchange between systems. 

Five human roles are associated to the functional goal \textit{manage roles} in Figure~\ref{fig:goalmanageroles} via the positiove emotional goal \textit{satisfactory} and negative emotional goal \textit{confused}. We omit detailing these emotional goals again as they are already explained in connection with Figure~\ref{fig:goalpreventsextortion}. These five human roles  who can manage roles are the psychologist, whitehat hacker, teenager, victim, legal aid provider who have already been explained in detail above as well. Furthermore, two AI agents are also associated with the functional goal \textit{manage roles} being the sextortion diagnoser and the legal aid diagnoser. Their purpose is to support the human roles with advice for well-targeted roles management.

The functional goal \textit{manage roles} compises sub-goals being the functions \textit{onboarding role}, \textit{offboarding role}, \textit{terminate participation}, and \textit{reward role} in case that the job performance is accordingly. In addition to all the hierarchically  inherited quality goals, emotional goals and roles, Figure ~\ref{fig:goalmanageroles} shows that additional associations exist that we explain next. 

The quality goal \textit{confidential} in Figure~\ref{fig:goalmanageroles} is associated to the functional goals \textit{onboarding role}, \textit{offboarding role}, and \textit{terminate participation}. The association of the quality goal \textit{confidential} with the functional goal \textit{onboarding role} emphasizes the need to ensure that the process of bringing new individuals or stakeholders into the system is conducted with the utmost confidentiality. Thus, When individuals, such as psychologists, whitehat hackers, teenagers, victims, or legal aid providers, join the system to fulfill their roles, their personal information, credentials, and any sensitive data must be handled in a confidential manner. This ensures that their privacy is respected and that their involvement in sextortion prevention or response is discreet. The quality goal \textit{confidential} associated with the functional goal \textit{offboarding role} highlights the importance of maintaining confidentiality when individuals or stakeholders leave their roles within the system. The same principles hold for the process of \textit{terminating participation}. 

Likewise,the functional goals \textit{onboarding role}, \textit{offboarding role}, and \textit{terminate participation} are associated with the human roles of \textit{friend} and family member via the positive emotional goal \textit{satisfactory}, and the negative emotional goal \textit{confused}. The human roles are consulted by an AI agent termed \textit{response support} and an oracle \textit{onboarder} to supply offchain data to the corresponding functions. Finally, the quality goal transparent in association with the functional goal \textit{reward role} signifies the importance of ensuring clarity and fairness in the process of recognizing and compensating individuals who perform roles within the sextortion prevention and response system. This association underscores the need to make the reward system transparent and easy to understand, fostering trust and motivation among participants. The AI agent \textit{reward support} associated with the functional goal \textit{reward role} is designed to assist in the fair and efficient distribution of rewards to individuals or stakeholders who have fulfilled roles within the system such as psychologists, white-hat hackers, or legal aid providers.   

\subsubsection{Functional Goal Providing Active Sextortion Aid}
\label{sec:sextaid}

\begin{figure*}[htpb]
    \vspace{0.2cm}
    \begin{center}
        \includegraphics[scale=0.55]{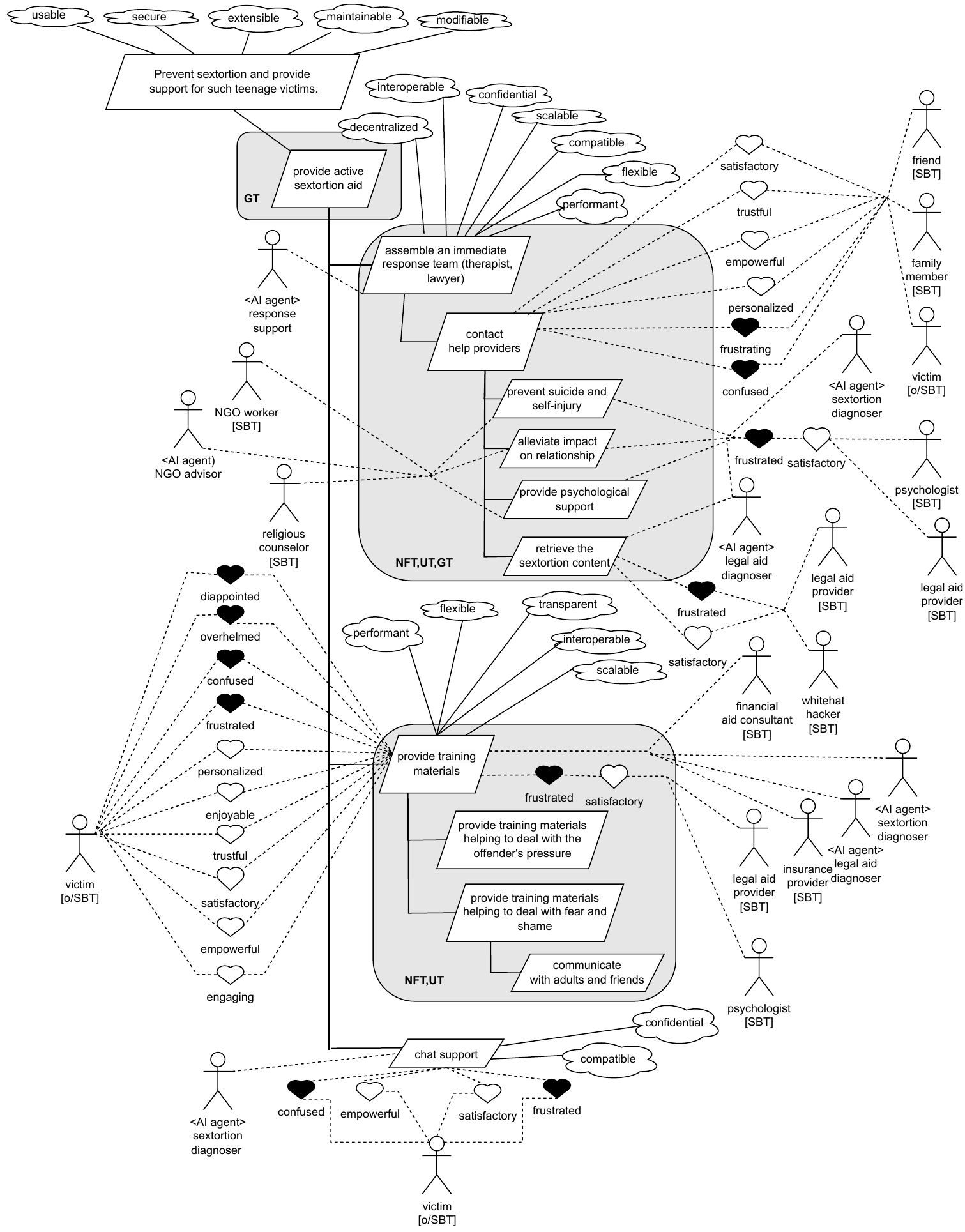}
        \caption{The goal-model refinement for providing active sextortion aid.}
        \label{fig:goalactivateaid}
    \end{center}
    \vspace{-0.5cm}
\end{figure*}

In Figure~\ref{fig:goalactivateaid}, the functional goal \textit{provide active sextorion aid} has three direct sub-goals being \textit{assemble an immediate response team}, \textit{provide training materials}, and \textit{chat support}. For the first goal of \textit{assemble an immediate response team}, the assumption is that the aim is to assemble a team of therapists and lawyers. An AI agent \textit{response support} to consult other roles in this goal that are associated to the further refining functional goals.

A set of quality goals is associated being \textit{decentralized}, \textit{interoperable}, \textit{confidential}, \textit{scalable}, \textit{compatible}, \textit{flexible}, and \textit{performant}.  First, the associated quality goal \textit{decentralized} signifies the intention to structure the process of assembling this team in a way that distributes decision-making and responsibilities across a network of stakeholders or entities. This underscores the objective of involving multiple stakeholders in the process of forming the immediate response team. Thus, decentralization implies that the responsibility for selecting and onboarding therapists and lawyers is not concentrated in a single entity or authority. Instead, it involves input from various participants, which is the case in Figure~\ref{fig:goalactivateaid} given the further refinement functional goals and their associated roles. 

The quality goal \textit{interoperable} associated with the functional goal \textit{assemble an immediate response team} signifies the intention to ensure that the components, systems, or entities involved in creating this team can seamlessly work together and exchange information effectively. Interoperability implies that the platforms, technologies, databases, and communication channels used by therapists, lawyers, AI agents, NGOs, and other stakeholders are compatible with each other. They can exchange data, communicate, and coordinate without encountering significant obstacles. Interoperability streamlines the process of assembling the team, enhances communication, and reduces friction in data sharing. 

The quality goal \textit{confidential} associated with the functional goal \textit{assemble an immediate response team} emphasizes the importance of maintaining the privacy and confidentiality of sensitive information during the process of forming this team. Confidentiality implies that any personal or sensitive data, such as the identities of therapists, lawyers, victims, or any other stakeholders, should be protected from unauthorized access or disclosure. By maintaining confidentiality, individuals who may be at risk due to their involvement in sextortion cases can feel safe participating in the team assembly process. It fosters trust and encourages professionals to offer their expertise without fear of exposure.

The quality goal \textit{scalable} associated with the functional goal \textit{assemble an immediate response team} highlights the need for the team assembly process to be designed and executed in a way that allows for easy expansion and adaptation to accommodate a growing number of participants and cases. Scalability ensures that the immediate response team can adapt to the changing needs and volumes of cases. It prevents bottlenecks in the team formation process and ensures that victims can access the help they need promptly.

The quality goal \textit{compatible} associated with the functional goal \textit{assemble an immediate response team} emphasizes the need for the processes and systems involved in forming the team of therapists and lawyers to be in alignment and harmonious with each other and with any existing technologies or platforms. Compatibility implies that the tools, technologies, and workflows used to identify, recruit, and onboard therapists and lawyers are designed to work together efficiently. It also implies that these processes align with any existing platforms or systems in use.

The quality goal \textit{flexible} associated with the functional goal \textit{assemble an immediate response team} highlights the need for adaptability and versatility in the processes and systems used to form the team of therapists and lawyers. Flexibility implies that the mechanisms for identifying, recruiting, and onboarding therapists and lawyers can adapt to evolving requirements, such as changes in the number of professionals needed or shifts in the types of expertise required. A flexible approach allows the immediate response team to respond promptly and efficiently to different sextortion cases and emerging needs. It ensures that the team can be assembled quickly and effectively, even when specific requirements vary from case to case.

The quality goal \textit{performant} associated with the functional goal \textit{assemble an immediate response team} emphasizes the need for the team formation process to operate efficiently and effectively. Performance in this context refers to the speed, accuracy, and overall efficiency of the team formation process. It ensures that the team can be put together quickly and that the selected therapists and lawyers are well-suited to the specific sextortion case. A performant team formation process is essential because it reduces response time, allowing the team to provide timely aid to victims of sextortion. It ensures that the right professionals are engaged promptly, enhancing the effectiveness of the aid provided.

The next functional goal in the hierarchy of Figure~\ref{fig:goalactivateaid} termed \textit{contact help providers} is associated with the human roles \textit{victim}, \textit{friend}, and \textit{family} via a set of positive emotional goals being \textit{satisfactory}, \textit{trustful}, \textit{empowered}, and \textit{personalized}. Furthermore, the negative emotional goals in Figure~\ref{fig:goalactivateaid} are \textit{frustrated} and \textit{confused}. Since these emotional goals have parallel semantics to their earlier introductions above, we omit explaining them now and from here on in detail. 

The leave refinmements if the functional-goal hierarchy in Figure~\ref{fig:goalactivateaid} are termed \textit{prevent suicide and self-justice}, \textit{alleviate impact on relationship}, \textit{provide psychological support}, and \textit{retrieve the sextortion content}. Associated to first thrtee of these functional goals are the human roles of \textit{psychologist} and \textit{legal aid provider} via the positive emotional goal \textit{satisfactory} and the negative emotional goal \textit{frustrated}. Additional human roles connected directly to these first three functional leave goals are the \textit{NGO worker} and \textit{religious counselor}. Finally, the non-human associated roles are the three AI agents termed \textit{NGO advisor}, \textit{sextortion diagnoser}, and \textit{legal aid diagnoser}.  To the final functional goal \textit{retrieve the sextortion content} are associated the AI agent \textit{legal aid provider} and the \textit{human roles} legal aid provider and \textit{whitehat hacker} via the positive and negative emotional goals \textit{satisfactory} and \textit{frustrated} respectively. 

Briefly, the whitehat hacker possesses advanced knowledge and skills in cybersecurity and ethical hacking. They are experts in identifying vulnerabilities and weaknesses in digital systems and applications. While their primary focus is on hacking and cybersecurity, whitehat hackers operate within the the law and ethical guidelines. In the context of sextortion cases, they ensure that all actions taken to retrieve content are legal and compliant with relevant regulations.

The Functional goal \textit{provide traning materials} in Figure~\ref{fig:goalactivateaid} has a set of quality goals associated being \textit{performant}, \textit{flexible}, \textit{transparent}, \textit{interoperable}, and \textit{scalable}, which have a parallel respective semantics compared to their earlier explanations. Furthermore, the human role \textit{victim} is associated with the functional goal \textit{provide traning materials} via a set of positive and negative emotional goals that are already explained above. To the right in Figure~\ref{fig:goalactivateaid}, additional associated huamn roles are the \textit{financial aid consultant}, the \textit{whitehat hacker}, the \textit{legal aid provider}, the \textit{insurance provider}, and a \textit{psychologist} via the positive emotional goal \textit{satisfactory} and the negative emotional goal \textit{frustrated}. Furthermore associated are at AI agents termed \textit{legal aid diagnoser}, and \textit{sextortion diagnoser}. Finally, the further refining sub-gols  are for providing training materials to help dealing with the offender's pressure, and the fear and shame respectively. The latter functional goal has yet another refinement goal to communicate with adults and friends as a skill that the victim should acquire.

The bottom leave functional goal in Figure~\ref{fig:goalactivateaid} is the \textit{chat support} for the human role \textit{victim}. The latter is connected to the functional goal via the positive emotional goals \textit{empowerful} and \textit{satisfactory} and the negative emotional goals \textit{confused} and \textit{trusted}. Associated to the chat support are the quality goals \textit{compatible} and \textit{confidential}. The associated AI agent \textit{sextortion diagnoser} is available for consulting the victim.

\subsubsection{Functional Goal Activating Help Seeking}
\label{sec:acthelpseek}

\begin{figure*}[htpb]
    \vspace{0.2cm}
    \begin{center}
        \includegraphics[scale=0.6]{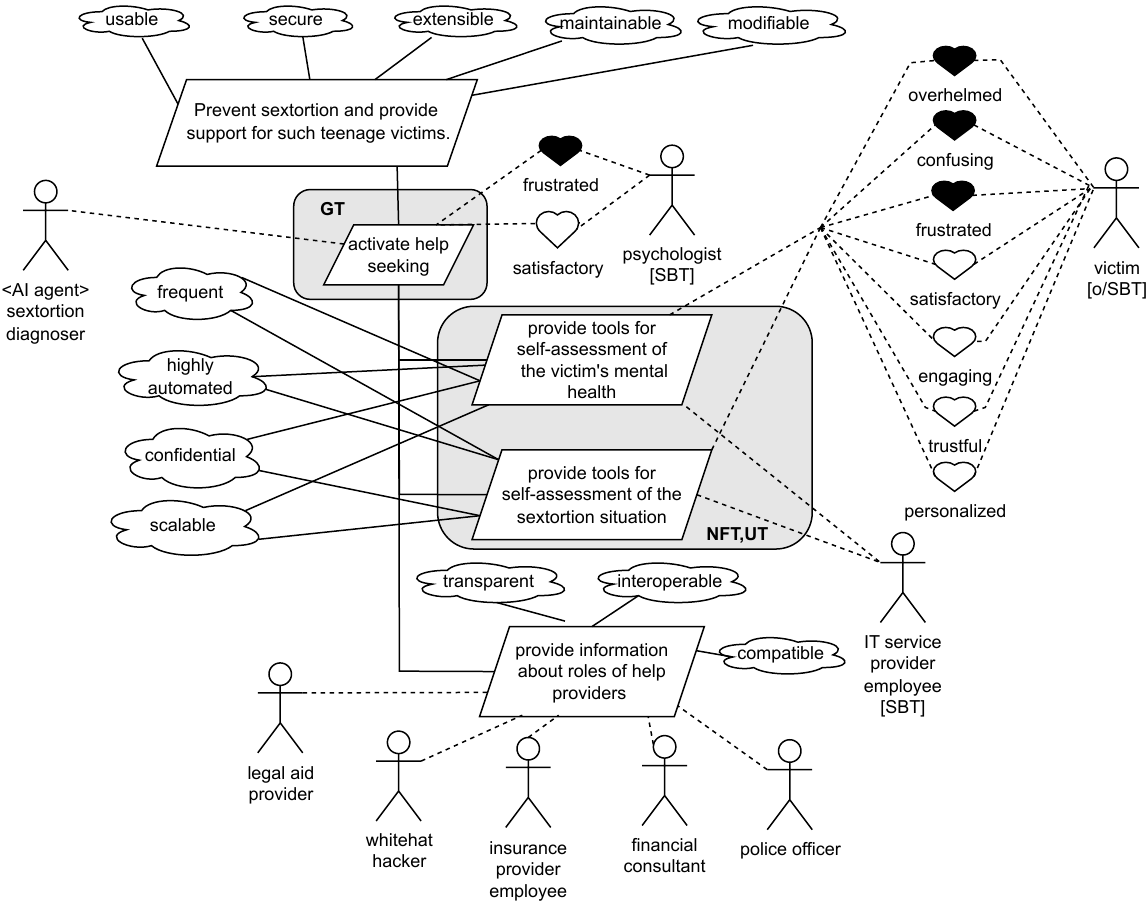}
        \caption{The goal-model refinement for activating help seeking.}
        \label{fig:goalhelpseek}
    \end{center}
    \vspace{-0.5cm}
\end{figure*}

The functional goal \textit{activate help seeking} in Figure~\ref{fig:goalhelpseek} has associated the AI agent \textit{sextortion diagnoser} and the human role \textit{psychologist} via the positive and emotional goals \textit{satisfactory} and \textit{frustrated} respectively. There are three further refining sub goals of which two have four quality goals assigned, namely the functional goals to provide tools for the self-assessment of the victim's mental health and self-assessment of the sextortion situation respectively. The associated quality goals to the two functional goals are \textit{frequent}, \textit{highly automated}, \textit{confidential}, and \textit{scalable}. Furthermore are two human roles associated to both functional goals being the \textit{victim} and the \textit{IT service provider employee}. The former is connected to these functional goals via a set of positive emotional goals being \textit{trustful}, \textit{engaging}, \textit{satisfactory}, \textit{frustrated}; and the negative emotional goals \textit{overwhelmed}, \textit{confusing}, \textit{frustrated}.

The bottom leave functional goal in Figure~\ref{fig:goalhelpseek} for \textit{providing information about the roles of the help providers} has three quality goals assigned being \textit{transparent}, \textit{interoperable}, and \textit{compatible}. The associated human roles to this functional goal are the \textit{legal aid provider}, the \textit{whitehat hacker}, the \textit{insurance provider employee}, the \textit{financial consultant}, and the \textit{police officer}.

\subsection{Token Economy Foundation}
\label{sec:tokeneconfound}

The four goal models in this section also lay the conceptual foundation for the token economy of the SocialDAO dApp. Thus, in Figures~\ref{fig:goalpreventsextortion}-\ref{fig:goalactivateaid} these blockchain zones with token economies are denoted as gray boxes that group functional goals and add the labels of the token types at the bottom. 

\subsubsection{Token types}
\label{sec:tokentypes}

We briefly list the token types that occur in the goal models above and describe their respective characteristics.  First, non-fungible tokens (NFT)~\cite{yaghy2023potential} are a type of digital asset that represent ownership or proof of authenticity of a unique item or piece of content using blockchain technology. Unlike cryptocurrencies like Bitcoin~\cite{nakamoto2008bitcoin} or Ethereum~\cite{buterin30ethereum}, which are fungible and can be exchanged on a one-to-one basis, NFTs are indivisible and each one is distinct from all others. NFTs have gained significant attention in the digital art, entertainment, and collectibles industries, but they can also have applications in various other domains. 

In the context of the sextortion SocialDAO dApp, NFTs find potential use in several ways. NFTs can be used to verify the authenticity of explicit multimedia content. When a user uploads content to the platform, it could be hashed, and the resulting hash could be stored as an NFT on the blockchain. This NFT would serve as proof that the content was indeed uploaded and can be used as evidence if required. NFTs manage and protect the privacy of users. For instance, certain NFTs could grant access to private chat rooms or specific features of the dApp. Users could control who has access to their content and when. NFTs create incentive structures for users and contributors to the platform. Users who actively participate in awareness campaigns or provide support to victims could earn NFT-based rewards or reputation tokens, enhancing the collaborative nature of the dApp. In the unfortunate event that sextortion occurs, NFTs collect and timestamp evidence. This evidence could then be stored on the blockchain, ensuring its integrity and making it admissible in legal proceedings. NFTs represent limited-edition digital assets related to awareness campaigns. These assets could be sold or auctioned, with the proceeds going to support victims or fund further anti-sexting education initiatives.

Utility tokens (UT)~\cite{benedetti2023utility}, also known as utility coins or app coins, are digital tokens or cryptocurrency units that provide access to specific services, features, or utilities within a particular platform or ecosystem. Unlike cryptocurrencies, utility tokens are designed to be used for specific purposes within a defined application or network. In the context of the sextortion SocialDAO dApp, UTs facilitate various functionalities and incentives. UTs allow access to premium features or services within the dApp, such as priority access to support, private chat rooms, or advanced search options. This creates a freemium model~\cite{benedetti2023utility}, where basic features are available for free, but enhanced capabilities require UTs. They serve as rewards for active participation in awareness campaigns, reporting sextortion cases, providing support to victims, or contributing content related to sextortion prevention. Users earn UTs as tokens of appreciation for their efforts. Content creators, such as educators or support providers, receive UTs as compensation for their contributions to the platform. This encourages experts and professionals to participate and share their expertise.  UTs enhance user privacy and security. For example, users spend UTs to access privacy-enhancing features, such as encrypted messaging or identity verification services. UTs are used for donations or crowdfunding campaigns aimed at supporting victims of sextortion or funding educational initiatives and awareness programs. UTs are used for voting on platform-related decisions, such as the allocation of funds, updates to platform rules, or the addition of new features. This would provide users with a say in the platform's direction. UTs are used to purchase or access exclusive NFTs related to awareness campaigns, collectibles, or special events. These NFTs could have real-world or in-app value.

Governance tokens (GT)~\cite{kozhan2022fundamentals} are a type of cryptocurrency token used in DAOs and blockchain-based platforms to facilitate decision-making and governance processes. Holders of governance tokens have the right to propose and vote on changes to the protocol, smart contracts, policies, and other aspects of the decentralized ecosystem. These tokens provide a way for participants to collectively manage and steer the direction of the platform.

In the context of the sextortion SocialDAO dApp, governance tokens (GT) are used in several ways to empower the community and drive decision-making. GT holders can vote on proposed changes to the dApp's policies, rules, and guidelines. This includes decisions related to user conduct, content moderation, and the overall governance framework. For example, the community can vote on updates to privacy policies or measures to enhance user safety. GT holders can participate in content moderation decisions. When explicit content or harmful behavior is reported, the community can collectively decide on the appropriate actions to take, such as removing content or suspending users. GTs can be used to prioritize and fund the development of new features and functionalities within the dApp. Users can propose ideas and initiatives and vote on which projects should receive resources and development efforts. GT can incentivize and reward active community members who contribute positively to the platform's objectives. This may include users who participate in awareness campaigns, provide support to victims, or contribute educational content. GT holders can influence the allocation of resources, such as funding for educational programs, support services, and legal initiatives aimed at preventing sextortion and helping victims. The community can use governance tokens to guide the long-term vision and strategic direction of the dApp. Decisions related to scaling, partnerships, and technological upgrades can be subject to community votes.  Governance tokens can be distributed as rewards to users who actively contribute to the prevention of sextortion and the support of victims. This incentivizes positive behavior and engagement within the community.

Finally, soulbound tokens (SBT) are a concept that extends from the idea of NFTs but with unique properties that emphasize their inextricable connection to a specific user or entity. We also use in the goal model \textit{[o/SBT]} to denote the optional use of this token for certain roles. Otherwise, the SBT token use is mandatory for human roles. Unlike traditional NFTs that represent ownership or proof of authenticity for digital or physical assets, soulbound tokens represent a unique bond between the token and the entity with which it is associated. This bond can be based on personal data, actions, or contributions within a community or ecosystem. In the context of the sextortion SocialDAO dApp, SBTs are employed in various ways to create personalized and meaningful interactions. In the case of victims of sextortion, SBT are used to personalize the support they receive. For example, victims could be matched with dedicated support agents or counselors based on their unique needs and experiences, as indicated by their SBT. SBT contribute to building trust and reputation within the dApp's community. Users with an SBT may be seen as trusted and respected members of the community. SBTs can also serve as a form of recognition and appreciation for users who actively contribute to the dApp's mission of preventing sextortion and supporting victims. These tokens are awarded to individuals or entities that have made a significant positive impact on the community. SBT are used to provide personalized incentives to users who engage in activities that align with the dApp's goals. For instance, users who create educational content, provide emotional support to victims, or report harmful content could earn SBT as a form of acknowledgment. SBT represent ownership or stewardship of specific content or resources within the dApp. Users who contribute valuable resources, such as educational materials or tools for self-protection, can receive SBT as a symbol of their dedication. SBT can foster a sense of belonging and community engagement. Users may receive SBT for participating in discussions, helping others, or sharing their personal experiences related to sextortion prevention. Depending on the design, SBT holders may have a stronger voice in the governance and decision-making processes of the dApp. They could participate in votes on policies, initiatives, or resource allocation. 

\section{Architecture Model for Sextortion Governance}
\label{sec:architecture}

In this section, we derive the architectural blueprint that underpins the sextortion SocialDAO. With an emphasis on user-centric design and ethical principles, this architecture model serves as the conceptual foundation for a blockchain-based dApp deployment that also integrates AI and decentralized governance to combat sextortion effectively. As we continuously detail this architectural framework, the goals of transparency and security are pursued to empower both victims and other actors within a trust-based ecosystem. The remainder of this section is structured as follows. 

The components in the models of this section are arranged and derived following the hierarchy of the corresponding goal models. Yet, the component diagrams do not refine beyond the second detail level. As Figure~\ref{fig:rootarchvalue} shows, the addition versus the goal mnodels is the presence of information-exchange interfaces between the components and actors.

\subsection{Top-Level SocialDAO Architecture}
\label{sec:toplevelarchitecture}

\begin{figure}[htpb]
    \vspace{0.2cm}
    \begin{center}
        \includegraphics[scale=0.9]{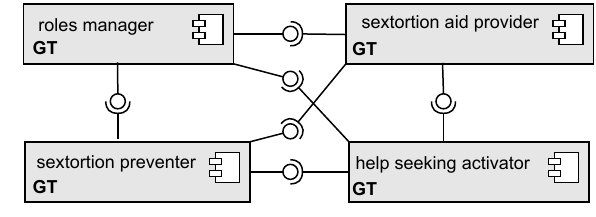}
        \caption{The highest-level architecture of the Social\textit{}DAO deduced from the first-level refinement quality goals.}
        \label{fig:rootarchvalue}
    \end{center}
    \vspace{-0.5cm}
\end{figure}

In Figure~\ref{fig:rootarchvalue}, the components depicted correspond to the first refinement level of the target models. These components are the \textit{roles manager}, the \textit{sextortion aid provider}, the \textit{sextortion preventer}, and the \textit{help seeking activator}. Each one of these components shows the use of GT governance tokens, which is typical for DAOs and their incentive management. Note that Figure~\ref{fig:rootarchvalue} abstracts from the actors associated to the components as they are depicted in the remaining figures of this section.

With respect to information exchanges, the \textit{roles manager} and the \textit{sextortion preventer} exchange as follows. The \textit{roles manager} relays information to the \textit{sextortion preventer} regarding the roles assigned to different actors within the DAO. This information is crucial for the \textit{sextortion preventer} to understand who is responsible for various aspects of sextortion prevention.  Information about actors, including their profiles, preferences, and permissions, are shared with the \textit{sextortion preventer}. This data is used by the \textit{sextortion preventer} to tailor its actions and responses based on user characteristics. If actors or other components of the dApp report incidents related to sextortion or suspicious activities, the \textit{roles manager} forwards these reports to the \textit{sextortion preventer} for further analysis and action. Any updates or changes in the governance policies or rules of the DAO are communicated from the \textit{roles manager} to the \textit{sextortion preventer} to ensure that the prevention measures are aligned with the latest policies. The \textit{sextortion preventer} requests real-time alerts or notifications from the \textit{roles manager} regarding specific actor activities or events that require immediate attention, such as potential threats to sextortion. Information related to actor authentication and access control, such as login credentials and permissions, are exchanged to ensure that only authorized individuals can engage with the \textit{sextortion preventer}. If the latter requires additional resources or permissions to carry out its functions effectively, it may request these from the \textit{roles manager}. The \textit{sextortion preventer} also provides feedback and analytical data to the \textit{roles manager}, helping to inform decisions about role assignments, policies, and system improvements. The \textit{roles manager} information exchanges with the \textit{sextortion aid provider} follows very similar principles as for the exchange with the \textit{sextortion preventer} so that the latter component can tailor the responses in an optimal way for the victim. 

The \textit{roles manager} and the \textit{help seeking activator} exchange information as follows. When an actor, such as a victim or someone in need of support, initiates a request for assistance, the \textit{help seeking activator} informs the \textit{roles manager} about the nature of the request, the actor's profile, and any specific requirements or preferences. The \textit{roles manager} can provide information to the \textit{help seeking activator} about the roles available within the system. This includes details about the types of assistance and support that can be offered, as well as the qualifications and expertise of individuals or entities occupying those roles. As assistance is provided to actors, the \textit{help seeking activator} can relay progress updates and feedback from users back to the \textit{roles manager}. This information can be valuable for assessing the effectiveness of assistance efforts and making any necessary adjustments. The \textit{roles manager} ensures that the assistance provided through the \textit{help seeking activator} complies with governance policies and ethical guidelines set by the sextortion SocialDAO. This maintains consistency and quality in the assistance offered. The \textit{help seeking activator} receives real-time alerts and notifications from the Roles Manager about available roles, new assistance requests, or any critical events that require attention. These alerts help streamline the assistance process. The \textit{help seeking activator} accesses user profiles and their assistance history, if applicable, to provide more personalized support and ensure that users receive assistance tailored to their needs.

The \textit{sextortion preventer} and \textit{sextortion aid provider} exchange information as follows. The \textit{sextortion preventer} provides threat intelligence and information on emerging threats, trends, and tactics of sextortion to the \textit{sextortion aid provider}. This information enables aid providers to stay informed about potential risks and tailor their support accordingly. The \textit{sextortion preventer} identifies potential victims of sextortion through various monitoring and detection mechanisms. It can relay information about these potential victims to the \textit{sextortion aid provider} so that they can reach out and offer assistance. The \textit{sextortion preventer} may refer victims or at-risk individuals to the \textit{sextortion aid provider} for further assistance. For example, if the \textit{sextortion preventer} identifies someone as a victim, it facilitates a direct connection to the aid provider for immediate support. The \textit{sextortion preventer} can send alerts and notifications to the \textit{sextortion aid provider} about critical incidents or high-risk situations. These alerts can trigger rapid responses and coordinated efforts to assist victims and prevent further harm. Both components share relevant data, such as victim profiles (with appropriate privacy measures), case histories, and behavioral patterns. These shared data help aid providers tailor their support and preventive measures based on individual circumstances. The \textit{sextortion aid provider} provides feedback to the \textit{sextortion preventer} about the effectiveness of preventive measures and early interventions. This feedback loop helps improve the overall prevention strategy. In cases where the \textit{sextortion aid provider} needs additional resources or expertise to assist victims, it requests support or collaboration from the \textit{sextortion preventer}. For example, if a legal issue arises, the \textit{sextortion aid provider} may seek legal aid through the \textit{sextortion preventer}. Both components ensure that their actions and interventions align with governance policies and ethical guidelines set by the sextortion SocialDAO. This ensures a coordinated and ethical approach to addressing sextortion. The \textit{sextortion preventer} receives incident reports from the \textit{sextortion aid provider}, providing details about cases, assistance provided, and outcomes. This reporting maintains transparency and accountability. Finally, the \textit{sextortion preventer} offers training materials and awareness campaigns to the \textit{sextortion aid provider} to keep the latter updated on best practices and the latest developments in sextortion prevention and victim support.

\subsection{Sextortion Preventer Architecture}
\label{sec:sexprevarchitecture}

\begin{figure*}[htpb]
    \vspace{0.2cm}
    \begin{center}
        \includegraphics[scale=0.8]{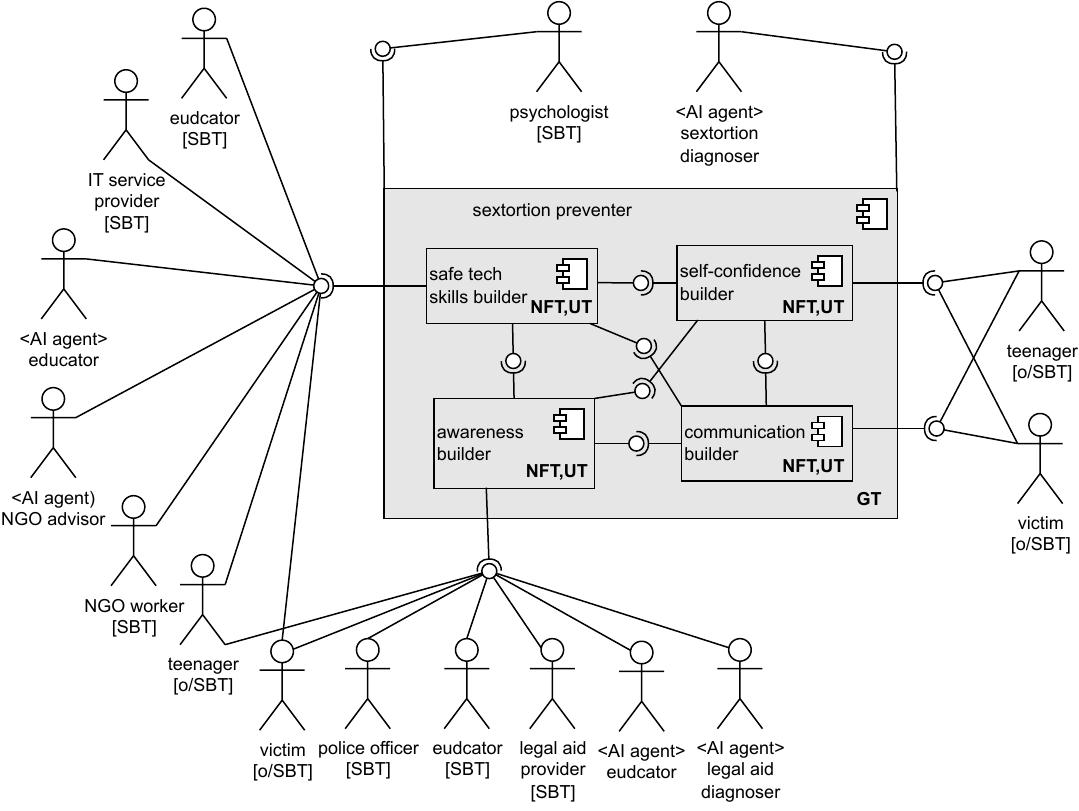}
        \caption{The architecture refinement for the \textit{sextortion preventer}.}
        \label{fig:archsexpreventer}
    \end{center}
    \vspace{-0.5cm}
\end{figure*}

In Figure~\ref{fig:archsexpreventer}, the \textit{sextortion preventer} architecture is depicted with the hierarchically contained components being the \textit{safe tech skills builder}, the \textit{self-confidence builder}, the \textit{awareness builder}, and the \textit{communication builder}. The \textit{sextortion preventer} has the GT assigned being the component responsible for governing the embedded components and the associated actors. 

We next explain the information-exchange channels between the embedded components in Figure~\ref{fig:archsexpreventer}. The \textit{safe tech skills builder} integrates elements of safe technology usage into the educational curriculum provided by the \textit{self-confidence builder}. This ensures that human actors not only develop self-confidence but also acquire practical skills for using technology safely. The two components share educational content and resources. For example, the \textit{safe tech skills builder} can provide resources on online safety, privacy, and cybersecurity that the \textit{self-confidence builder} incorporates into its confidence-building materials. The components collaborate on workshops or training sessions where individuals learn both self-confidence-building techniques and safe technology practices concurrently. This integrated approach allows participants to develop a holistic understanding of staying safe online. The components can exchange data related to actor progress. For instance, the \textit{safe tech skills builder} can share information about how well participants are adopting safe tech practices, and the \textit{self-confidence builder} can assess changes in actor self-confidence levels over time. They can establish a feedback loop to enhance their respective programs. The \textit{self-confidence builder} collects feedback from participants about their experiences with online safety, which informs improvements in the \textit{safe tech skills builder} curriculum, and vice versa. Depending on an actor-specific needs and vulnerabilities, the components share information to customize their support. For instance, if someone is particularly vulnerable to online threats, the \textit{safe tech skills builder} can inform the \textit{self-confidence builder} to provide additional confidence-building support. Both components align their objectives to ensure that actors are not only equipped with safe tech skills but also empowered with self-confidence to assert their boundaries and seek help when faced with sextortion risks. If one component identifies a need for specialized resources or expertise related to its area of focus, it requests support or collaboration from the other component. For example, if the \textit{self-confidence builder} identifies a case of sextortion, it can seek guidance from the \textit{safe tech skills builder} on addressing the technological aspects of the issue.

The \textit{safe tech skills builder} and \textit{awareness builder} exchange information in Figure~\ref{fig:archsexpreventer} as follows. The components can share educational content and resources. The \textit{safe tech skills builder} provides materials related to online safety, privacy, and cybersecurity, while the \textit{awareness builder} incorporates this content into its awareness campaigns and messaging. They collaborate on awareness campaigns and initiatives. For example, the \textit{safe tech skills builder} can contribute expertise on the technical aspects of online safety to the \textit{awareness builder}'s campaigns, ensuring that individuals are not only informed but also equipped with practical skills. The \textit{awareness builder} can integrate safe tech skills into its awareness materials. This ensures that individuals who are exposed to awareness campaigns also receive guidance on how to implement safe practices in their online activities. Establishing a feedback loop allows both components to improve their programs. The \textit{awareness builder} may collect feedback from awareness campaign participants about their experiences with online safety, which can inform updates to the \textit{safe tech skills builder}'s curriculum and vice versa. Information exchange can support customization based on individuals' needs. For instance, if someone is particularly receptive to awareness messages but lacks technical skills, the \textit{safe tech skills builder} can provide additional support tailored to their requirements. Both components can align their objectives to ensure that individuals not only gain awareness of sextortion risks but also have the technical skills to protect themselves. This alignment ensures a holistic approach to prevention. If one component identifies a need for specific resources or expertise related to its area of focus, it can request support or collaboration from the other component. For example, if the \textit{awareness builder} identifies a need for technical guidance, it can seek input from the \textit{safe tech skills builder}. The components can exchange information about emerging online threats or changes in technology that may affect actor safety. This enables them to provide timely updates and guidance to users.

The \textit{safe tech skills builder} and \textit{communication builder} exchange information in Figure~\ref{fig:archsexpreventer} as follows. The \textit{communication builder} collects feedback from awareness campaigns, including the effectiveness of messaging and communication channels. This feedback can be shared with the \textit{safe tech skills builder} to improve the content and delivery of prevention materials. The \textit{communication builder} provides insights into the preferences and behaviors of the target audience to help the \textit{safe tech skills builder} tailor its prevention strategies. For example, if certain communication channels or messaging styles are more effective, this information can inform the prevention approach. Information exchange can involve the integration of prevention content into awareness materials. The \textit{communication builder} incorporates tips and guidance from the \textit{safe tech skills builder} into its messaging to ensure that individuals not only become aware of sextortion risks but also receive actionable advice. Both components can share information about emerging threats or trends related to sextortion and online safety. This ensures that prevention and communication strategies remain up-to-date and relevant. The components collaborate on awareness campaigns that focus on communication skills and strategies. For instance, they jointly develop campaigns that educate individuals on effective communication to prevent sextortion attempts. The \textit{safe tech skills builder} provides insights into the psychological and behavioral aspects of sextortion, helping the \textit{communication builder} craft messages that resonate with the target audience and encourage desired behaviors. If one component requires additional resources or expertise related to its area of focus, it requests support or collaboration from the other component. This ensures that both prevention and communication efforts are adequately supported. Information exchange enables customization of prevention and communication strategies based on individual needs and preferences. This personalisation ensures that actors receive tailored support and guidance. 

The \textit{awareness builder} and \textit{self-confidence builder} exchange information in Figure~\ref{fig:archsexpreventer} as follows. The \textit{awareness builder} incorporates information and guidance from the \textit{self-confidence builder} into its awareness campaigns. For example, awareness materials include content that promotes self-confidence, resilience, and the importance of setting boundaries. Information exchange ensures that the messages conveyed through awareness campaigns align with the goal of building self-confidence. The \textit{self-confidence builder} provides input on messaging that reinforces positive self-image and empowerment. The \textit{awareness builder} shares insights about the actor's awareness levels and their receptiveness to messages related to self-confidence. This information supports the \textit{self-confidence builder} tailor its content to effectively reach and resonate with the audience. The \textit{self-confidence builder} provides insights into the psychological and behavioral aspects of self-confidence and self-esteem. This information can guide the \textit{awareness builder} in crafting messages that address specific challenges related to self-confidence. Both components can share resources and materials that support their respective goals. For instance, the \textit{self-confidence builder} may provide self-help resources that can be included in awareness campaigns to empower individuals. The components collaborate on campaigns that focus on both awareness and self-confidence. These campaigns may promote the idea that self-confidence is a key element in protecting oneself from sextortion and other online risks. Continuous feedback and data exchange supports both components to refine their strategies. For example, if awareness campaigns positively impact self-confidence levels, this feedback informs future campaigns. Information exchange enables the customization of awareness materials to address the specific needs of individuals in terms of building self-confidence. Personalized content is more effective in empowering users. Both components collaborate on monitoring and evaluating the impact of their efforts. They share data on awareness levels, self-confidence improvements, and other relevant metrics to assess the effectiveness of their joint initiatives.

The \textit{communication builder} and \textit{self-confidence builder} exchange information in Figure~\ref{fig:archsexpreventer} as follows. The \textit{communication builder} incorporates content and guidance from the \textit{self-confidence builder} into its communication materials. For instance, communication resources include messaging that encourages assertiveness, setting boundaries, and building self-confidence. The \textit{self-confidence builder} provides insights into the development of communication skills, such as assertiveness, active listening, and conflict resolution. The \textit{communication builder} incorporates these insights into its resources and training materials. The \textit{self-confidence builder} provides insights into the development of communication skills, such as assertiveness, active listening, and conflict resolution. The \textit{communication builder} incorporates these insights into its resources and training materials. The \textit{self-confidence builder} offers insights into the psychological and behavioral aspects of self-confidence and how they relate to effective communication. This information guides the \textit{communication builder} in crafting messages and materials that empower individuals in their communication efforts. The components collaborate on role-playing scenarios and training exercises that combine self-confidence-building activities with effective communication techniques. This approach supports actors develop practical skills while boosting their confidence. Both components share resources, such as articles, videos, and interactive tools, that support their respective goals. These resources are integrated into a unified platform for actors to access. The components organize joint workshops or webinars that focus on both self-confidence and effective communication. These events provide participants with the opportunity to practice communication skills in a supportive environment. Continuous feedback and data exchange supports both components refine their strategies. For example, if communication training positively impacts self-confidence levels, this feedback informs future training sessions and materials. Information exchange enables the customization of communication resources to meet the specific needs of individuals in terms of self-confidence. Personalized content empowers users to communicate more confidently in their unique situations. Both components collaborate on monitoring and evaluating the impact of their efforts. They share data on communication effectiveness, self-confidence improvements, and other relevant metrics to assess the success of their joint initiatives.

The \textit{awareness builder} and \textit{communication builder} exchange information in Figure~\ref{fig:archsexpreventer} as follows. The \textit{communication builder} incorporates content and messaging from the \textit{awareness builder} into its communication materials. This ensures that communication resources align with the broader goal of building awareness about sextortion and its prevention. Information exchange is instrumental to coordinate messaging efforts. The \textit{awareness builder} provides insights into the most effective ways to raise awareness about sextortion, which influence the content and style of communication materials created by the \textit{communication builder}. The \textit{awareness builder} collaborates with the \textit{communication builder} to develop and execute awareness campaigns. These campaigns include educational materials, social media posts, and other content that simultaneously informs and empowers individuals to prevent sextortion. Both components cross-promote each other's initiatives. For example, awareness-building content includes calls to action for users to access communication resources, creating a seamless transition from awareness to action. The \textit{awareness builder} offers insights into the psychological and behavioral aspects of sextortion awareness and prevention. This information guides the \textit{communication builder} in crafting messages and materials that resonate with actors and motivate them to take preventive actions. Collaborative efforts lead to the creation of interactive content, such as quizzes, surveys, or interactive webinars, that combines awareness-raising activities with communication skill-building. This engagement enhances the learning experience. Continuous feedback and data exchange support both components refine their strategies. User engagement and response data informs adjustments to awareness campaigns and communication materials for greater effectiveness. The components share relevant resources, such as statistics, case studies, or testimonials, that reinforce both awareness and communication efforts. Sharing compelling stories humanizes the issue and motivate individuals to take preventive actions. Collaborative events, such as seminars or workshops, can be organized to address both awareness and communication needs. These events include sessions on recognizing sextortion risks, followed by communication skill-building exercises. Both components collaborate on monitoring and evaluating the impact of their joint initiatives. They share data on awareness levels, communication effectiveness, and other relevant metrics to assess the success of their efforts. Information exchange enables the customization of awareness and communication resources to meet the specific needs and concerns of individuals. Personalized content increases engagement and relevance. The components work together in advocacy efforts aimed at policy changes or legal actions to combat sextortion. Effective communication and increased awareness supports these advocacy campaigns. 

The human actor \textit{psychologist} and the \textit{sextortion diagnoser} AI agent exchange information in Figure~\ref{fig:archsexpreventer} with the \textit{sextortion preventer} component as follows. The \textit{sextortion diagnoser} AI agent analyzes and detects potential sextortion cases or risks by processing data from various sources, such as online communications or behavior patterns. It then share this analyzed data with the \textit{sextortion preventer} component. When the \textit{sextortion diagnoser}identifies a potential sextortion case or a user at risk, it generates alerts or notifications. These alerts are forwarded to both the \textit{sextortion preventer} and the human \textit{psychologist} for immediate attention and intervention. The \textit{sextortion diagnoser} provides risk assessment reports and insights regarding the severity of potential sextortion cases. This information guides the \textit{psychologist} and the \textit{sextortion preventer} in prioritizing and addressing cases based on their urgency. The \textit{psychologist} and the \textit{sextortion preventer} communicate with the \textit{sextortion diagnoser} for additional information, clarification, or updates related to identified cases. This communication channel ensures that all parties involved have access to the most up-to-date information. Based on the information provided by the \textit{sextortion diagnoser}, the \textit{psychologist} and the \textit{sextortion preventer} collaborate to develop intervention strategies tailored to the specific cases. This involves deciding on appropriate actions, such as providing support, reporting to authorities, or initiating educational programs. The \textit{psychologist} and the \textit{sextortion preventer} provide feedback to the \textit{sextortion diagnoser} on the accuracy of its assessments and the outcomes of interventions. This feedback loop support improving the AI agent's detection capabilities over time. Given the sensitive nature of sextortion cases, it is crucial to ensure that information exchange respects privacy and confidentiality. Encryption and secure communication protocols are implemented to safeguard sensitive data. Information from the \textit{sextortion diagnoser} clarifies resource allocation decisions. For example, if the AI agent identifies a surge in sextortion cases among a particular demographic, resources are directed toward targeted awareness campaigns or support services. The \textit{sextortion diagnoser} learns from the expertise and experiences of the human \textit{psychologist}. Insights and strategies shared by the \textit{psychologist} are used to enhance the AI agent's capabilities and decision-making processes.

The component \textit{safe tech skills people}  exchanges information in Figure~\ref{fig:archsexpreventer} with the human actors termed \textit{educator}, \textit{IT service provider}, \textit{NGO worker}, \textit{teenager}, and \textit{victim}. Additional non-human actors are the AI agents \textit{educator} and \textit{NGO advisor}. The exchange of information between the component \textit{safe tech skills people} and the various human actors, as well as the AI agents, serves the purpose of building safe technology skills among individuals, especially \textit{teenagers} and potential \textit{victims} of sextortion. The \textit{educator} provides educational content, materials, and guidance to the \textit{safe tech skills people} component. This information includes lessons, tutorials, and resources related to safe technology use. The purpose is to equip individuals, especially \textit{teenagers}, with the knowledge and skills needed to protect themselves from sextortion and online threats. The \textit{IT service provider} offers technical expertise, updates on the latest cybersecurity practices, and advice on safe technology usage. This information supports individuals to stay informed about potential risks and protective measures related to their online activities. \textit{NGO workers} contribute by sharing awareness campaigns, educational programs, and support resources with the \textit{safe tech skills builder}. This information assists in creating a well-rounded educational approach that addresses both awareness and practical skills. \textit{Teenagers}  provide feedback, questions, and insights regarding their experiences and challenges in using technology safely. This feedback loop allows the \textit{safe tech skills people} to tailor its educational content to the specific needs and concerns of teenagers. Victims of sextortion share their experiences and provide insights into the tactics used by perpetrators. While sensitive, this information is input for creating educational content that highlights real-world risks and prevention strategies. The AI agent with the role of an \textit{educator} assists in generating and disseminating educational content efficiently. It adapts content based on the needs and progress of learners, providing personalized guidance and feedback. The AI agent with the role of an \textit{NGO advisor} provides recommendations and insights on the most effective educational approaches, awareness campaigns, and support resources. It analyzes data to identify trends and areas of focus.

Next, the \textit{awareness builder} exchanges information in Figure~\ref{fig:archsexpreventer} with the human actors \textit{teenager}, \textit{victim}, \textit{police officer}, \textit{educator}, and the \textit{legal aid provider}. Additional non-human actors are the AI agents \textit{educator} and \textit{legal aid diagnoser}. \textit{Teenagers} provide insights into their online experiences, challenges, and concerns related to sextortion and online safety. They also share their preferences for receiving awareness messages and educational content. This information  supports tailoring awareness campaigns to resonate with the \textit{teenage} audience and address their specific needs. \textit{Victims} of sextortion share their personal experiences, including how they were targeted, the emotional impact, and the challenges they faced. This firsthand information is anonymized and used to create awareness materials that highlight the real consequences of sextortion and the importance of prevention. \textit{Police officers} provide data and insights into sextortion cases they have encountered in their work. This information includes common patterns, trends, and challenges in handling such cases. It supports creating awareness content that educates the public about reporting sextortion and seeking legal support. \textit{Educators} collaborate by providing input on the most effective methods of delivering awareness messages to students and integrating them into educational curricula. They also offer feedback on the relevance and impact of awareness campaigns in school settings. \textit{Legal aid providers} contribute legal expertise and insights into the legal aspects of sextortion cases. They share anonymized case studies and legal resources that are used to inform the public about their rights and available legal support. The AI agent with the role of an \textit{educator} assists in creating awareness content, quizzes, and interactive materials that engage and educate users effectively. It adapts content based on user feedback and learning preferences, ensuring that awareness campaigns are engaging and informative. The AI agent with the role of a \textit{legal aid diagnoser} provides information on legal resources, procedures, and rights related to sextortion cases. It generates legal guidance documents and frequently asked questions to assist \textit{victims} and the public in understanding their legal options.

Finally, the human actors \textit{teenager} and \textit{victim} in Figure~\ref{fig:archsexpreventer} exchange information with the two components \textit{self-confidence builder} and \textit{communication builder}. \textit{Teenagers} provide insights with the \textit{self-confidence builder} into their self-esteem, self-confidence, and any challenges they face in dealing with peer pressure, online interactions, and self-image. They may share their experiences, concerns, and areas where they feel they lack confidence. \textit{Teenagers} provide feedback with the \textit{communication builder} on their communication skills, including their ability to express themselves, ask for help, or assert their boundaries. They share their preferences for communication styles and channels. \textit{Victims} of sextortion share with the \textit{self-confidence builder} their experiences and the emotional toll of being victimized. They provide information about their self-confidence levels, which may be affected by the incident. The component offers guidance and support to help victims rebuild their self-confidence. \textit{Victims} communicate their difficulties with the \textit{communication builder} in reaching out for help or sharing their experiences with others. The \textit{communication builder} provides resources and strategies to improve their communication skills, helping them express their needs and emotions effectively.

\subsection{Roles Manager Architecture}
\label{sec:rolesmanagerarchitecture}

\begin{figure*}[htpb]
    \vspace{0.2cm}
    \begin{center}
        \includegraphics[scale=0.8]{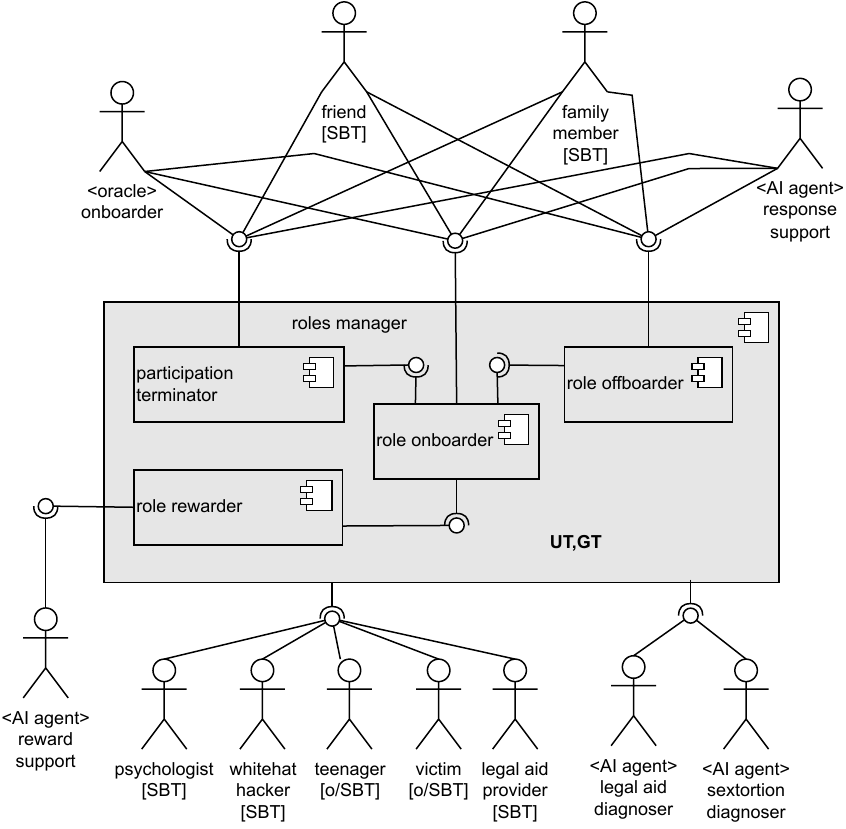}
        \caption{The architecture refinement for the \textit{roles manager}.}
        \label{fig:archrolesmanage}
    \end{center}
    \vspace{-0.5cm}
\end{figure*}

In Figure~\ref{fig:archrolesmanage}, the \textit{roles manager} component has hierarchically embedded the components labeled \textit{participation terminator}, \textit{role rewarded}, \textit{role onboarder}, and \textit{role offboarder}. The embedded components exchange information with each other. The \textit{role onboarder} and the \textit{participation terminator} components exchange information as follows. This exchange of information serves the purpose of notifying the \textit{participation terminator} component about the successful onboarding of a role. It includes details about the role, such as its responsibilities, permissions, and the actor assigned to it. For example, when a new role is created or assigned within the system, the \textit{role onboarder} component communicates this information to the \textit{participation terminator}. This notification supports the ensurance that the \textit{participation terminator} is aware of the role's existence and can track its participation status.

The \textit{role onboarder} and the \textit{role rewarder} components exchange information as follows. fairly similarly to the previous case. Thus, information flows from the \textit{role onboarder} component to the \textit{role rewarder} component. Also this exchange of information serves the purpose of notifying the \textit{role rewarder} component about the successful onboarding of a new role. It includes details about the role, such as its responsibilities, permissions, and the actor assigned to it. Thus, when a new role is created or assigned within the system, the \textit{role onboarder} component communicates this information to the \textit{role rewarder}. This notification supports the \textit{role rewarder} component identify the addition of a new role and take any actions related to rewarding or incentivizing the actor assigned to that role.

The \textit{role onboarder} and the \textit{role offboarder} components exchange information as follows. Information flows from the \textit{role offboarder} component to the \textit{role onboarder} component. This exchange of information serves the purpose of notifying the \textit{role onboarder} component about the offboarding or removal of a role from the system. It informs the \textit{role onboarder} that a particular role should no longer be considered active or available within the system. For instance, when a role needs to be removed or offboarded from the system, the \textit{role offboarder} component communicates this information to the \textit{role onboarder}. This notification supports the \textit{role onboarder} component in updating its records and ensure that the offboarded role is no longer active or accessible within the system.

The \textit{roles manager} exchanges information with human actors being the \textit{psychologist}, \textit{whitewater hacker}, \textit{teenager}, \textit{victim}, and the \textit{legal aid provider}. Additionally, the \textit{roles manager} alao exchanges information with non-human actors being the AI agents \textit{legal aid diagnoser} and \textit{sextortion diagnoser}. The \textit{psychologist} exchanges information with the \textit{roles manager} to provide insights, recommendations, or assessments related to the roles and their management. This include feedback on role definitions, role performance, or any psychological aspects relevant to role management. The \textit{whitehat hacker} communicates with the \textit{roles manager} to report security vulnerabilities, potential threats, or issues related to role-based access control within the system. This exchange ensures the security of role management. These roles, represented by the \textit{teenager} and \textit{victim}, interact with the \textit{roles manager} to request role-related actions. For example, a \textit{victim} seeks assistance or a change in their role status, and the \textit{roles manager} facilitates these requests. The \textit{legal aid provider} communicates with the \textit{roles manager} to coordinate legal assistance for victims or provide updates on legal matters related to role management. These AI agents interact with the \textit{roles manager} to assist in diagnosing and managing legal or sextortion-related issues. They provide recommendations, alerts, or data analysis to support role-related decisions and actions. The purpose of these information exchanges are as follows. The \textit{roles manager} ensures that roles are properly defined, assigned, and managed within the system. It receive requests for role changes, updates, or removals from human actors and AI agents. Interaction with the \textit{whitehat hacker} maintains the security of role-based access control by addressing vulnerabilities and threats. The \textit{legal aid provider} and AI agents assist in legal matters, providing guidance, recommendations, and updates related to role management in the context of sextortion prevention. The \textit{psychologist}, \textit{teenager}, and \textit{victim} provide input and requests related to their roles, and the \textit{roles manager} ensures these requests are appropriately processed. 

The components \textit{participation terminator} and \textit{roles offboarder} in Figure~\ref{fig:archrolesmanage} exchange information with the following human actors being the \textit{friend} and \textit{family member}. Additional non-human actors the components exchange information with are an \textit{onboarding} oracle and a \textit{response support} AI agent. The \textit{participation terminator} and \textit{roles offboarder} receive information from friends and family members of a participant (e.g., a victim or teenager) regarding their well-being or concerns related to their participation in the SocialDAO. The purpose of this information exchange is to ensure the safety and well-being of participants. Friends and family members provide valuable insights or raise concerns that require appropriate actions, such as participant support or, in some cases, termination of participation if it is in the best interest of the participant. The \textit{participation terminator} and \textit{roles offboarder} interact with non-human actors such as the \textit{onboarding oracle} and \textit{response support} AI agent to gather relevant information about the participant's onboarding process, status, or any ongoing support activities. This information exchange serves several purposes. The \textit{onboarding} oracle provides information about the participant's initial onboarding, which helps determine if they have completed necessary training or orientation. The \textit{response support} AI agent provides insights into ongoing support or counseling sessions with participants. Information from these non-human actors aids in assessing the participant's status, including any risk factors or necessary interventions.

Finally, the \textit{reward support} AI agent exchanges information with the \textit{role rewarder} component. Thus, the \textit{reward support} AI agent communicates with the \textit{role rewarder} component to share data related to participant rewards, incentives, or recognition. This information exchange serves the following purposes. The AI agent provides information about rewards that are allocated to actors based on their contributions, achievements, or adherence to the SocialDAO's guidelines and objectives. It involves updates on the recognition or acknowledgment that actors are entitled to receive for their active and meaningful contributions to the SocialDAO. The information exchange also includes details about any incentives or benefits that actors have earned as part of their engagement with the platform.

\subsection{Sextortion Aid Provider Architecture}
\label{sec:sexaidprovarchitecture} 

\begin{figure*}[htpb]
    \vspace{0.2cm}
    \begin{center}
        \includegraphics[scale=0.8]{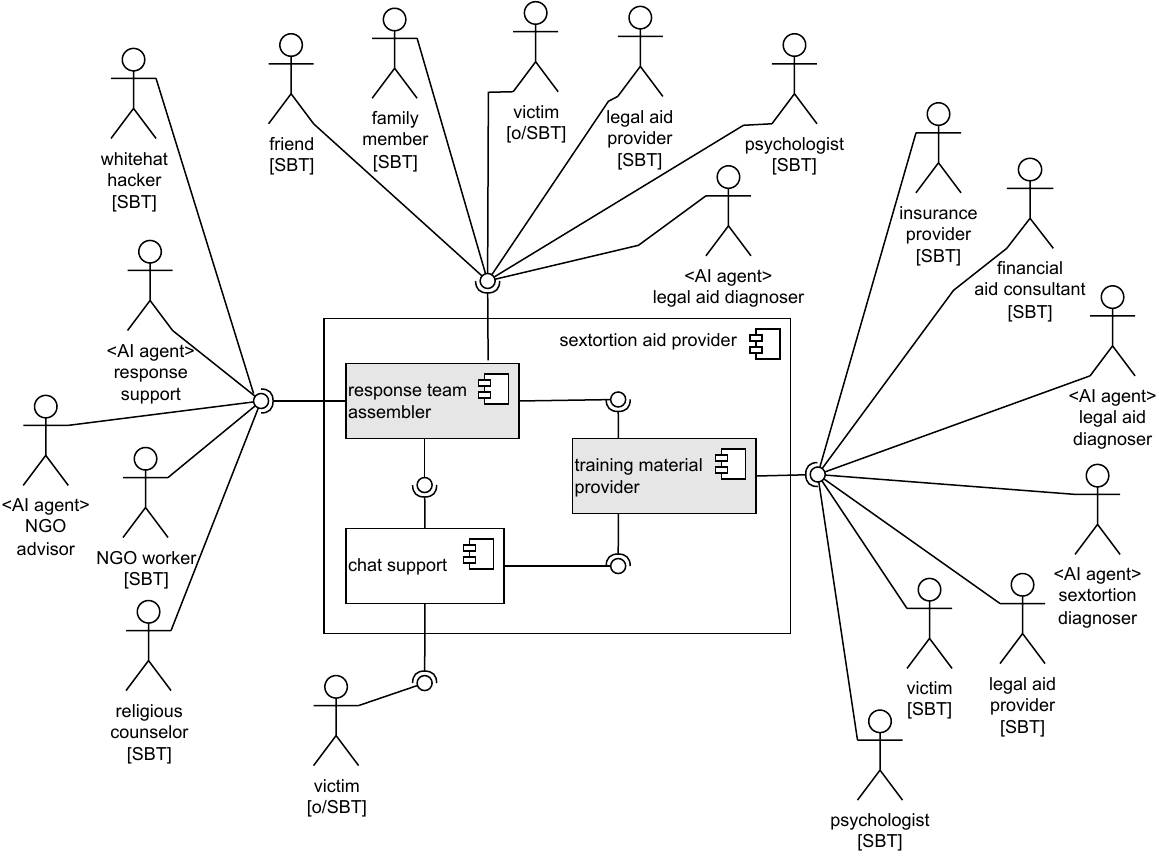}
        \caption{The architecture refinement for the \textit{sextortion aid provider}.}
        \label{fig:archsexaidprov}
    \end{center}
    \vspace{-0.5cm}
\end{figure*}

In Figure~\ref{fig:archsexaidprov}, the \textit{sextortion aid provider} component contains embedded additional components termed \textit{response team assembler}, \textit{training material provider}, and the \textit{chat support}. Thus, the \textit{response team assembler} and the \textit{training material provider} components exchange information as follows. These two embedded components exchange information to facilitate the seamless operation of the \textit{sextortion aid provider} and enhance its effectiveness in combating sextortion. The \textit{response team assembler} communicates its resource needs to the \textit{training material provider} to ensure that the necessary materials, such as training modules, educational content, or documentation, are available for the response team. Information related to the composition of the response team, including the roles, skills, and expertise required, is shared between these components to ensure that the team is adequately prepared. If the \textit{training material provider} has updates or improvements to training materials, it relays this information to the \textit{response team assembler} for the team's continuous improvement. There may be a feedback loop where the \textit{response team assembler} provides feedback on the effectiveness of the training materials, and this information is used by the \textit{training material provider} to refine and update the content. 

The \textit{response team assembler} and the \textit{training material provider} components exchange information with the \textit{chat support} component too. The exchange of information between these components is crucial for creating a cohesive and effective support system for victims of sextortion. The \textit{response team assembler} communicates with the \textit{chat support} to ensure that the response team is available and ready to provide immediate assistance through chat. The \textit{training material provider} may provide relevant training materials or guidance to the \textit{chat support} component to ensure that support agents are well-informed and equipped to handle sextortion cases. The \textit{chat support} component interacts directly with victims, offering immediate assistance, guidance, and emotional support. It shares anonymized or generalized feedback with the \textit{response team assembler} to help improve the overall response process. There could be a feedback loop where the \textit{chat support} component provides feedback on the effectiveness of training materials or the response team's performance. This information is used by both the \textit{training material provider} and the \textit{response team assembler} for continuous improvement.

Furthermore, the component \textit{response team assembler} in Figure~\ref{fig:archsexaidprov}  exchanges information with the human actors \textit{whitehat hacker}, \textit{NGO worker}, \textit{religious counselor}; and the non-human actors being the AI agents \textit{response support} and \textit{NGO advisor}. The component and the actors exchange information as follows. The \textit{response team assembler} communicates with the \textit{whitehat hacker} to coordinate technical aspects of sextortion prevention, cyber-security, or digital forensics. This involves sharing information about ongoing cases, technical vulnerabilities, or collaborative efforts to combat sextortion. Interaction with the \textit{NGO worker} involves coordination on various aspects of victim support, awareness campaigns, and educational initiatives. The \textit{response team assembler} shares information about victim cases, assistance required, or collaborate on awareness-building strategies. Communication with the \textit{religious counselor} focuses on providing emotional and psychological support to victims from a faith-based perspective. The \textit{response team assembler} exchanges information regarding the emotional state and needs of victims, seeking guidance or counseling support. The AI agent \textit{response support} assists the \textit{response team assembler} by providing real-time information, recommendations, or suggested responses to sextortion cases. It analyzes the incoming data and suggests appropriate actions or resources to the human response team. The AI agent \textit{NGO advisor} provides guidance and recommendations to the \textit{response team assembler} on NGO-related activities, legal matters, or ethical considerations. It analyzes data and offers insights to improve the effectiveness of the response team.

Additional information exchanges of the component \textit{response team assembler} in Figure~\ref{fig:archsexaidprov} occur with the human actors \textit{friend}, \textit{family member},  \textit{victim}, \textit{legal aid provider}, \textit{psychologist}, and the non-human AI agent \textit{legal aid diagnoser}. \textit{Friends} and \textit{family members} of the \textit{victim} share information about the \textit{victim}'s situation, concerns, and any sextortion-related incidents they are aware of. The \textit{response team assembler} communicates with friends and family members to coordinate emotional support and gather insights into the victim's well-being. The \textit{victim} reports sextortion incidents, share evidence, and provide information about the extortionist or the situation. The \textit{victim} reaches out to request assistance, counseling, or legal support. The \textit{response team assembler} provides the \textit{victim} with updates on the progress of the response efforts and available support resources. The \textit{legal aid provider} informs the \textit{response team assembler} about the availability of legal assistance, counseling services, or other forms of support. Legal professionals offer insight into legal procedures, potential actions against the extortionist, and victims' rights. \textit{Psychologists} provide emotional and psychological support to victims. The \textit{response team assembler} shares relevant information about the \textit{victim}'s emotional state and progress in counseling. The \textit{legal aid diagnoser} AI agent analyzes data related to sextortion cases, identifies potential legal issues, and recommends appropriate legal actions. It provides \textit{legal aid providers} with information on legal strategies, precedents, or relevant laws to support victims effectively. The purposes of these information exchanges are multifaceted. The \textit{response team assembler} acts as a central coordinator, gathering information from various sources to assess the severity of the sextortion case and prioritize response efforts.  The information shared by friends, family members and \textit{victim} supports tailoring emotional support, counseling, and legal assistance to the specific needs of \textit{victim}. The component ensures that the right resources, such as legal aid and counseling, are allocated to effectively address the sextortion incident. Providing status updates to the \textit{victim} and involved actors keep everyone informed about the progress of the response and support efforts. The \textit{legal aid diagnoser} AI agent contributes data-driven insights, supporting \textit{legal aid providers} to make informed decisions about legal actions and strategies. Information exchange respects the privacy and confidentiality of victims, adhering to ethical and legal standards.

The component \textit{training material provider} exchanges information in Figure~\ref{fig:archsexaidprov} with the human actors \textit{insurance provider}, \textit{financial aid consultant}, \textit{legal aid provider}, \textit{victim}, and \textit{psychologist}. Additional information exchanges occur from the component to the non-human AI agents \textit{legal aid diagnoser} and the \textit{sextortion diagnoser}. The concrete information exchanges occur as follows. The \textit{training material provider} shares educational materials and resources related to sextortion prevention, awareness, and support with the \textit{insurance provider}. This information exchange supports the \textit{insurance provider} to understand and promote preventive measures and resources. Information related to financial assistance options, resources for \textit{victims}, and preventive measures are shared with the \textit{financial aid consultant}. This exchange provides financial guidance to \textit{victims} and raises awareness of available support. Educational materials about legal rights, procedures, and resources are exchanged with \textit{legal aid providers}. This supports them in offering tailored legal guidance to \textit{victims}. \textit{Victims} receive training materials, guides, and resources for self-help and awareness. These materials help victims understand their situation better, make informed decisions, and access support. \textit{Psychologists} receive educational materials and resources related to psychological support for \textit{victims} of sextortion. This exchange ensures that \textit{psychologists} have access to relevant materials for counseling. The \textit{training material provider} shares data and information related to legal aspects, cases, and strategies with the \textit{legal aid diagnoser} AI agent. This exchange supports the AI agent in providing legal insights and guidance to \textit{legal aid providers}. Data and insights related to sextortion incidents and trends are exchanged with the \textit{sextortion diagnoser} AI agent. This exchange helps the AI agent analyze and diagnose sextortion cases effectively. The purposes of these information exchanges are as follows. Sharing training materials with human actors raises awareness about sextortion, preventive measures, and available resources. It empowers them to provide informed assistance to \textit{victims}. \textit{Victims} receive educational materials that empower them to understand their situation, make informed decisions, and access the necessary support services. AI agents, such as the \textit{legal aid diagnoser} and \textit{sextortion diagnoser}, use exchanged data and information to provide data-driven insights and recommendations to human actors. \textit{Legal aid providers}, \textit{financial aid consultants}, and \textit{psychologists} benefit from tailored training materials that align with their respective areas of expertise. Information sharing fosters coordination among different actors, ensuring a holistic approach to supporting \textit{victims} of sextortion.

Finally, the human actor \textit{victim} exchanges information with the \textit{chat support} component in the following way and for the purpose of receiving immediate assistance and guidance. \textit{Victims} initiate a chat session with the \textit{chat support} component through a user interface or platform. They share details of their sextortion experience, concerns, and questions. The primary purpose of this information exchange is to provide immediate support and assistance to \textit{victims}. The \textit{chat support} component can engage in real-time conversations with \textit{victims}, offering empathetic listening, emotional support, and guidance. \textit{Victims} receive information, resources, and guidance on steps to take when faced with sextortion. The \textit{chat support} component provides information on reporting the incident, seeking legal help, emotional coping strategies, and self-protection measures. \textit{Victims} often experience fear, anxiety, and distress. The \textit{chat support} component offers reassurance, validate their feelings, and aupport them feel heard and understood. Depending on the severity of the case, the \textit{chat support} component refers \textit{victims} to appropriate professionals, such as psychologists, legal aid providers, or law enforcement agencies. 

\subsection{Help Seeking Activator Architecture}
\label{sec:helpseekactarchitecture}

\begin{figure*}[htpb]
    \vspace{0.2cm}
    \begin{center}
        \includegraphics[scale=0.8]{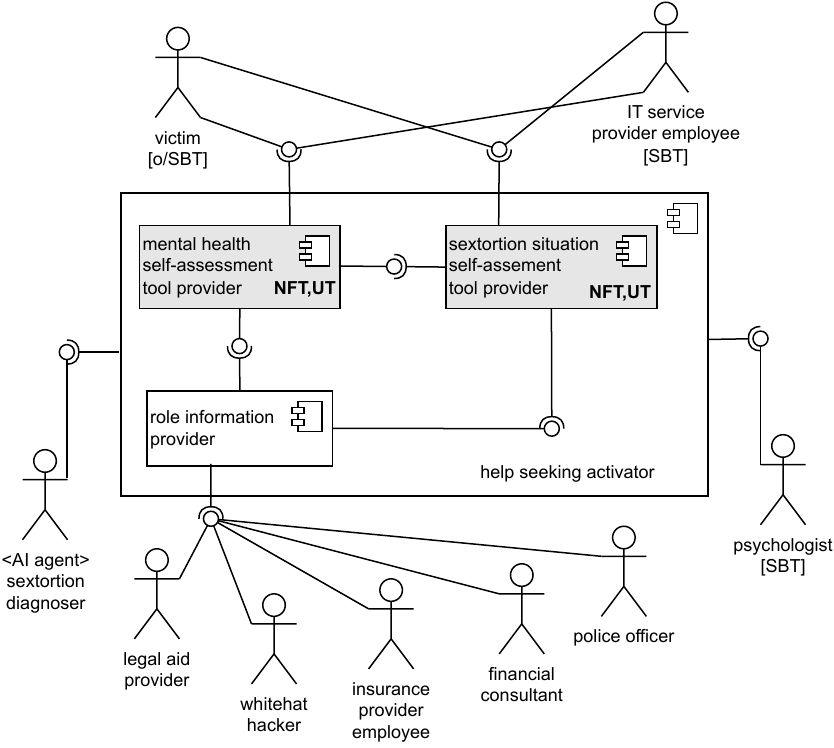}
        \caption{The architecture refinement for the help seeking activator.}
        \label{fig:archhelpactivator}
    \end{center}
    \vspace{-0.5cm}
\end{figure*}

In Figure~\ref{fig:archhelpactivator}, the component \textit{help seeking activator} contains several further embedded components named \textit{mental health self-assessment tool provider}, \textit{sextortion situation self-assessment tool provider}, and the \textit{role information provider}. The \textit{mental health self-assessment tool provider} and \textit{sextortion situation self-assessment tool provider} components exchange information as follows. Actors, including potential \textit{victims} and individuals seeking information, access the self-assessment tools provided by these components through a user interface or platform. Actors interact with these tools by answering questions or providing relevant information about their mental health status and the specific situation related to sextortion. The primary purpose of this information exchange is to enable actors to self-assess their mental health and evaluate the severity or impact of the sextortion situation they are facing. Actors may answer questions related to their emotional well-being, stress levels, anxiety, and other mental health factors. Additionally, they may assess the details of the sextortion incident, such as the extent of harassment, threats, or emotional distress. Based on the information provided during the self-assessment, these components offer guidance and recommendations to actors. For example, if an actor's self-assessment indicates high levels of distress, they may receive guidance on seeking immediate professional help or contacting support services. Similarly, actors receive recommendations on self-help strategies, coping mechanisms, or next steps based on their assessment results. The goal of this exchange is to empower actors to make informed decisions about their mental health and the sextortion situation they are dealing with. By providing self-assessment tools, actors gain a better understanding of their own well-being and take proactive steps toward seeking help or support. It is important to ensure that the information exchanged through these self-assessment tools is kept confidential and secure. Actors should feel comfortable sharing their thoughts and feelings without fear of privacy breaches.

Next, the components \textit{mental health self-assessment tool provider} and \textit{role information provider} exchange information in Figure~\ref{fig:archhelpactivator} as follows. Actors who are accessing the \textit{mental health self-assessment tool provider} also have questions or require information about the roles and resources available to them within the sextortion SocialDAO. This information exchange enables actors to access role-specific guidance and resources in addition to assessing their mental health. Based on the actor's self-assessment and any specific questions or needs related to roles and resources, the \textit{mental health self-assessment tool provider} requests relevant information from the \textit{role information provider}. For example, an actor expresses concerns about their role as a \textit{victim} and seek guidance on available support services or legal assistance. In response, the \textit{mental health self-assessment tool provider} requests role-specific information from the \textit{role information provider}. The \textit{role information provider} supplies role-specific resources, such as contact details of \textit{legal aid providers}, \textit{psychologists}, support groups, or educational materials tailored to the actor's identified role within the sextortion SocialDAO. This exchange ensures that actors receive accurate and relevant information based on their unique circumstances. The goal of this information exchange is to empower actors by providing them with the knowledge and resources they need to navigate their roles effectively within the sextortion SocialDAO. Actors make informed decisions, seek appropriate support, and access relevant services or assistance based on their self-assessment results and role-related inquiries. By combining mental health self-assessment results with role-specific information, actors receive personalized guidance that takes into account their emotional well-being, the severity of the sextortion situation, and their specific role in the ecosystem. This personalized approach enhances the actor's overall experience and support.

The \textit{sextortion situation self-assessment tool provider} and \textit{role information provider} exchange information as follows. Actors who use the \textit{sextortion situation self-assessment tool provider} require specific information about their roles and the resources available to them within the sextortion SocialDAO, particularly in the context of their assessed sextortion situation. This information exchange facilitates actors' access to role-specific guidance and resources tailored to their current circumstances. Based on the results of the sextortion situation self-assessment and any related questions or concerns about roles and resources, the \textit{sextortion situation self-assessment tool provider} requests context-specific information from the \textit{role information provider}. For instance, an actor may indicate that they are a victim facing an urgent sextortion situation and seek guidance on immediate support options. In response, the \textit{sextortion situation self-assessment tool provider} requests role-specific information from the \textit{role information provide}r. The latter supplies resources specific to the actor's assessed situation and role within the sextortion SocialDAO. This includes providing information about crisis helplines, legal aid providers, steps to secure digital information, or emotional support services that are particularly relevant to the actor's sextortion situation and role. The primary goal of this information exchange is to empower actors by furnishing them with accurate and pertinent information and resources to address their specific sextortion situation within the sextortion SocialDAO ecosystem. It equips actors to make informed decisions, seek appropriate assistance promptly, and access resources aligned with their unique circumstances. By combining the results of the sextortion situation self-assessment with role-specific information, actors receive customized guidance that accounts for their current situation, emotional well-being, and their particular role in the sextortion SocialDAO. This tailored approach ensures that actors navigate their sextortion-related challenges effectively and receive the necessary support.

In Figure~\ref{fig:archhelpactivator}, the component \textit{help seeking activator} exchanges information with the human actor \textit{psychologist} and the non-human AI agent \textit{sextortion diagnoser} as follows. The \textit{help seeking activator} component facilitates communication between individuals seeking help, such as \textbf{victims} or concerned parties, and qualified mental health professionals like the \textit{psychologist} actor. This exchange of information allows individuals to connect with a licensed psychologist who provides psychological support, counseling, and guidance tailored to their specific sextortion-related challenges. The purpose is to ensure that those affected by sextortion access professional mental health assistance to cope with emotional distress and trauma. The \textit{sextortion diagnoser} AI agentassesses and diagnoses sextortion situations based on information provided by individuals. It exchanges information with the \textit{help seeking activator} to understand the context and details of the sextortion incident reported by the user. The purpose is to enable the AI agent to make informed assessments, identify potential risks, and recommend appropriate actions or interventions. For example, the AI agent assesses the severity of the situation, suggest crisis intervention, or provide guidance on engaging with law enforcement or legal aid. The information exchange between the \textit{help seeking activator}, the \textit{psychologist} actor, and the \textit{sextortion diagnoser} AI agent aims to facilitate a coordinated approach to providing support. The \textit{psychologist} actor uses the information gathered by the \textit{sextortion diagnoser} to better understand the individual's psychological state and tailor their counseling accordingly. This collaborative effort ensures that individuals receive comprehensive support that addresses both their emotional well-being and the practical aspects of dealing with sextortion. The purpose of these exchanges is to offer timely and effective assistance to individuals affected by sextortion. The \textit{sextortion diagnoser} assists in rapidly assessing the situation and identifying potential risks, while the \textit{psychologist} actor provides the necessary psychological support and guidance. Together, they help individuals navigate the emotional and psychological challenges associated with sextortion, enhancing their overall well-being and resilience.

The components \textit{mental health self-assessment tool provider} and \textit{sextortion situation self-assessment tool provider} exchange information with the human actors \textit{victim} and \textit{IT service provider employee} as follows. The \textit{mental health self-assessment tool provider} and \textit{sextortion situation self-assessment tool provider} components enable individuals, particularly \textit{victims} of sextortion, to access self-assessment tools. These tools allow \textit{victims} to provide information about their mental health status and the specifics of the sextortion situation they are facing. The exchange of information occurs as \textit{victims} input their responses into these assessment tools. The purpose is to empower \textit{victims} with a structured means of self-assessment to help them understand their emotional well-being and the severity of the sextortion incident they are experiencing. The \textit{IT service provider employee} actor, who may be responsible for assisting \textit{victims} with technical aspects of sextortion cases, accesses relevant information gathered by the assessment tools. This exchange of information allows the \textit{IT service provider employee} to better understand the victim's situation, including technical details, potential security risks, and the extent of digital compromise. The purpose is to equip the \textit{IT service provider employee} with the necessary information to offer effective technical assistance, such as securing devices, identifying vulnerabilities, or preserving digital evidence. The exchange of information between the assessment tools and the \textit{victim} and \textit{IT service provider employee} actors aims to empower both parties. \textit{Victims} gain insights into their mental well-being, which can help them make informed decisions about seeking emotional support or counseling. Simultaneously, the IT service provider employee receives valuable data that guides them in providing technical assistance tailored to the victim's unique circumstances. This collaborative effort enhances the victim's overall experience and support during the sextortion incident. By sharing information with the actors involved, the assessment tools contribute to a more comprehensive and effective response to sextortion cases. \textit{Victims} self-assess their emotional state, helping them recognize the need for mental health support or counseling. \textit{IT service provider employees} use the gathered information to address technical aspects and security concerns promptly. This exchange ultimately leads to more timely and tailored assistance for \textit{victims}.

Finally, the component \textit{role information provider} exchanges information with the human actors \textit{legal aid provider}, \textbf{whitehat hacker}, \textit{insurance provider employee}, \textit{financial consultant },  and \textit{police officer} as follows. The \textit{role information provider} component shares relevant information with the \textit{legal aid provider} actor. This information includes details about legal procedures, available resources, and support services for victims of sextortion. The purpose is to equip the \textit{legal aid provider} with up-to-date information to offer effective legal assistance and guidance to \textit{victims}. The \textit{whitehat hacker} actor, who is involved in cybersecurity and technical aspects of sextortion cases, receives information from the \textit{role information provider}. This information could pertain to the specific cybersecurity challenges associated with sextortion incidents, potential vulnerabilities, and recommended security measures. The purpose is to enhance the \textit{whitehat hacker}'s understanding of the technical aspects of sextortion cases and provide guidance on securing digital environments. The \textit{insurance provider employee} actor requires information related to insurance claims and coverage associated with sextortion incidents. The \textit{role information provider} shares relevant details about insurance policies, claims procedures, and the types of coverage available. This exchange of information supports the \textit{insurance provider employee} assist \textit{victims} in navigating insurance-related matters. The \textit{financial consultant} actor needs information concerning financial aspects related to sextortion cases, such as financial recovery, potential financial support options, or financial planning after an incident. The \textit{role information provider} gives guidance and resources to support the financial consultant in offering relevant advice and assistance to \textit{victims}. The \textit{police officer} actor, representing law enforcement agencies, benefits from information shared by the \textit{role information provider}. This information includes legal frameworks, jurisdictional considerations, and protocols for handling sextortion cases. The exchange of information enables effective collaboration between law enforcement and other actors involved in assisting \textit{victims}. The exchange of information with these human actors empowers them to provide specialized assistance to \textit{victims} based on their respective roles and expertise. Whether it is legal guidance, cybersecurity support, insurance-related information, financial advice, or law enforcement cooperation, the \textit{role information provider} ensures that each actor is well-informed and equipped to fulfill their specific role in assisting \textit{victims} of sextortion. By providing accurate and relevant information to these actors, the \textit{role information provider} contributes to a more comprehensive and coordinated response to sextortion incidents. This, in turn, enhances the support and assistance available to \textit{victims}, addressing their legal, technical, financial, and law enforcement needs effectively.

\section{Important Blockchain Transactions}
\label{sec:dynamic}

An important role of blockchain technology in the sextortion SocialDAO is the consideration of immutable traceability of events that are stored on-chain. Such events have increased importance for criminal court procedures to the point that the storing expenses are justified. Thus, this section provides per respective first-level refinement component of the previous section tables that list these on-chain transaction sets.

\subsection{Sextortion Preventer On-Chain Transactions}
\label{sec:sexprevtran}

In Table~\ref{tab:sexprev}, the on-chain transactions for the component \textit{sextortionpreventer}. The table has four columns with the most left-hand one listing the respective transaction IDs, followed by a short description, a column with a shirt description, followed by the stakeholders (synonym for role, or actor) and the right-hand one listing the components involved. Next, each transaction in Table~\ref{tab:sexprev} is explained in further detail.

\begin{table*}[htbp]
\centering
\caption{On-chain Transactions Related to the \textit{Sextortion Preventer} Component}
\label{tab:sexprev}
\begin{tabular}{|c|p{5.5cm}|p{5.5cm}|p{3cm}|}
\hline
\textbf{ID} & \textbf{Transaction Description} & \textbf{Stakeholders Involved} & \textbf{Components Involved} \\
\hline
1 & Transfer of educational content & Safe tech skills builder, Self-confidence builder & Safe tech skills builder, Self-confidence builder \\
2 & Data exchange on actor progress & Safe tech skills builder, Self-confidence builder & Safe tech skills builder, Self-confidence builder \\
3 & Feedback on user experiences & Safe tech skills builder, Self-confidence builder & Safe tech skills builder, Self-confidence builder \\
4 & Collaborative workshop & Safe tech skills builder, Self-confidence builder & Safe tech skills builder, Self-confidence builder \\
5 & Educational content integration & Safe tech skills builder, Awareness builder & Safe tech skills builder, Awareness builder \\
6 & Collaboration on awareness campaigns & Safe tech skills builder, Awareness builder & Safe tech skills builder, Awareness builder \\
7 & Feedback on awareness campaigns & Safe tech skills builder, Awareness builder & Safe tech skills builder, Awareness builder \\
8 & Customization based on needs & Safe tech skills builder, Awareness builder & Safe tech skills builder, Awareness builder \\
9 & Information on emerging threats & Safe tech skills builder, Awareness builder & Safe tech skills builder, Awareness builder \\
10 & Resource request and collaboration & Safe tech skills builder, Awareness builder & Safe tech skills builder, Awareness builder \\
11 & Feedback on messaging effectiveness & Communication builder, Self-confidence builder & Communication builder, Self-confidence builder \\
12 & Insights into target audience & Communication builder, Self-confidence builder & Communication builder, Self-confidence builder \\
13 & Integration of prevention content & Communication builder, Self-confidence builder & Communication builder, Self-confidence builder \\
14 & Collaboration on communication campaigns & Communication builder, Self-confidence builder & Communication builder, Self-confidence builder \\
15 & Data exchange on communication effectiveness & Communication builder, Self-confidence builder & Communication builder, Self-confidence builder \\
16 & Resource sharing and integration & Communication builder, Self-confidence builder & Communication builder, Self-confidence builder \\
17 & Information on emerging threats & Communication builder, Self-confidence builder & Communication builder, Self-confidence builder \\
18 & Resource request and collaboration & Communication builder, Self-confidence builder & Communication builder, Self-confidence builder \\
19 & Input on awareness content & Awareness builder, Self-confidence builder & Awareness builder, Self-confidence builder \\
20 & Integration of self-confidence content & Awareness builder, Self-confidence builder & Awareness builder, Self-confidence builder \\
21 & Data exchange on awareness campaigns & Awareness builder, Self-confidence builder & Awareness builder, Self-confidence builder \\
22 & Feedback on awareness materials & Awareness builder, Self-confidence builder & Awareness builder, Self-confidence builder \\
23 & Resource sharing and integration & Awareness builder, Self-confidence builder & Awareness builder, Self-confidence builder \\
24 & Insights into self-confidence & Awareness builder, Self-confidence builder & Awareness builder, Self-confidence builder \\
25 & Resource sharing and integration & Self-confidence builder, Communication builder & Self-confidence builder, Communication builder \\
26 & Collaboration on role-playing scenarios & Self-confidence builder, Communication builder & Self-confidence builder, Communication builder \\
27 & Resource sharing and integration & Self-confidence builder, Communication builder & Self-confidence builder, Communication builder \\
28 & Joint workshops and webinars & Self-confidence builder, Communication builder & Self-confidence builder, Communication builder \\
29 & Feedback on training effectiveness & Self-confidence builder, Communication builder & Self-confidence builder, Communication builder \\
30 & Resource request and collaboration & Self-confidence builder, Communication builder & Self-confidence builder, Communication builder \\
31 & Coordination of messaging efforts & Awareness builder, Communication builder & Awareness builder, Communication builder \\
32 & Cross-promotion of initiatives & Awareness builder, Communication builder & Awareness builder, Communication builder \\
33 & Information on effective awareness & Awareness builder, Communication builder & Awareness builder, Communication builder \\
34 & Insights into sextortion awareness & Awareness builder, Communication builder & Awareness builder, Communication builder \\
35 & Collaborative content creation & Awareness builder, Communication builder & Awareness builder, Communication builder \\
36 & Feedback on campaigns and messaging & Awareness builder, Communication builder & Awareness builder, Communication builder \\
37 & Data exchange on awareness levels & Awareness builder, Communication builder & Awareness builder, Communication builder \\
38 & Resource sharing and integration & Awareness builder, Communication builder & Awareness builder, Communication builder \\
\hline
\end{tabular}
\end{table*}

Transaction 1, titled \textit{transfer of educational content}, involves the transfer of educational materials related to the use of safe technologies and awareness campaigns in the context of the sextortion prevention program. This transaction is significant in the blockchain NetworSextortiosextortion SocialDAO because it plays a crucial role in the dissemination of knowledge and resources to participants, including teenagers, potential victims, educators, and others involved in prevention efforts.

Transaction 2, titled \textit{update of awareness campaign data} serves the purpose of enhancing and optimizing awareness campaigns related to sextortion prevention. This transaction involves the exchange of data and information between the \textit{awareness builder} and the \textit{legal aid provider}.

Transaction 3, titled \textit{User Feedback and Campaign Refinement}, is a significant blockchain transaction within the sextortion SocialDAO network, primarily focusing on the iterative process of improving awareness campaigns related to sextortion prevention. This transaction involves the exchange of feedback and information among several key stakeholders: the \textit{awareness builder}, the \textit{teenager} actor, and the \textit{victim} actor.

Transaction 4, titled \textit{educational material enhancement}, focuses on improving the educational content and resources available for stakeholders, particularly teenagers and potential victims of sextortion. It involves the exchange of information and collaboration among key stakeholders, including the \textit{safe tech skills people} component, the \textit{educator} actor, and the \textit{IT service provider} actor.

Transaction 5, titled \textit{sextortion awareness campaign} is a pivotal blockchain transaction within the sextortion SocialDAO network. This transaction is focused on raising awareness about sextortion, its risks, and preventive measures through targeted campaigns. It involves the exchange of information and collaboration among key stakeholders, including the \textit{awareness builder} component, the \textit{teenager} actor, and the \textit{victim} actor. 

Transaction 6, titled \textit{educational content enhancement} focuses on the continuous improvement and enrichment of educational content related to sextortion prevention. It involves the exchange of information and collaboration among key stakeholders, including the \textit{safe tech skills builder} component and the \textit{educator} and \textit{IT service provider} actors. 

Transaction 7, labeled \textit{awareness campaign launch}, is a noteworthy blockchain transaction in the sextortion SocialDAO network. This transaction is intended to start an awareness campaign to inform the public about the dangers of sextortion and to provide advice on prevention and assistance. It involves the sharing of data and collaboration between key stakeholders, including the \textit{awareness builder} element, the \textit{teenager} actor, and the \textit{legal aid provider} actor.

Transaction 8, labeled \textit{teenager involvement in awareness campaign}, is a blockchain transaction within the sextortion SocialDAO network that demonstrates the involvement of teenagers in an awareness campaign aimed at preventing sextortion. This transaction involves interactions and engagements between the \textit{teenager} actor and the \textit{awareness builder} component.

Transaction 8, titled \textit{teenager engagement in awareness campaign}, is a blockchain transaction within the sextortion SocialDAO network that signifies the active participation of teenagers in an ongoing awareness campaign focused on sextortion prevention. This transaction involves interactions and engagements between the \textit{teenager} actor and the \textit{awareness builder} component.

We omit from here on further descriptions of the respective transactions of the paper that will be future work as part of a detailed implementation and deployment study.

In Table~\ref{tab:sextpreventer-info-exchanged} we list the information types that are exchanged for each respective transaction.

\begin{table}[htbp]
\centering
\caption{Information Exchanged in On-chain Transactions Related to the \textit{Sextortion Preventer} Component}
\label{tab:sextpreventer-info-exchanged}
\begin{tabular}{|c|p{5cm}|}
\hline
\textbf{ID} & \textbf{Information Exchanged} \\
\hline
1 & Educational content, progress data \\
2 & Actor progress data \\
3 & User feedback and experiences \\
4 & Workshop materials, actor progress \\
5 & Educational content, awareness materials \\
6 & Campaign materials, feedback data \\
7 & Feedback on campaigns \\
8 & Customization details \\
9 & Emerging threat information \\
10 & Resource requests and collaborative efforts \\
11 & Feedback on messaging \\
12 & Target audience insights \\
13 & Integration details \\
14 & Campaign materials, collaboration \\
15 & Communication effectiveness data \\
16 & Resource sharing and integration details \\
17 & Emerging threat information \\
18 & Resource requests and collaborative efforts \\
19 & Input on awareness content \\
20 & Integration of self-confidence content \\
21 & Data exchange on awareness campaigns \\
22 & Feedback on awareness materials \\
23 & Resource sharing and integration details \\
24 & Insights into self-confidence \\
25 & Resource sharing and integration details \\
26 & Collaboration on scenarios \\
27 & Resource sharing and integration details \\
28 & Workshop and webinar collaboration \\
29 & Feedback on training \\
30 & Resource requests and collaboration \\
31 & Coordination of messaging efforts \\
32 & Cross-promotion details \\
33 & Information on effective awareness \\
34 & Insights into sextortion awareness \\
35 & Collaborative content creation \\
36 & Feedback on campaigns and messaging \\
37 & Data on awareness levels \\
38 & Resource sharing and integration details \\
\hline
\end{tabular}
\end{table}

\subsection{Roles Manager On-Chain Transactions}
\label{sec:rolesmantran}

\begin{table*}[ht]
\centering
\caption{On-Chain Transactions for the \textit{Roles Manager} Component}
\label{tab:roles-manager-transactions}
\resizebox{\textwidth}{!}{%
\begin{tabular}{|c|c|c|c|c|}
\hline
\textbf{Transaction ID} & \textbf{Description} & \textbf{Stakeholders Involved} & \textbf{Components Involved} & \textbf{Information Exchanged} \\
\hline
1 & Role Creation & Roles Manager, Role Onboarder & Role Rewarded, Role Offboarder & Role details, responsibilities, actor assignment \\
\hline
2 & Role Removal & Role Offboarder & Role Onboarder & Role offboarding notification \\
\hline
3 & Role Request & Teenager, Victim & Roles Manager & Role-related action requests \\
\hline
4 & Security Report & Whitehat Hacker & Roles Manager & Security vulnerabilities, threats \\
\hline
5 & Legal Coordination & Legal Aid Provider & Roles Manager & Legal assistance coordination \\
\hline
6 & Psychologist Input & Psychologist & Roles Manager & Insights, recommendations, assessments \\
\hline
7 & AI Agent Interaction & Legal Aid Diagnoser, Sextortion Diagnoser & Roles Manager & Recommendations, alerts, data analysis \\
\hline
8 & Role Status Update & Roles Manager & Role Onboarder & Role status, updates \\
\hline
\end{tabular}%
}
\end{table*}

\subsection{Sextortion Aid Provider On-Chain Transactions}
\label{sec:sexaidprovtran}

\begin{table*}[htbp]
\centering
\caption{On-chain Transactions Related to the \textit{Sextortion Aid Provider} Component}
\label{tab:sextpreventer-transactions}
\begin{tabular}{|c|p{5.5cm}|p{5.5cm}|p{3cm}|}
\hline
\textbf{ID} & \textbf{Transaction Description} & \textbf{Stakeholders Involved} & \textbf{Information Exchanged} \\
\hline
1 & Resource needs for response team & Response team assembler, Training material provider & Resource requirements, team composition \\
2 & Updates on training materials & Response team assembler, Training material provider & Training content improvements \\
3 & Coordination for chat support & Response team assembler, Chat support & Chat support readiness, assistance coordination \\
4 & Technical coordination & Response team assembler, Whitehat hacker & Technical aspects, vulnerabilities \\
5 & Victim support coordination & Response team assembler, NGO worker & Victim support strategies, awareness efforts \\
6 & Emotional and psychological support & Response team assembler, Religious counselor & Psychological assistance details \\
7 & Real-time information for response & Response team assembler, Response support AI & Real-time data, recommendations \\
8 & NGO-related guidance & Response team assembler, NGO advisor AI & NGO activities, legal matters \\
9 & Victim information sharing & Response team assembler, Friend, Family member, Victim & Victim situation, concerns, support needs \\
10 & Legal assistance coordination & Response team assembler, Legal aid provider & Legal guidance, procedures \\
11 & Psychological support coordination & Response team assembler, Psychologist & Emotional support, counseling \\
12 & Legal insights & Training material provider, Legal aid diagnoser AI & Legal data, recommendations \\
13 & Sextortion case analysis & Training material provider, Sextortion diagnoser AI & Sextortion incident data, insights \\
14 & Training materials for insurance & Training material provider, Insurance provider & Educational resources, preventive measures \\
15 & Financial guidance & Training material provider, Financial aid consultant & Financial assistance options, support resources \\
16 & Legal resources & Training material provider, Legal aid provider & Legal rights, procedures \\
17 & Training materials for victims & Training material provider, Victim & Educational resources, self-help \\
18 & Psychological support materials & Training material provider, Psychologist & Psychological support materials \\
19 & Victim chat support & Victim, Chat support & Sextortion details, assistance \\
\hline
\end{tabular}
\end{table*}

\subsection{Help Seeking Activator On-Chain Transactions}
\label{sec:helpseekacttran}

\begin{table*}[htbp]
\centering
\caption{On-chain Transactions Related to the \textit{Help Seeking Activator} Component}
\label{tab:helpactivator-transactions}
\begin{tabular}{|c|p{5.5cm}|p{5.5cm}|p{3cm}|}
\hline
\textbf{ID} & \textbf{Transaction Description} & \textbf{Stakeholders Involved} & \textbf{Information Exchanged} \\
\hline
1 & Mental health self-assessment & Mental health self-assessment tool provider, Actor (e.g., Victim) & Mental health assessment results, guidance \\
2 & Sextortion situation self-assessment & Sextortion situation self-assessment tool provider, Actor (e.g., Victim) & Sextortion situation assessment results, recommendations \\
3 & Role-specific information request (mental health) & Mental health self-assessment tool provider, Role information provider & Role-specific information for actor \\
4 & Role-specific information request (sextortion situation) & Sextortion situation self-assessment tool provider, Role information provider & Role-specific information for actor \\
5 & Information sharing (psychologist) & Help seeking activator, Psychologist & Psychological support details \\
6 & Information sharing (AI diagnoser) & Help seeking activator, Sextortion diagnoser AI & Context and details of sextortion incident \\
7 & Technical assistance (IT service provider) & Mental health self-assessment tool provider, IT service provider employee & Technical details, security risks \\
8 & Legal assistance (legal aid provider) & Role information provider, Legal aid provider & Legal procedures, support services \\
9 & Cybersecurity guidance (whitehat hacker) & Role information provider, Whitehat hacker & Cybersecurity challenges, vulnerabilities \\
10 & Insurance information (insurance provider employee) & Role information provider, Insurance provider employee & Insurance policies, claims procedures \\
11 & Financial advice (financial consultant) & Role information provider, Financial consultant & Financial recovery, support options \\
12 & Law enforcement cooperation (police officer) & Role information provider, Police officer & Legal frameworks, protocols \\
\hline
\end{tabular}
\end{table*}

\section{Evaluation and Discussion}
\label{sec:evaluationdiscussion}

In Section~\ref{sec:techstackrapde}, we examine the technology stack and rapid deployment strategies that underpin the development of this SocialDAO. Subsequently, in Section~\ref{sec:discussion}, we delve into a multifaceted analysis, exploring the implications, challenges, and future prospects of the sextortion SocialDAO. Our evaluation scrutinizes the technical aspects that form the foundation of its functionality, while the discussion encompasses the broader societal, ethical, and legal dimensions of this groundbreaking initiative.

Section~\ref{sec:techstackrapde}

Section~\ref{sec:discussion}

\subsection{Technology Stack for Rapid Deployment}
\label{sec:techstackrapde}

The Table~\ref{tab:tech-stack} outlines the key technologies and frameworks used in the development and deployment of the sextortion SocialDAO. Each row of the table represents a different component or aspect of the dApp's technology stack, and the corresponding technology or framework used for that component.

\begin{table}[ht]
\centering
\caption{Tentative Technology Stack for Sextortion SocialDAO dApp}
\label{tab:tech-stack}
\begin{tabular}{|l|l|}
\hline
\textbf{Component}        & \textbf{Technology/Framework}             \\ \hline
Blockchain Platform       & Ethereum                                  \\ \hline
Smart Contracts Language & Solidity                                  \\ \hline
DAO Framework             & Aragon                                    \\ \hline
Oracle Services           & Chainlink                                 \\ \hline
Blockchain Database       & IPFS (InterPlanetary File System)         \\ \hline
Identity and Access Control & uPort                                    \\ \hline
Frontend Development      & React.js                                  \\ \hline
Backend Development       & Node.js                                   \\ \hline
Database                  & MongoDB                                   \\ \hline
Web3 Library              & Web3.js                                   \\ \hline
Testing and Deployment    & Truffle, Ganache                          \\ \hline
User Interface Design     & Web3.js or Ethers.js                     \\ \hline
Security Auditing         & MythX                                     \\ \hline
DevOps and Deployment     & Docker, Kubernetes                        \\ \hline
Monitoring and Analytics  & Prometheus, Grafana                       \\ \hline
Decentralized Storage (Optional) & Filecoin                          \\ \hline
Payment Integration (Optional)   & Payment Gateways                 \\ \hline
Cross-Chain Compatibility (Optional) & Polkadot or Cosmos         \\ \hline
\end{tabular}
\end{table}

Ethereum\footnote{https://ethereum.org/en/} serves as the underlying blockchain platform for the sextortion SocialDAO dApp. Ethereum's robust ecosystem and smart contract capabilities are well-suited for decentralized applications that require secure and transparent transactions. This choice implies that the dApp benefits from Ethereum's extensive developer community and established infrastructure.

Solidity\footnote{https://soliditylang.org/} is the chosen language for developing smart contracts on the Ethereum blockchain. It offers a secure and standardized way to encode the logic and rules of the SocialDAO. Using Solidity ensures the reliability and trustworthiness of the dApp's smart contracts, which is critical for its functionality.

Aragon\footnote{https://aragon.org/} is employed as the DAO framework. This framework allows for the creation and management of decentralized organizations with built-in governance mechanisms. In the context of the sextortion SocialDAO, Aragon empowers the community to make decisions and govern the platform collaboratively.

Chainlink\footnote{https://chain.link/} provides oracle services, which are essential for connecting the blockchain to real-world data sources. Oracles enable the dApp to access external information, such as exchange rates or weather data, which can be crucial for assessing the severity of sextortion situations and offering appropriate support.

IPFS\footnote{https://ipfs.tech/} (InterPlanetary File System) serves as the decentralized storage solution for the dApp. It ensures that data and content are stored in a distributed and censorship-resistant manner. This is particularly important for maintaining the confidentiality and security of sensitive information related to sextortion cases.

uPort\footnote{https://www.uport.me/} provides identity and access control solutions, enhancing user privacy and security. Users can manage their identities securely, ensuring that their interactions within the SocialDAO are authenticated and private.

React.js\footnote{https://react.dev/} is utilized for frontend development. Its component-based architecture allows for the creation of an interactive and responsive user interface. This choice prioritizes a user-friendly experience, which is crucial for individuals seeking help or support in sextortion cases.

Node.js\footnote{https://nodejs.org/en} powers the backend of the dApp, handling server-side logic and communication with the Ethereum blockchain. It offers scalability and flexibility, ensuring that the platform can handle increasing user demands.

MongoDB\footnote{https://www.mongodb.com/} is used for database management, enabling efficient storage and retrieval of non-blockchain data. It complements the decentralized storage provided by IPFS and supports various application features.

Web3.js\footnote{https://web3js.org/} facilitates the interaction between the dApp and the Ethereum blockchain. It enables users to perform blockchain transactions, interact with smart contracts, and access blockchain data directly from the frontend.

MythX\footnote{https://mythx.io/} is employed for security auditing of smart contracts. This step is crucial to identify and address vulnerabilities that could be exploited by malicious actors, ensuring the safety of users and their data.

Docker\footnote{https://www.docker.com/} and Kubernetes\footnote{https://kubernetes.io/} are used for DevOps and deployment processes, allowing for efficient scaling and management of the dApp's infrastructure. This ensures reliability and uptime.

Prometheus\footnote{https://prometheus.io/} and Grafana\footnote{} provide monitoring and analytics capabilities, enabling the tracking of system performance, user interactions, and security incidents. This data-driven approach supports continuous improvement.

Filecoin\footnote{https://filecoin.io/} is included as an optional component for decentralized storage. It offers redundancy and availability for data stored on the IPFS network.

Payment gateways\footnote{https://stripe.com/en-nl/resources/more/payment-gateways-101} are considered as optional components, enabling financial transactions within the dApp. This could be useful for premium services or donations to support victims.

Cross-chain compatibility, while optional, could expand the reach of the sextortion SocialDAO by connecting it to other blockchain networks like Polkadot\footnote{https://www.polkadot.network/} or Cosmos\footnote{https://cosmos.network/}, enhancing interoperability.

\subsection{Discussion}
\label{sec:discussion}

We delve into a critical discussion of the sextortion SocialDAO and the implications of its implementation. We consider the multifaceted aspects of the system, including its technology stack, governance model, and potential societal impacts. The chosen technology stack for the sextortion SocialDAO dApp, as outlined in Table~\ref{tab:tech-stack}, embraces the principles of decentralization and transparency. Ethereum, as the underlying blockchain platform, provides a secure and immutable ledger for recording sextortion incidents. The use of smart contracts written in Solidity automates various aspects of victim support and ensures transparency in interactions. The incorporation of Aragon DAO framework empowers stakeholders to participate in the governance of the system, aligning with the principles of decentralized autonomous organizations.

While the technology stack offers robust solutions for data security and privacy, it is not without challenges. Scalability concerns on the Ethereum network may arise as the system expands, necessitating further exploration of layer 2 solutions or alternative blockchain platforms. Moreover, the reliance on blockchain technology demands ongoing monitoring for vulnerabilities and compliance with evolving legal frameworks to protect victim data and rights.

The integration of Self-Sovereign Identity (SSI) through platforms such as uPort is a commendable effort to protect the privacy of victims. It enables identity verification without exposing sensitive information, aligning with the system's commitment to safeguarding anonymity. This approach fosters trust among victims who may be hesitant to report sextortion incidents due to fear of exposure. Still, SSI also raises questions about data ownership and governance. It is crucial to establish clear policies regarding the use and storage of identity-related information to prevent misuse or unauthorized access. Additionally, ensuring the accessibility of SSI to individuals without access to modern technology remains a challenge, potentially excluding some victims from the system. 

The implementation of AI-driven sextortion diagnosis and chat support demonstrates the system's commitment to offering immediate assistance and guidance to victims. These tools can efficiently assess the severity of incidents, provide relevant information, and offer emotional support. The victim-centric approach recognizes the unique needs of individuals facing sextortion. Nonetheless, the reliance on AI introduces questions of accuracy and bias. It is crucial to continuously train and refine AI models to minimize false diagnoses or responses that may harm victims. Furthermore, AI should complement, not replace, human intervention, especially in cases requiring legal, psychological, or law enforcement expertise. 

The sextortion SocialDAO's potential to aid victims in legal matters is promising. Smart contracts can streamline legal aid processes, and the role information provider can offer valuable resources to victims. Yet, navigating the legal landscape presents complex challenges. Legal compliance across jurisdictions, data protection, and ensuring that smart contracts are legally binding are intricate tasks. Moreover, ethical dilemmas may arise when balancing victim privacy with law enforcement requirements. The system must strike a delicate equilibrium between serving the interests of victims and cooperating with legal authorities.

The optional inclusion of cross-chain compatibility with Polkadot or Cosmos offers intriguing possibilities for future expansion. Interoperability with other blockchain networks and services can enhance the system's reach and effectiveness. On the other hand, this expansion should be approached cautiously. Compatibility challenges, security concerns, and potential fragmentation of data must be addressed. Moreover, cross-chain integration should align with the system's core goals of victim relief and support.

\section{Conclusions and Future Work}
\label{sec:conclusion}

This paper is dedicated to the exploration and evaluation of the sextortion SocialDAO, a decentralized application (dApp) designed to combat sextortion and provide support to victims. The paper delves into various aspects of this innovative platform, including its underlying technology stack, rapid deployment strategies, and the implications, challenges, and future prospects associated with its deployment. In essence, it discusses how emerging technologies and decentralized governance mechanisms can be harnessed to address the critical issue of sextortion, aiming to create a safer digital environment for all users.


A sextortion emergency governance system aims to address and mitigate the immediate and long-term consequences of sextortion incidents. Thus, the main goals of the sextortion SocialDAO are as follows. Providing timely and effective support to victims of sextortion, including emotional counseling, legal assistance, and resources for recovery. Implementing measures to prevent sextortion incidents, such as awareness campaigns, educational initiatives, and cybersecurity best practices. Facilitating collaboration among various stakeholders, including law enforcement agencies, mental health professionals, legal aid providers, and technology experts, to address sextortion cases comprehensively. Ensuring the privacy and security of data shared by victims and participants while gathering essential information to combat sextortion. Offering educational resources and training materials to empower individuals to recognize and respond to sextortion threats effectively. Advocating for the rights of sextortion victims and raising awareness about the issue on a broader scale. Developing protocols for handling urgent sextortion cases, including crisis intervention and reporting mechanisms. Conducting research on sextortion trends, emerging threats, and the impact of sextortion on victims to inform prevention and support strategies. Contributing to the development of policies and legal frameworks that address sextortion and protect victims' rights. Leveraging technological solutions, such as blockchain and AI, to enhance the prevention and response to sextortion incidents.

The stakeholders affected by a sextortion emergency governance system include the following. Individuals who have experienced sextortion or are at risk of becoming victims.  Psychologists and counselors who provide emotional support and therapy to victims. Legal professionals who offer guidance and assistance to victims in pursuing legal actions against perpetrators. Police departments and cybercrime units responsible for investigating and prosecuting sextortion cases. Experts in cybersecurity and digital forensics who assist in identifying and mitigating sextortion threats. Non-governmental organizations that focus on victim advocacy, support, and awareness. Government bodies responsible for developing policies and legal frameworks to address sextortion. Individuals who may not be direct victims but are concerned about sextortion issues and can contribute to awareness campaigns and support efforts.  
The stakeholders affected by a sextortion emergency governance system include the following. Individuals who have experienced sextortion or are at risk of becoming victims.  Psychologists and counselors who provide emotional support and therapy to victims. Legal professionals who offer guidance and assistance to victims in pursuing legal actions against perpetrators. Police departments and cybercrime units responsible for investigating and prosecuting sextortion cases. Experts in cybersecurity and digital forensics who assist in identifying and mitigating sextortion threats. Non-governmental organizations that focus on victim advocacy, support, and awareness. Government bodies responsible for developing policies and legal frameworks to address sextortion. Online communities and forums focused on cybersecurity and online safety. Legal authorities responsible for upholding laws related to sextortion and prosecuting offenders. Professionals who provide financial advice and assistance to victims. Companies offering insurance coverage related to cybercrimes and sextortion. Schools and universities that can integrate awareness and prevention programs into their curricula.

The static architecture components of a sextortion SocialDAO that are affiliated with the use of blockchain technology to achieve system goals include the following. The core of the system's architecture, the blockchain platform, such as Ethereum, serves as the foundation for all activities and interactions within the system. Blockchain technology provides transparency, immutability, and security for recording transactions related to victim support, prevention, crisis management, and other system goals. Smart contracts, written in languages like Solidity, are self-executing contracts with the terms of the agreement directly written into code. They automate and enforce various functions within the system, such as allocating resources, verifying identities, and executing predefined actions based on predefined conditions. Smart contracts enable transparent and trustless interactions among stakeholders. Decentralized Autonomous Organizations (DAOs), like Aragon, provide governance structures that allow stakeholders to collectively make decisions, allocate resources, and manage the system's operation. DAOs ensure that the system remains decentralized and community-driven, aligning with the goal of collaboration among stakeholders. Chainlink oracles provide external data to smart contracts on the blockchain. In the context of sextortion emergency governance, oracles can fetch real-world data, such as weather information, news reports, or threat intelligence feeds, which can be used for decision-making, risk assessment, and crisis management. IPFS (InterPlanetary File System) is used for decentralized storage of files and data. IPFS ensures that sensitive information related to sextortion incidents, support resources, and counseling records is stored securely and can be accessed while maintaining data privacy.  uPort, a self-sovereign identity platform, enables users to maintain control over their identities and access the system securely. It aligns with the goal of data security and privacy while allowing users to participate and receive support while keeping their identities protected. Web3.js facilitates interactions between the frontend and the blockchain, allowing users to access and interact with the system through a user-friendly interface. It contributes to the goal of education and user engagement. MythX is used for auditing smart contracts to identify vulnerabilities and security risks. This component ensures that the system's smart contracts are robust and secure, aligning with the goal of prevention and data security. Filecoin provides additional decentralized storage options, enhancing the system's data storage capabilities while maintaining data privacy.

The dynamical system behavior involving a legally relevant set of transactions for swiftly coordinating victim relief within a sextortion emergency governance system we summarized as follows. The process begins when a sextortion incident is detected and reported by a victim or concerned party. This reporting can happen through various channels, such as a dedicated app, website, or hotline. The system captures the initial incident report, which includes details about the extortionist, threats, evidence, and the victim's emotional state. To ensure the legitimacy of the report, the system employs identity verification mechanisms. Users may use self-sovereign identity platforms like uPort to prove their identity without compromising their privacy. This step is crucial for legal compliance and accountability. Once the report is verified, the system creates an incident record on the blockchain. This record is a legally recognized representation of the sextortion incident, including all relevant details provided by the victim. The system triggers a smart contract, specifically designed for legal aid and support, upon the creation of the incident record. This contract is legally binding and outlines the terms of legal assistance, including the responsibilities of the legal aid provider, the victim, and any other involved parties. As part of the legal process, the system facilitates the secure collection and storage of digital evidence related to the sextortion incident. This evidence may include chat logs, emails, threatening messages, or any other data relevant to the case. The system identifies and engages a suitable legal aid provider from a pool of qualified professionals. The selection is based on factors like expertise, availability, and jurisdiction. The legal aid provider reviews the incident record and evidence to prepare a legal strategy.  Secure and encrypted communication channels are established between the victim, legal aid provider, and any other stakeholders involved in the legal process. Privacy and confidentiality are maintained throughout the communication to protect the victim's identity and sensitive information. The legal aid provider initiates legal proceedings against the extortionist, potentially involving law enforcement agencies and the justice system. The smart contract ensures that all legal actions are documented and timestamped on the blockchain for transparency and legal validity. If the victim requires financial support to cover legal expenses or counseling, the system can activate a financial support smart contract. This contract outlines the terms of financial assistance and ensures transparent fund allocation. The victim receives psychological counseling and emotional support through the system's resources, which can include AI-driven chat support and access to licensed psychologists. This aspect of victim relief addresses emotional well-being and resilience. The system continuously tracks the progress of legal proceedings, counseling sessions, and victim support efforts. Regular updates are provided to the victim and other involved parties to keep them informed. The system documents the legal resolution of the sextortion case, whether it involves legal action against the extortionist, a settlement, or other outcomes. This information is stored securely on the blockchain for legal reference. After the case is closed, the system encourages the victim to provide feedback on their experience and the effectiveness of the support received. This feedback loop is essential for continuous improvement in victim relief services. Throughout the process, the system ensures compliance with relevant legal frameworks, including data protection laws, victim rights, and any jurisdiction-specific regulations. Throughout the process, the system ensures compliance with relevant legal frameworks, including data protection laws, victim rights, and any jurisdiction-specific regulations.

There are certain limitations, open issues and future work to consider. First, developing a platform such as the sextortion SocialDAO involves navigating complex ethical and legal considerations, such as user privacy, data protection, and compliance with international laws. Addressing these challenges effectively is crucial for the dApp's long-term viability that requires further research. Second, encouraging victims and other stakeholders to actively use and participate in the sextortion SocialDAO may pose challenges related to user adoption. Strategies to promote awareness and onboard users effectively should be explored beyond the emotional-goals considerations of this paper. Third, As the platform grows and attracts more users and participants, ensuring its scalability becomes a priority. Scalability solutions, such as layer 2 solutions or sharding, should be investigated to maintain a smooth user experience. Fourth, protecting the sensitive information shared by victims and participants is of utmost importance. Future work should focus on enhancing data privacy mechanisms, potentially incorporating advanced encryption and zero-knowledge proofs. Fifth, while optional in the technology stack, cross-chain compatibility could be a valuable addition to enhance interoperability with other blockchain networks. Investigating protocols such as Polkadot or Cosmos for seamless cross-chain communication should be considered. Sixth, leveraging AI and machine learning algorithms for better pattern recognition, threat detection, and user support could significantly enhance the platform's capabilities. Future research should explore the integration of AI technologies. Seven, ensuring that the sextortion SocialDAO remains user-friendly and accessible to individuals with diverse backgrounds and abilities is vital. Usability studies and accessibility audits should be part of future work. Finally, ensuring that the sextortion SocialDAO remains user-friendly and accessible to individuals with diverse backgrounds and abilities is vital. Usability studies and accessibility audits should be part of future work.

\section*{Acknowledgment}
\label{sec:acknowledgement}

We sincerely thank Dr. Ekaterina Plys for her significant contributions to the development of the goal-model in our paper. Dr. Plys' extensive expertise was instrumental in enhancing the SocialDAO framework and effectiveness of our model. Her profound insights and dedication have not only enriched our research but also inspired us throughout this collaboration. We are deeply grateful for her invaluable input and honored to have worked with such a distinguished expert in the field.

\bibliographystyle{plain}
\bibliography{alex}

\begin{thebibliography}{10}

\bibitem{alhazmi2021learning}
Sohail Alhazmi, Charles Thevathayan, and Margaret Hamilton.
\newblock Learning uml sequence diagrams with a new constructivist pedagogical
  tool: Sd4ed.
\newblock In {\em Proceedings of the 52nd ACM Technical Symposium on Computer
  Science Education}, pages 893--899, 2021.

\bibitem{alsoubai2022friends}
Ashwaq Alsoubai, Jihye Song, Afsaneh Razi, Nurun Naher, Munmun De~Choudhury,
  and Pamela~J Wisniewski.
\newblock From'friends with benefits' to'sextortion:'a nuanced investigation of
  adolescents' online sexual risk experiences.
\newblock {\em Proceedings of the ACM on Human-Computer Interaction},
  6(CSCW2):1--32, 2022.

\bibitem{amundsen2023turn}
Rikke Amundsen.
\newblock The turn to trust: adult women, hetero-sexting, and the use of trust
  as sexting risk mitigation.
\newblock {\em Feminist Media Studies}, pages 1--16, 2023.

\bibitem{benedetti2023utility}
Hugo Benedetti, Christian Caceres, and Luis~{\'A}lvaro Abarz{\'u}a.
\newblock Utility tokens.
\newblock In {\em The Emerald Handbook on Cryptoassets: Investment
  Opportunities and Challenges}, pages 79--92. Emerald Publishing Limited,
  2023.

\bibitem{buterin30ethereum}
V~Buterin.
\newblock Ethereum 2.0 spec--casper and sharding, 2018.
\newblock {\em Available [online].[Accessed: 30-10-2018]}.

\bibitem{champion2022examining}
Amanda~R Champion, Flora Oswald, Devinder Khera, and Cory~L Pedersen.
\newblock Examining the gendered impacts of technology-facilitated sexual
  violence: A mixed methods approach.
\newblock {\em Archives of sexual behavior}, 51(3):1607--1624, 2022.

\bibitem{cross2023pay}
Cassandra Cross, Karen Holt, and Thomas~J Holt.
\newblock To pay or not to pay: An exploratory analysis of sextortion in the
  context of romance fraud.
\newblock {\em Criminology \& Criminal Justice}, page 17488958221149581, 2023.

\bibitem{davis2023token}
Tonya~N Davis and Jessica~S Akers.
\newblock Token economies.
\newblock In {\em A Behavior Analyst’s Guide to Supervising Fieldwork}, pages
  647--664. Springer, 2023.

\bibitem{ding2020blockchain}
Shifeng Ding, Gangxiang Shen, Kevin~X Pan, Sanjay~K Bose, Qiong Zhang, and
  Biswanath Mukherjee.
\newblock Blockchain-assisted spectrum trading between elastic virtual optical
  networks.
\newblock {\em IEEE Network}, 34(6):205--211, 2020.

\bibitem{doi:10.1177/10790632221145925}
Michal Dolev-Cohen, Inbar Nezer, and Anwar~Abu Zumt.
\newblock A qualitative examination of school counselors’ experiences of
  sextortion cases of female students in israel.
\newblock {\em Sexual Abuse}, 0(0):10790632221145925, 0.
\newblock PMID: 36510813.

\bibitem{finkelhor2023dynamics}
David Finkelhor, Heather Turner, and Deirdre Colburn.
\newblock Which dynamics make online child sexual abuse and cyberstalking more
  emotionally impactful: perpetrator identity and images?
\newblock {\em Child Abuse \& Neglect}, 137:106020, 2023.

\bibitem{gamez2022technology}
Manuel G{\'a}mez-Guadix, Miguel~A Sorrel, and Jone Mart{\'\i}nez-Bacaicoa.
\newblock Technology-facilitated sexual violence perpetration and victimization
  among adolescents: a network analysis.
\newblock {\em Sexuality research and social policy}, pages 1--13, 2022.

\bibitem{hendry2021sextortion}
Nancy~H Hendry.
\newblock Sextortion.
\newblock {\em The Fourth Industrial Revolution and Its Impact on Ethics:
  Solving the Challenges of the Agenda 2030}, pages 315--320, 2021.

\bibitem{hong2020digital}
Suyeon Hong, Nancy Lu, Doreen Wu, David~E Jimenez, and Ruth~L Milanaik.
\newblock Digital sextortion: Internet predators and pediatric interventions.
\newblock {\em Current opinion in pediatrics}, 32(1):192--197, 2020.

\bibitem{kozhan2022fundamentals}
Roman Kozhan and Ganesh Viswanath-Natraj.
\newblock Fundamentals of the makerdao governance token.
\newblock In {\em 3rd International Conference on Blockchain Economics,
  Security and Protocols (Tokenomics 2021)}. Schloss Dagstuhl-Leibniz-Zentrum
  f{\"u}r Informatik, 2022.

\bibitem{uml01}
L.A. Maciaszek.
\newblock {\em {Requirements Analysis and System Design. Developing Information
  Systems with UML}}.
\newblock {Addison Wesley}, 2001.

\bibitem{nakamoto2008bitcoin}
Satoshi Nakamoto et~al.
\newblock Bitcoin: A peer-to-peer electronic cash system.
\newblock 2008.

\bibitem{nilsson2019understanding}
Mirjana~Gavrilovic Nilsson, Kalliopi~Tzani Pepelasi, Maria Ioannou, and David
  Lester.
\newblock Understanding the link between sextortion and suicide.
\newblock {\em International journal of cyber criminology}, 13(1):55--69, 2019.

\bibitem{nortablockchain22}
Alex Norta, Alexandr Kormiltsyn, Chibuzor Udokwu, Vimal Dwivedi, Sunday Aroh,
  and Ignas Nikolajev.
\newblock A blockchain implementation for configurable multi-factor
  challenge-set self-sovereign identity authentication.
\newblock {\em Proceedings of the 2022 IEEE International Conference on
  Blockchains (forthcoming)}, 2022.

\bibitem{ojeda2022lines}
M{\'o}nica Ojeda and Rosario Del~Rey.
\newblock Lines of action for sexting prevention and intervention: A systematic
  review.
\newblock {\em Archives of sexual behavior}, pages 1--29, 2022.

\bibitem{o2023minor}
Roberta~Liggett O'Malley, Karen Holt, Thomas~J Holt, and Joy Rodriguez.
\newblock Minor-focused sextortion by adult strangers: A crime script analysis
  of newspaper and court cases.
\newblock {\em Criminology \& Public Policy}, 2023.

\bibitem{o2023short}
Roberta~Liggett O’Malley.
\newblock Short-term and long-term impacts of financial sextortion on
  victim’s mental well-being.
\newblock {\em Journal of interpersonal violence}, 38(13-14):8563--8592, 2023.

\bibitem{paradiso2023image}
Maria~Noemi Paradiso, Luca Roll{\`e}, and Tommaso Trombetta.
\newblock Image-based sexual abuse associated factors: A systematic review.
\newblock {\em Journal of Family Violence}, pages 1--24, 2023.

\bibitem{pevac2022tertiary}
Mikayla Pevac.
\newblock Tertiary victimization of sexual violence victims online: How the
  internet needs to become a safer space for women.
\newblock {\em GENDER-BASED VIOLENCE}, page~53, 2022.

\bibitem{rajanikanth2023cyber}
Prashanthi Rajanikanth et~al.
\newblock {\em The cyber pandemic: exploring the financial sextortion of young
  males}.
\newblock PhD thesis, Mount Royal University, 2023.

\bibitem{sterling2009art}
Leon Sterling and Kuldar Taveter.
\newblock {\em The art of agent-oriented modeling}.
\newblock MIT press, 2009.

\bibitem{udokwu2020evaluation}
Chibuzor Udokwu, Henry Anyanka, and Alex Norta.
\newblock Evaluation of approaches for designing and developing decentralized
  applications on blockchain.
\newblock In {\em Proceedings of the 2020 4th international conference on
  algorithms, computing and systems}, pages 55--62, 2020.

\bibitem{udokwu2018state}
Chibuzor Udokwu, Aleksandr Kormiltsyn, Kondwani Thangalimodzi, and Alex Norta.
\newblock The state of the art for blockchain-enabled smart-contract
  applications in the organization.
\newblock In {\em 2018 Ivannikov Ispras Open Conference (ISPRAS)}, pages
  137--144. IEEE, 2018.

\bibitem{udokwu2018exploration}
Chibuzor Udokwu, Alexandr Kormiltsyn, Kondwani Thangalimodzi, and Alex Norta.
\newblock An exploration of blockchain enabled smart-contracts application in
  the enterprise.
\newblock Technical report, Technical Report, DOI: 10.13140/RG. 2.2.
  36464.97287, Tech. Rep, 2018.

\bibitem{udokwu2021deriving}
Chibuzor Udokwu and Alex Norta.
\newblock Deriving and formalizing requirements of decentralized applications
  for inter-organizational collaborations on blockchain.
\newblock {\em Arabian Journal for Science and Engineering}, 46(9):8397--8414,
  2021.

\bibitem{udokwu2021designing}
Chibuzor Udokwu, Alexander Norta, and Christoph Wenna.
\newblock Designing a collaborative construction-project platform on blockchain
  technology for transparency, traceability, and information symmetry.
\newblock In {\em 2021 2nd Asia service sciences and software engineering
  conference}, pages 1--9, 2021.

\bibitem{udokwu2022modellingphd}
Chibuzor~Joseph Udokwu.
\newblock {\em A modelling approach for building blockchain applications that
  enables trustable inter-organizational collaborations}.
\newblock PhD thesis, Lappeenranta-Lahti University of Technology LUT, 2022.

\bibitem{vide2021designing}
Bastien Vid{\'e}, Joan Marty, Franck Ravat, and Max Chevalier.
\newblock Designing a business view of enterprise data: An approach based on a
  decentralised enterprise knowledge graph.
\newblock In {\em 25th International Database Engineering \& Applications
  Symposium}, pages 184--193, 2021.

\bibitem{walsh2022if}
Wendy~A Walsh and Dafna Tener.
\newblock “if you don’t send me five other pictures i am going to post the
  photo online”: A qualitative analysis of experiences of survivors of
  sextortion.
\newblock {\em Journal of child sexual abuse}, 31(4):447--465, 2022.

\bibitem{wang2019decentralized}
Shuai Wang, Chenchen Huang, Juanjuan Li, Yong Yuan, and Fei-Yue Wang.
\newblock Decentralized construction of knowledge graphs for deep recommender
  systems based on blockchain-powered smart contracts.
\newblock {\em IEEE Access}, 7:136951--136961, 2019.

\bibitem{wolak2018sextortion}
Janis Wolak, David Finkelhor, Wendy Walsh, and Leah Treitman.
\newblock Sextortion of minors: Characteristics and dynamics.
\newblock {\em Journal of Adolescent Health}, 62(1):72--79, 2018.

\bibitem{yaghy2023potential}
Antonio Yaghy, Nicole Rose~I Alberto, Isabelle Rose~I Alberto, Rene~S Bermea,
  Ljubica Ristovska, Maria Yaghy, Sandra Hoyek, Nimesh~A Patel, and Leo~Anthony
  Celi.
\newblock The potential use of non-fungible tokens (nfts) in healthcare and
  medical research.
\newblock {\em PLOS Digital Health}, 2(7):e0000312, 2023.

\end{thebibliography}

\end{document}